\documentclass[12pt]{article}
\usepackage{amssymb,amsmath,bm}
\usepackage{epsf}
\usepackage{epsfig}
\usepackage{afterpage}
\usepackage{longtable}
\usepackage{cite}

\def\dfrac#1#2{{\displaystyle {#1 \over #2}}}

\def\simge{\mathrel{\rlap{\raise 0.511ex \hbox{$>$}}{\lower 0.511ex \hbox{$\sim$}}}}
\def\simle{\mathrel{\rlap{\raise 0.511ex \hbox{$<$}}{\lower 0.511ex \hbox{$\sim$}}}}
\def\slash#1{\setbox0=\hbox{$#1$}\dimen0=\wd0
      \setbox1=\hbox{/} \dimen1=\wd1 \ifdim\dimen0>\dimen1
      \rlap{\hbox to \dimen0{\hfil/\hfil}} #1                        \else
      \rlap{\hbox to \dimen1{\hfil$#1$\hfil}}
      /   \fi}

\newcommand{\lsim}{
\mathrel{\hbox{\rlap{\hbox{\lower4pt\hbox{$\sim$}}}\hbox{$<$}}}}

\newcommand{\gsim}{
\mathrel{\hbox{\rlap{\hbox{\lower4pt\hbox{$\sim$}}}\hbox{$>$}}}}

\newcommand{\vcb}{|V_{cb}|}
\newcommand{\vtd}{|V_{td}|}

\newcommand{\vts}{|V_{ts}|}

\def\eps{\varepsilon}

\newcommand{\tev}{\, {\rm TeV}}
\newcommand{\gev}{\, {\rm GeV}}
\newcommand{\mev}{\, {\rm MeV}}

\newcommand{\mtb}{\overline{m}_{\rm t}}
\newcommand{\mcb}{\overline{m}_{\rm c}}

\newcommand{\mw}{M_{\rm W}}

\allowdisplaybreaks[2]

\newcommand{\be}{\begin{equation}}
\newcommand{\ee}{\end{equation}}
\newcommand{\bea}{\begin{eqnarray}}
\newcommand{\eea}{\end{eqnarray}}
\newcommand{\nn}{\nonumber}
\newcommand{\bi}{\begin{itemize}}
\newcommand{\ei}{\end{itemize}}
\newcommand{\ord}{{\cal O}}

\newcommand{\RE}{{\rm Re}}
\newcommand{\IM}{{\rm Im}}

\def\kpn{K^+\rightarrow\pi^+\nu\bar\nu}

\def\klpn{K_{L}\rightarrow\pi^0\nu\bar\nu}

\renewcommand{\baselinestretch}{1.2}
\newcommand{\newsection}[1]{\section{#1}\setcounter{equation}{0}}

\begin{document}
\begin{titlepage}
\vspace*{-0.5truecm}

\begin{flushright}
TUM-HEP-649/06\\
MPP-2006-130
\end{flushright}

\vfill

\begin{center}
\boldmath

{\Large\textbf{Rare and CP-Violating $K$ and $B$ Decays
\vspace{0.3truecm}\\in
    the Littlest Higgs Model with T-Parity}}

\unboldmath
\end{center}

\vspace{0.4truecm}

\begin{center}
{\bf Monika Blanke$^{a,b}$, Andrzej J.~Buras$^a$, Anton Poschenrieder$^a$,\\
  Stefan Recksiegel$^a$,
 Cecilia  Tarantino$^a$, Selma Uhlig$^a$ and Andreas Weiler$^a$
}
\vspace{0.4truecm}

 $^a${\sl Physik Department, Technische Universit\"at M\"unchen,
D-85748 Garching, Germany}

 {\sl $^b$Max-Planck-Institut f{\"u}r Physik (Werner-Heisenberg-Institut), \\
D-80805 M{\"u}nchen, Germany}
\end{center}
\renewcommand{\baselinestretch}{1.1}
\vspace{0.6cm}
\begin{abstract}
\vspace{0.2cm}
\noindent 
We calculate the most interesting rare and CP-violating
$K$ and $B$ decays in the Littlest Higgs model with T-parity.
We give a collection of Feynman rules including $v^2/f^2$
contributions that are presented here for the first time and could
turn out to be useful also for applications outside flavour physics. We
adopt a model-independent parameterization of rare decays in
terms of the gauge independent functions $X_i,\,Y_i,\,Z_i\; (i=K,d,s)$,
which is in particular useful for the study of the breaking of the
universality between $K$, $B_d$ and $B_s$ systems through non-MFV
interactions.
Performing the calculation in the unitary and 't Hooft-Feynman
gauge, we find that the final result contains a divergence which
signals some sensitivity to the ultraviolet completion of the
theory. Including an estimate of this contribution, we calculate
the branching ratios for the decays $K^+\to\pi^+\nu\bar\nu$,
$K_L\to\pi^0\nu\bar\nu$, $B_{s,d}\to \mu^+\mu^-$, $B\to
X_{s,d}\nu\bar\nu$, $K_L \to\pi^0\ell^+\ell^-$ and $B\to
X_{s,d}\ell^+\ell^-$, paying particular attention to
non-MFV contributions present in the model. 

The main feature of mirror fermion effects is the possibility of large
modifications in rare $K$ decay branching ratios and in those $B$ decay
observables, like $S_{\psi \phi}$ and $A^s_\text{SL}$, that are very small in
the SM.
Imposing all available constraints we find that the decay rates for $B_{s,d}
\to \mu^+ \mu^-$ and $B \to X_{s,d} \nu \bar \nu$ can be
enhanced by at most $50\%$  and $35\%$ relative to the SM values, while
$Br(K^+\to\pi^+\nu\bar\nu)$ and $Br(K_L\to\pi^0\nu\bar\nu)$ can be both as
high as $5 \cdot 10^{-10}$.
Significant enhancements of the decay rates $K_L \to\pi^0\ell^+\ell^-$ are
also possible.
Simultaneously, the CP-asymmetries $S_{\psi \phi}$ and $A^s_\text{SL}$ can be
enhanced by an order of magnitude, while the electroweak penguin effects in $B
\to \pi K$ turn out to be small, in agreement with the recent data. 
\end{abstract}\renewcommand{\baselinestretch}{1.2}
\end{titlepage}

\thispagestyle{empty}

\begin{center}
{\Large\bf Note added}
\end{center}

\noindent
An additional contribution to the $Z$ penguin in the Littlest Higgs model with T-parity has been pointed out in \cite{Goto:2008fj,delAguila:2008zu}, which has been overlooked in the present analysis. This contribution leads to the cancellation of the left-over quadratic divergence in the calculation of some rare decay amplitudes. Instead of presenting separate errata to the present work and our papers \cite{Blanke:2007db,Blanke:2007ee,Blanke:2007wr,Blanke:2008ac} partially affected by this omission, we have presented a corrected and updated analysis of flavour changing neutral current processes in the Littlest Higgs model with T-parity in \cite{Blanke:2009am}.

\newpage

\setcounter{page}{1}
\pagenumbering{arabic}

\newsection{Introduction}
\label{sec:intro}

Rare and CP-violating $K$ and $B$ meson decays will hopefully
provide in the coming years a new insight into the origin of the
hierarchy of quark masses and their hierarchical flavour and
CP-violating interactions. While the presently available data on
these decays give a strong indication that the CKM matrix
\cite{ckm} and more generally minimal flavour violation (MFV)
\cite{mfv1,mfv2,mfvearly}, encoded entirely in Yukawa couplings of
quarks and leptons, is likely to be the dominant source of flavour
and CP violation, there is clearly still room left for
contributions governed by non-MFV interactions,
in particular those with new CP-violating phases.

A prime example of a model with non-MFV interactions is a general
MSSM in which non-MFV contributions originate in squark mass
matrices that are not aligned with the quark mass matrices.
Extensive analyses of the impact of such non-MFV contributions on
particle-antiparticle mixing and rare $K$ and $B$ decays in the
MSSM with and without $R$-parity have been presented in the
literature. The basic strategy proposed in \cite{genMSSM} is to constrain 
the non-diagonal entries
$(\delta_{ij})_{\rm AB}$ of the squark mass matrices given in the
so-called super-CKM basis, in which all neutral gauge interactions
and the quark and lepton mass matrices are flavour diagonal. These
studies are complicated by the fact that in addition to
$(\delta_{ij})_{\rm AB}$ also new fermion and scalar particle
masses appear as free parameters. The situation in this respect
may improve significantly in the coming years provided the
supersymmetric particles will be discovered at LHC and their
masses measured at LHC and later at ILC.

While supersymmetry~\cite{susy} appears at present to be the leading
candidate for new physics beyond the Standard Model (SM),
Little Higgs models~\cite{oldLH,LHreview} appear as an interesting
alternative.
Also here, new particles below $1 \tev$,
 or slightly above, with significant contributions to electroweak precision 
studies and FCNC
processes are present. Among the most popular Little Higgs models
is the so-called Littlest Higgs model~\cite{l2h}. In this model in
addition to the SM particles, new charged heavy vector bosons
($W_H^\pm$), a neutral heavy vector boson ($Z_H^0$), a heavy
photon ($A_H$), a heavy top quark ($T_+$) and a triplet of scalar
heavy particles ($\Phi$) are present.

In the original Littlest Higgs model (LH) \cite{l2h}, the
custodial $SU(2)$ symmetry, of fundamental importance for electroweak precision
studies, is unfortunately broken already at tree level, implying
that the relevant scale of new physics, $f$, must be at least $2-3
\tev$ in order to be consistent with electroweak precision data
\cite{Logan,PHEN1,PHEN2,PHEN3,PHEN4,PHEN5,PHEN6}. As a consequence, the contributions of the new
particles to FCNC processes turn out to be at most $10-20\%$
\cite{BPU,Choudhury:2004bh,BPU2,Huo:2003vd,BPUB}, which will not be easy to distinguish from the
SM due to experimental and theoretical uncertainties.  In particular, 
a detailed analysis of particle-antiparticle mixing in the LH model has been
published in \cite{BPU} and the corresponding analysis of rare $K$
and $B$ decays has recently been presented in \cite{BPUB}.

More promising and more interesting from the point of view of FCNC
processes is the Littlest Higgs model with a discrete symmetry
(T-parity)~\cite{tparity} under which all new particles listed
above, except $T_+$, are odd and do not contribute to processes
with external SM quarks (T-even) at tree level. As a
consequence, the new physics scale $f$ can be lowered down to $1
\tev$ and even below it, without violating electroweak precision
constraints \cite{mH}.

A consistent and phenomenologically viable Littlest Higgs model
with T-parity (LHT) requires the introduction of three doublets of
``mirror quarks'' and three doublets of ``mirror leptons'' which
are odd under T-parity, transform vectorially under $SU(2)_L$ and
can be given a large mass. Moreover, there is an additional
heavy $T_-$ quark that is odd under T-parity \cite{mirror}.

In the first phenomenological study of FCNC processes in the LHT model
by Hubisz et al.~\cite{Hubisz}, the impact of mirror fermions on the
neutral meson mixing in the $K$, $B$ and $D$ systems has been studied
in some detail. As described in that paper, in the LHT model
there are new flavour violating interactions in the
       mirror quark sector, parameterized by two CKM-like mixing
       matrices $V_{Hd}$ and $V_{Hu}$, relevant for processes with
       external light down-type quarks and up-type quarks, respectively.
       It turns out that the spectrum of mirror quarks must be generally
      quasi-degenerate, but
there exist regions of parameter space, where
only a loose degeneracy is necessary in order to satisfy
constraints coming from particle-antiparticle mixing.

In \cite{BBPTUW} we have confirmed the analytic expressions for
the effective Hamiltonians for $K^0-\bar K^0$, $B_d^0-\bar B_d^0$
and $B_s^0-\bar B_s^0$ mixings presented in~\cite{Hubisz} and we
have generalized the analysis of these authors to other quantities
that allowed a deeper insight into the flavour structure of the
LHT model.
 While the authors of~\cite{Hubisz} analyzed only the mass
       differences $\Delta M_K$, $\Delta M_d$, $\Delta M_s$, $\Delta
       M_D$ and the CP-violating parameter $\varepsilon_K$, we included
       in our analysis also other theoretically cleaner quantities: 
       the CP asymmetries $A_{\rm CP}(B_d
       \to \psi K_S)$, $A_{\rm CP}(B_s \to \psi \phi)$
       and $A^{d,s}_{\rm SL}$, and the width differences $\Delta\Gamma_{d,s}$. 
       We have also calculated
       $Br(B\to X_{s,d}\gamma)$ and the corresponding CP asymmetries.
       Moreover we
       emphasized that
while the original LH model belongs to the class of models with
MFV, this is certainly not the case of the LHT model where the
presence of the matrices $V_{Hd}$ and $V_{Hu}$ in the mirror quark
sector introduces new flavour and CP-violating interactions that
could have a very different pattern from the ones present in the SM.

The main messages of our analysis in \cite{BBPTUW} are as follows:
\begin{itemize}
\item The analysis of the mixing induced CP asymmetries
$A_{\rm CP}(B_d \to \psi K_S)$ and $A_{\rm CP}(B_s \linebreak \to
\psi \phi)$ illustrates very clearly that with mirror fermions at
work these asymmetries do \emph{not} measure the phases $-\beta$
and $-\beta_s$ of the CKM elements $V_{td}$ and $V_{ts}$,
respectively.
\item  This has two interesting consequences: first, the
possible ``discrepancy'' between the values of $\sin 2\beta $
following directly from $A_{\rm CP}(B_d \to \psi K_S)$ and
indirectly from the usual analysis of the unitarity triangle
involving $\Delta M_{d,s}$, $\varepsilon_K$ and $|V_{ub}/V_{cb}|$ can
be avoided within the LHT model. Second, the asymmetry $A_{\rm
CP}(B_s \to \psi \phi)$ can be significantly enhanced over
the SM expectation $\sim 0.04$, so that values as high as $0.3$  are
possible. Moreover, the asymmetry $A^s_\text{SL}$ can be enhanced by an
order of magnitude.
\item   We also found that the usual relation between $\Delta M_d/\Delta M_s$ and $|V_{td}/V_{ts}|$
characteristic for all models with MFV is no longer
satisfied.
\item 
 Calculating $Br(B\to X_s\gamma)$ in the LHT model for the first time 
 we have found that in spite of large effects in $\Delta F=2$ 
processes considered in \cite{Hubisz,BBPTUW}, 
it can be modified by at most $\pm 4\%$ 
 relative to the SM value \cite{bsgSM1,bsgSM2}. 
This is welcome as the SM branching ratio
 agrees well with the data. Also new physics effects in $B\to X_d\gamma$ 
 and in the CP asymmetries in both decays are at the level of a few percent
 of the SM values. 
\end{itemize}

The beauty of the LHT model, when compared with other models with non-minimal 
flavour violating interactions, like general MSSM, is a relatively small
number of new parameters and the fact that the local operators
involved are the same as in the SM. Therefore the non-perturbative
uncertainties, present in certain quantities already in the SM,
are the same in the LHT model. Consequently the departures from
the SM are entirely due to short distance physics that can be
calculated within perturbation theory. In stating this we are
aware of the fact that we deal here with an effective field theory
whose ultraviolet completion has not been specified, with the
consequence that at a certain level of accuracy one has to worry
about the effects coming from the cut-off scale $\Lambda\sim 4\pi
f$.

As pointed out recently in \cite{BPUB}, such effects are signaled
by left-over logarithmic divergences in the final result for FCNC
amplitudes and they appear as poles $1/\varepsilon$ when
dimensional regularization is used. It turns out that such
divergences are absent in particle-antiparticle mixing and $B\to
X_{s,d}\gamma$ decays both in the
LH and the LHT model. On the other hand, they are present in $Z^0$-penguin diagrams
in the LH model considered in \cite{BPUB}. As we will see in the
present paper, the imposition of T-parity eliminates such
divergences from the T-even sector as the diagrams containing
these divergences are forbidden by T-parity. However, we find that
the mirror fermion contributions to $\Delta F=1$ processes contain
such logarithmic divergences and their effects on rare decays has
to be taken into account following the steps of \cite{BPUB}.

In the present paper we extend our analysis of \cite{BBPTUW} to
include all prominent rare $K$ and $B$ decays. In particular we
calculate the LHT contributions from both  T-even and T-odd
sectors to the one-loop functions $X$, $Y$ and $Z$ \cite{BBH90} that encode the
short distance contributions to the rare decays in question. The
new properties of these functions relative to MFV are:
\begin{itemize}
\item
They are complex quantities, as in the phenomenological analysis
in \cite{BFRS}, while in contrast to the real $X$, $Y$ and $Z$ of MFV.
\item
They are different for $K$, $B_d$ and $B_s$ systems in contrast to
\cite{BFRS}, where these functions were the same for the three
systems in question. Thus in the present model we deal really with
nine short distance functions. But as we calculate them in a
specific model, interesting correlations between observables in
rare $K$, $B_d$ and $B_s$ decays are still present, although their
structure differs from the correlations in MFV models
\cite{MFVrel,Bergmann,BBGT} and in the phenomenological non-MFV analysis
in \cite{BFRS}.
\end{itemize}

Our paper is organized as follows. In Section~\ref{sec:model} we
review those ingredients of the LHT model that are of relevance
for our analysis. In particular we summarize in an appendix the
relevant Feynman rules that go beyond those presented in
\cite{Hubisz,LHTphen,0609179}. Section~\ref{sec:XYZ} is devoted to the master
formulae for rare decays in models with non-MFV interactions. This
presentation goes beyond the LHT model and the formulae given in
this section are general enough to be used in other
models that contain non-MFV contributions. This section is
basically  a generalization of the formulae in \cite{BFRS} to
include the breakdown of universality in the functions $X$, $Y$ and $Z$
 as mentioned above.
 In Section~\ref{sec:Teven} we
calculate the new contributions to $X$, $Y$ and $Z$ coming from
the T-even sector. The results from the analysis of the LH model
in \cite{BPUB} were very helpful in obtaining these results. In
Section~\ref{sec:Todd}, the most important section of our paper, we
calculate the T-odd contributions to $X$ and $Y$ in the
't Hooft-Feynman and unitary gauge, obtaining a gauge invariant result for
these functions. Similarly to \cite{BPUB} we identify logarithmic
divergences that are gauge independent. We recall the
interpretation of these divergences and we analyze their impact on
the functions $X$ and $Y$ in Section \ref{sec:sing}.
In Section \ref{sec:BSLL} we calculate the function $Z$ that
contributes to the $B\to X_s \ell^+\ell^-$ decay. In Section
\ref{sec:KpiLL} we study the decay $K_L\to\pi^0\ell^+\ell^-$.
The benchmark  scenarios for our numerical analysis are described in
Section \ref{sec:benchmark}.
In Section~\ref{sec:numerics}, the numerical analysis of the relevant
branching ratios including T-even and T-odd contributions is
presented in detail. Of particular interest is the pattern of the
violations of several MFV relations \cite{MFVrel,Bergmann,BBGT}
that can be tested experimentally one day. This analysis takes
into account the constraints found from the analysis of $\Delta
F=2$ processes and $B\to X_{s}\gamma$ presented in \cite{BBPTUW} and from
$Br(B \to X_s \ell^+ \ell^-)$ calculated here. It can be considered as the first 
global analysis of FCNC processes in the LHT model. 
In this section we also comment on $B \to \pi K$ decays.
In Section~\ref{sec:summary} we conclude
our paper with a list of messages resulting from our analysis and
with a brief outlook. Few technical details and the Feynman rules
can be found in the appendices.

\newsection{General Structure of the LHT Model}
\label{sec:model}
\subsection{Gauge and Scalar Sector}
\label{subsec:2.1}

The original Littlest Higgs model \cite{l2h} is based on a
non-linear sigma model describing the spontaneous breaking of a
global $SU(5)$ down to a global $SO(5)$. This symmetry breaking
takes place at the scale $f\sim\ord(\text{TeV})$ and originates
from the vacuum expectation value (VEV) of an $SU(5)$ symmetric
tensor $\Sigma$, given by
\be\label{Sigma0}\addtolength{\arraycolsep}{3pt} \Sigma_0 \equiv
\langle \Sigma \rangle =
\left(\begin{array}{ccc} \mathbf{0}_{2\times2} & 0 & \mathbf{1}_{2\times2} \\
0 & 1 & 0\\
\mathbf{1}_{2\times2} & 0 & \mathbf{0}_{2\times2}
\end{array}\right).\addtolength{\arraycolsep}{-3pt}
\ee

The generators $T^a$ of the unbroken $SO(5)$ symmetry fulfill
$T^a\Sigma_0+\Sigma_0 (T^a)^T=0$, whereas the broken generators
$X^a$ of $SU(5)/SO(5)$ satisfy $X^a\Sigma_0-\Sigma_0 (X^a)^T=0$.
The symmetry breaking can thus be considered as a $\mathbb{Z}_2$
automorphism under which $T^a\mapsto T^a$ and $X^a\mapsto -X^a$.
This automorphism is the basic motivation for the way T-parity is
implemented, as discussed later.

The mechanism for this symmetry breaking is unspecified, therefore
the Littlest Higgs model is an effective theory, valid up to a
scale $\Lambda\sim 4\pi f$. From the $SU(5)/SO(5)$ breaking, there
arise 14 Nambu-Goldstone bosons $\chi^a$ which are described by
the ``pion'' matrix $\Pi$, given explicitly by
\be\label{Pi}\addtolength{\arraycolsep}{3pt}\renewcommand{\arraystretch}{1.3}
 \Pi=\chi^a
X^a=\left(\begin{array}{ccccc}
-\frac{\omega^0}{2}-\frac{\eta}{\sqrt{20}} &
-\frac{\omega^+}{\sqrt{2}} &
  -i\frac{\pi^+}{\sqrt{2}} & -i\phi^{++} & -i\frac{\phi^+}{\sqrt{2}}\\
-\frac{\omega^-}{\sqrt{2}} &
\frac{\omega^0}{2}-\frac{\eta}{\sqrt{20}} &
\frac{v+h+i\pi^0}{2} & -i\frac{\phi^+}{\sqrt{2}} & \frac{-i\phi^0+\phi^P}{\sqrt{2}}\\
i\frac{\pi^-}{\sqrt{2}} & \frac{v+h-i\pi^0}{2} &\sqrt{4/5}\eta &
-i\frac{\pi^+}{\sqrt{2}} & \frac{v+h+i\pi^0}{2}\\
i\phi^{--} & i\frac{\phi^-}{\sqrt{2}} & i\frac{\pi^-}{\sqrt{2}} &
-\frac{\omega^0}{2}-\frac{\eta}{\sqrt{20}} & -\frac{\omega^-}{\sqrt{2}}\\
i\frac{\phi^-}{\sqrt{2}} &  \frac{i\phi^0+\phi^P}{\sqrt{2}} &
\frac{v+h-i\pi^0}{2} & -\frac{\omega^+}{\sqrt{2}} &
\frac{\omega^0}{2}-\frac{\eta}{\sqrt{20}}
\end{array}\right).\renewcommand{\arraystretch}{1.0}
\ee
Here, $H=(-i{\pi^+}/{\sqrt{2}},(v+h+i\pi^0)/{2})^T$ plays the
role of the SM Higgs doublet, i.\,e. $h$ is the usual Higgs
field, $v=246\,$GeV the Higgs VEV, and $\pi^\pm,\,\pi^0$ are the
Goldstone bosons associated with the spontaneous symmetry breaking
$SU(2)_L\times U(1)_Y \to U(1)_\text{em}$. The fields $\eta$ and
$\omega$ are additional Goldstone bosons, and $\Phi$ is a physical scalar
triplet, as described below.

The low
energy dynamics of the symmetric tensor $\Sigma$ is then described by
\be\label{Sigma}
\Sigma=e^{i\Pi/f}\Sigma_0e^{i\Pi^T/f}\equiv e^{2i\Pi/f}\Sigma_0\,.
\ee

An $[SU(2)\times U(1)]_1\times[SU(2)\times U(1)]_2$ subgroup of the
$SU(5)$ global symmetry is gauged, with the generators
\begin{gather}\addtolength{\arraycolsep}{3pt}
Q_1^a=\frac{1}{2}\begin{pmatrix}\sigma^a&0&0\\
0&0&0\\
0&0&0\end{pmatrix},\quad
Y_1=\frac{1}{10}\text{diag}(3,3,-2,-2,-2)\,,\nonumber
\\\addtolength{\arraycolsep}{3pt}
Q_2^a=\frac{1}{2}\begin{pmatrix}0&0&0\\
0&0&0\\
0&0&-\sigma^{a*}\end{pmatrix},\quad
Y_2=\frac{1}{10}\text{diag}(2,2,2,-3,-3)\,,\label{generators}
\end{gather}
and the corresponding gauge bosons
$W_1^{a\mu},\;B_1^\mu,\;W_2^{a\mu},\;B_2^\mu$. Having the
$\mathbb{Z}_2$ automorphism $T^a\mapsto T^a$ and $X^a\mapsto -X^a$
in mind, one implements T-parity by assigning a symmetry
transformation to all fields, such that the theory is symmetric
under T-parity. A natural way to define the action of T-parity on
the gauge fields is
\be\label{Tgauge}
W_1^a \leftrightarrow
W_2^a\,,\qquad B_1 \leftrightarrow B_2\,.
\ee
In other words, the
action of T-parity exchanges the two $SU(2)\times U(1)$ factors.
An immediate consequence of this definition is that the gauge
couplings of the two $SU(2)\times U(1)$ factors have to be equal.

The gauge boson T-parity eigenstates are given by \bea
W^a_L=\frac{W^a_1+W^a_2}{\sqrt{2}}\,,\qquad
B_L=\frac{B_1+B_2}{\sqrt{2}}\qquad\text{(T-even)},\\
W^a_H=\frac{W^a_1-W^a_2}{\sqrt{2}}\,,\qquad
B_H=\frac{B_1-B_2}{\sqrt{2}}\qquad\text{(T-odd)}, \eea
where ``$L$'' and ``$H$'' denote ``light'' and ``heavy'', respectively.

Now, the VEV $\Sigma_0$ breaks the gauge symmetry to the diagonal T-even
$SU(2)\times U(1)$, which is identified with the SM gauge group.
The breaking $SU(2)_L\times U(1)_Y\to U(1)_\text{em}$ then takes
place via the usual Higgs mechanism. Finally, the mass eigenstates
are given at $\ord(v^2/f^2)$ by
\begin{gather}
W^\pm_L=\frac{W^1_L\mp i W^2_L}{\sqrt{2}}\,,\qquad
W^\pm_H=\frac{W^1_H\mp i W^2_H}{\sqrt{2}}\,,\\
Z_L=\cos\theta_W W^3_L-\sin\theta_WB_L\,,\qquad
Z_H=W^3_H+x_H\frac{v^2}{f^2} B_H\,,\\
A_L=\sin\theta_W W^3_L+\cos\theta_WB_L\,,\qquad
A_H=-x_H\frac{v^2}{f^2} W^3_H+B_H\,,
\end{gather}
where $\theta_W$ is the usual weak mixing angle and \be
x_H=\frac{5gg'}{4(5g^2-g'^2)}\,. \ee

From these breakings, the T-odd set of $SU(2)\times U(1)$ gauge
bosons acquires masses, given at $\ord(v^2/f^2)$ by
\be\label{heavy_gauge}
M_{W_H}=fg\left(1-\frac{v^2}{8f^2}\right),\qquad
M_{Z_H}\equiv M_{W_H} \,,\qquad M_{A_H}=\frac{f
g'}{\sqrt{5}}\left(1-\frac{5v^2}{8f^2}\right). \ee
 The masses of
the T-even gauge bosons are generated only through the second step
of symmetry breaking. They are given by
\be\label{SM_gauge}
M_{W_L}=\frac{gv}{2}\left(1-\frac{v^2}{12f^2}\right),\quad
M_{Z_L}=\frac{gv}{2\cos\theta_W}\left(1-\frac{v^2}{12f^2}\right),\quad
M_{A_L}=0\,. \ee Note that T-parity ensures that the custodial
relation $M_{W_L}={M_{Z_L}}{\cos\theta_W}$ is exactly satisfied at
tree level.

In order to ensure that $\Phi\,,\omega\,,\eta$ are odd under
T-parity, whereas the SM Higgs doublet $H$ is even, one makes the
following T-parity assignment:
\be \Pi \mapsto -\Omega
\Pi\Omega\,,\qquad\text{where}\quad\Omega=\text{diag}(1,1,-1,1,1)\,.
\ee

As mentioned above, $\Phi$ is an additional scalar triplet in a
symmetric representation of $SU(2)_L$. Its mass is given  by  
\be
m_{\Phi}=\sqrt{2}m_H\frac{f}{v}\,, \ee
with $m_H$ being the mass of the SM Higgs scalar. 
As pointed out in~\cite{mH}, $m_H$ in the LHT model can be significantly higher
than in supersymmetry. In Appendix \ref{sec:appA} we show that
$\Phi$ has only negligible effects on the decays studied in the
present paper.

The fields $\omega^\pm,\,\omega^0$ and $\eta$ are the Goldstone bosons
associated with the breaking of $[SU(2)\times
U(1)]_1\times[SU(2)\times U(1)]_2$ to its diagonal subgroup. They
are thus eaten by the heavy gauge bosons $W_H^\pm,\,Z_H$ and
$A_H$, respectively.\footnote{To be exact, $\Phi,\,\omega$
and $\eta$ mix with each other at $\ord(v^2/f^2)$ \cite{mH} and it
is a linear combination of the fields that is eaten. See Appendix
\ref{sec:appB} for details.}

\subsection{Fermion Sector}

A consistent and phenomenologically viable implementation of
T-parity in the fermion sector requires the introduction of mirror
fermions \cite{mirror}. We embed the fermion doublets into the
following incomplete representations of $SU(5)$, and introduce a
right-handed $SO(5)$ multiplet $\Psi_R$: \be
\Psi_1=\begin{pmatrix}i\psi_1\\0\\0\end{pmatrix},\qquad
\Psi_2=\begin{pmatrix}0\\0\\i\psi_2\end{pmatrix},\qquad
\Psi_R=\begin{pmatrix}\tilde\psi_R\\\chi_R\\\psi_R\end{pmatrix},
\ee
with
\be \psi_i=-\sigma^2 q_i=
-\sigma^2\begin{pmatrix}u_i\\d_i\end{pmatrix}\qquad(i=1,2)\,,\qquad
\psi_R=-i\sigma^2\begin{pmatrix}u_{HR}\\d_{HR}\end{pmatrix}. \ee
Under T-parity these fields transform as \be \Psi_1\mapsto
-\Sigma_0\Psi_2\,,\qquad \Psi_2\mapsto -\Sigma_0\Psi_1\,,\qquad
\Psi_R\mapsto -\Psi_R\,.
\ee
Thus, the T-parity eigenstates of the
fermion doublets are given by
\be
q_\text{SM}=\frac{q_1-q_2}{\sqrt{2}}\,,\qquad
q_H=\frac{q_1+q_2}{\sqrt{2}}\,.
\ee
$q_\text{SM}$ are
the left-handed SM fermion doublets (T-even), and
$q_H$ are the left-handed mirror fermion doublets (T-odd). The
right-handed mirror fermion doublet is given by $\psi_R$.

The mirror fermions can be given $\ord(f)$ masses via a mass
term \be\label{eq:Dirac}
\mathcal{L}_\text{mirror}=-\kappa_{ij}f\left(\bar\Psi_2^i\xi +
  \bar\Psi_1^i\Sigma_0\Omega\xi^\dagger\Omega\right)\Psi_R^j\,,
\ee
where we sum over the generation indices $i,j=1,2,3$ and  $\xi=e^{i\Pi/f}$ is needed to make
$\mathcal{L}_\text{mirror}$ $SU(5)$ invariant.

The mirror fermions thus acquire masses, given by
\cite{Hubisz}
\bea m^u_{Hi}&=&\sqrt{2}\kappa_i
f\left(1-\frac{v^2}{8f^2}\right)\equiv
m_{Hi} \left(1-\frac{v^2}{8f^2}\right),\\
m^d_{Hi}&=&\sqrt{2}\kappa_i f \equiv m_{Hi}\,,
\eea
where $\kappa_i$ are the eigenvalues of the mass matrix $\kappa$.

The additional fermions $\tilde\psi_R$ and $\chi_R$ can be given
large Dirac masses by introducing additional fermions, as
described in detail in \cite{mirror,LHTphen}. In what follows, we will simply
assume that they are decoupled from the theory.

\subsection{Yukawa Sector}

In order to cancel the quadratic divergence of the Higgs mass
coming from top loops, an additional heavy quark $T_+$ is
introduced, which is even under T-parity and transforms, to leading
order in $v/f$, as a singlet under $SU(2)_L$. The implementation of
T-parity then requires also a T-odd partner $T_-$, which is
an exact singlet under $SU(2)_1\times SU(2)_2$  and therefore does
not contribute to the decays in question (see \cite{BBPTUW} for
details).

The Yukawa coupling for the top sector is then given by \cite{LHTphen,Chen}
\bea\nn
\mathcal{L}_\text{top}&=&-\frac{1}{2\sqrt{2}}\lambda_1 f
  \epsilon_{ijk}\epsilon_{xy}\left[(\bar
    Q_1)_i(\Sigma)_{jx}(\Sigma)_{ky} - (\bar Q_2 \Sigma_0)_i (\tilde
    \Sigma)_{jx}(\tilde \Sigma)_{ky}\right] u^3_R \\
&&-\; \lambda_2 f (\bar
  t'_1 t'_{1R}+ \bar t'_2 t'_{2R}) +h.c.\label{eq:Ltop}\,,
\eea
where
\be
Q_1=\begin{pmatrix}\psi_1\\t'_1\\0\end{pmatrix}\,,
\qquad Q_2=
\begin{pmatrix}0\\t'_2\\\psi_2\end{pmatrix}\,,
\ee
the superscript in $u^3_R$ denotes the third quark generation, and $\tilde\Sigma=\Sigma_0\Omega\Sigma^\dagger\Omega\Sigma_0$ is
the image of $\Sigma$ under T-parity.

Note that under T-parity
\be
Q_1 \leftrightarrow -\Sigma_0
Q_2\,,\qquad  t'_{1R}\leftrightarrow -t'_{2R}\,,\qquad u^3_R \mapsto
u^3_R\,, 
\ee 
so that the T-parity eigenstates are given by 
\be
t'_\pm =\frac{t'_1\mp
t'_2}{\sqrt{2}}\,,\qquad t'_{\pm R} =\frac{t'_{1R}\mp
t'_{2R}}{\sqrt{2}}\,.
\ee

As $t'_-$ and $t'_{-R}$ do not mix with the mirror fermions at tree level, the
mass eigenstates, denoted by $(T_-)_L$ and $(T_-)_R$ for the left
and right-handed part, respectively, are simply given by
\be
(T_-)_L\equiv t'_-\,,\qquad (T_-)_R\equiv t'_{-R}\,.
\ee

However, the T-even eigenstates mix with each other, so that the
mass eigenstates of the top quark $t$ and its heavy partner $T_+$
are given by
\bea
t_L&=&c_L (q_\text{SM})_1 - s_L t'_+ \,,\qquad  (T_+)_L= s_L (q_\text{SM})_1 + c_L t'_+\,,\\
t_R&=&c_R u^3_R - s_R t'_{+R}\,,\qquad \quad (T_+)_R=s_R u^3_R + c_R
t'_{+R}\,, \eea 
 where $(q_\text{SM})_1$ denotes the upper component of
the left-handed SM quark doublet, and 
\bea 
s_L&=& x_L
\frac{v}{f}\left[1+\frac{v^2}{f^2}\,d_2\right]\,,
\\
c_L&=& 1- \frac{x_L^2}{2}\frac{v^2}{f^2}\,,\\
s_R&=& {\sqrt{x_L}}\left[1-\frac{v^2}{f^2}(1-x_L) \left(\frac{1}{2}-x_L\right)\right]\,,\\
c_R&=& {\sqrt{1-x_L}}\left[1+\frac{v^2}{f^2}x_L\left(\frac{1}{2}-x_L\right)\right]\,,
\eea
with
\be\label{eq:xLd2}
x_L=\frac{\lambda_1^2}{\lambda_1^2+\lambda_2^2}\,,\qquad d_2=-\frac{5}{6}+\frac{1}{2}x_L^2+2x_L(1-x_L)\,.
\ee
This mixing
leads to a modification of the top quark couplings relatively to the SM, as can be seen
in the Feynman rules given in Appendix \ref{sec:appB}.

The masses are then given by
\bea
m_t &=&
\frac{\lambda_1\lambda_2
v}{\sqrt{\lambda_1^2+\lambda_2^2}}\left[1+
\frac{v^2}{f^2}\left(-\frac{1}{3}+\frac{1}{2}x_L(1-x_L)\right)\right],\\
m_{T_+} &=& \frac{f}{v}\frac{m_t}{\sqrt{x_L(1-x_L)}}\left[
  1+\frac{v^2}{f^2}\left(\frac{1}{3}-x_L(1-x_L)\right) \right],\\
m_{T_-} &=& \frac{f}{v}\frac{m_t}{\sqrt{x_L}}\left[
  1+\frac{v^2}{f^2}\left(\frac{1}{3}-\frac{1}{2}x_L(1-x_L)\right) \right].
\eea

As the Yukawa couplings of the other SM quarks are small, there is
no need to introduce additional heavy partners to cancel their
quadratically divergent contribution to the Higgs mass. Thus the
Yukawa coupling for the other up-type fermions is simply given by\footnote{Strictly speaking, all three
generations have to be included in the top Yukawa term, where
$\lambda_1$ then becomes the usual $3\times 3$ Yukawa coupling
matrix, as discussed in detail in \cite{Lee}. However, as found
there, the mixing of $T_+$ with $u,\,c$ quarks is experimentally
highly constrained, so we can safely neglect it. For simplicity,
we thus write the Yukawa term for each generation separately, and
later include the CKM mixing ``by hand'' in the Feynman rules
given in Appendix \ref{sec:appB}.}
\be \label{eq:Lup}
\mathcal{L}_\text{up} =
-\frac{1}{2\sqrt{2}}\lambda_u f
  \epsilon_{ijk}\epsilon_{xy}\left[(\bar
    Q_1)_i(\Sigma)_{jx}(\Sigma)_{ky} - (\bar Q_2 \Sigma_0)_i (\tilde
    \Sigma)_{jx}(\tilde \Sigma)_{ky}\right] u_R+h.c.\,,
\ee
and their masses are given by
\be
m_u^i = \lambda^i_u
v\left(1-\frac{v^2}{3f^2}\right)\qquad(i=1,2)\,.
\ee

For the down-type Yukawa term, we take \cite{Chen}
\be\label{eq:Ldown}
\mathcal{L}_\text{down}=\frac{i\lambda_d}{2\sqrt{2}}f
\epsilon_{ij}\epsilon_{xyz}\left[(\bar \Psi_2 )_x (\Sigma)_{iy}
  (\Sigma)_{jz}X-(\bar \Psi_1\Sigma_0)_x (\tilde \Sigma)_{iy}(\tilde
 \Sigma)_{jz}\tilde X\right] d_R + h.c.\,,
\ee
where we sum over $i,j=1,2$ and $x,y,z=3,4,5$, and $X\equiv (\Sigma_{33})^{-1/4}$ has
been introduced in order to make $\mathcal{L}_\text{down}$ gauge invariant.
Note that here
\be
\Psi_1=\begin{pmatrix}q_1\\0\\0\end{pmatrix}\,,\qquad 
\Psi_2=\begin{pmatrix}0\\0\\q_2\end{pmatrix}\,,
\ee
i.\,e. no insertion of $\sigma^2$ is needed.
From this Yukawa term, we obtain the down-type quark masses to be
\be
m_d^i = \lambda^i_d v\left(1-\frac{v^2}{12f^2}\right)\qquad(i=1,2,3)\,.
\ee

Lepton masses are generated in a completely analogous way.

\subsection{Weak Mixing in the Mirror Sector} \label{subsec:2.4}

As discussed in detail in~\cite{Hubisz,BBPTUW}, one of the
important ingredients of the mirror sector is the existence of
four CKM-like unitary mixing matrices, two for mirror quarks and
two for mirror leptons:
\begin{equation}\label{2.10}
V_{Hu}\,,\quad V_{Hd}\,,\qquad V_{H\ell}\,,\quad V_{H\nu}\,.
\end{equation}
They satisfy
\begin{equation}\label{2.11}
V_{Hu}^\dagger V_{Hd}=V_\text{CKM}\,,\qquad V_{H\nu}^\dagger
V_{H\ell}=V_\text{PMNS}\,,
\end{equation}
where in $V_\text{PMNS}$~\cite{pmns} the Majorana phases are set
to zero as no Majorana mass term has
been introduced for the right-handed neutrinos. The mirror
mixing matrices in \eqref{2.10} parameterize flavour violating
interactions between SM fermions and mirror fermions that are
mediated by the heavy gauge bosons $W_H$, $Z_H$ and $A_H$. The
notation in \eqref{2.10} indicates which of the light fermions of
a given electric charge participates in the interaction. Feynman rules for these interactions
are given in Appendix \ref{sec:appB}.

In the course of our analysis it will be useful to introduce the
following quantities \cite{BBPTUW}:
\begin{equation}\label{2.12}
\xi_i^{(K)}=V^{*is}_{Hd}V^{id}_{Hd}\,,\qquad
\xi_i^{(d)}=V^{*ib}_{Hd}V^{id}_{Hd}\,,\qquad
\xi_i^{(s)}=V^{*ib}_{Hd}V^{is}_{Hd}\, \qquad(i=1,2,3)\,,
\end{equation}
that govern $K$, $B_d$ and $B_s$ decays, respectively.

Following~\cite{SHORT} we will parameterize $V_{Hd}$ generalizing
the usual CKM parameterization, as a product of three rotations, and
introducing a complex phase in each of them, thus obtaining
\be
\addtolength{\arraycolsep}{3pt}
V_{Hd}= \begin{pmatrix}
1 & 0 & 0\\
0 & c_{23}^d & s_{23}^d e^{- i\delta^d_{23}}\\
0 & -s_{23}^d e^{i\delta^d_{23}} & c_{23}^d\\
\end{pmatrix}\,\cdot
 \begin{pmatrix}
c_{13}^d & 0 & s_{13}^d e^{- i\delta^d_{13}}\\
0 & 1 & 0\\
-s_{13}^d e^{ i\delta^d_{13}} & 0 & c_{13}^d\\
\end{pmatrix}\,\cdot
 \begin{pmatrix}
c_{12}^d & s_{12}^d e^{- i\delta^d_{12}} & 0\\
-s_{12}^d e^{i\delta^d_{12}} & c_{12}^d & 0\\
0 & 0 & 1\\
\end{pmatrix}\ee
Performing the product one obtains the expression
\be\label{2.12a}
\addtolength{\arraycolsep}{3pt}
V_{Hd}= \begin{pmatrix}
c_{12}^d c_{13}^d & s_{12}^d c_{13}^d e^{-i\delta^d_{12}}& s_{13}^d e^{-i\delta^d_{13}}\\
-s_{12}^d c_{23}^d e^{i\delta^d_{12}}-c_{12}^d s_{23}^ds_{13}^d e^{i(\delta^d_{13}-\delta^d_{23})} &
c_{12}^d c_{23}^d-s_{12}^d s_{23}^ds_{13}^d e^{i(\delta^d_{13}-\delta^d_{12}-\delta^d_{23})} &
s_{23}^dc_{13}^d e^{-i\delta^d_{23}}\\
s_{12}^d s_{23}^d e^{i(\delta^d_{12}+\delta^d_{23})}-c_{12}^d c_{23}^ds_{13}^d e^{i\delta^d_{13}} &
-c_{12}^d s_{23}^d e^{i\delta^d_{23}}-s_{12}^d c_{23}^d s_{13}^d
e^{i(\delta^d_{13}-\delta^d_{12})} & c_{23}^d c_{13}^d\\
\end{pmatrix}
\ee
As in the case of the CKM matrix 
the angles $\theta_{ij}^d$ can all be made to lie in the first quadrant 
with $0\le \delta^d_{12}, \delta^d_{23}, \delta^d_{13}< 2\pi$.
The matrix $V_{Hu}$ is then determined through
$V_{Hu}=V_{Hd}V_\text{CKM}^\dagger$. 

We point out that in \cite{Hubisz} and in the first version of
\cite{BBPTUW} $V_{Hd}$ was parameterized in terms of three mixing angles and
only one phase like $V_\text{CKM}$, overlooking the presence of two additional
phases.
The presence of three phases was first pointed out in \cite{SHORT}.
In short, the reason for the appearance of two additional phases relative to
the CKM matrix is as follows.
 $V_\text{CKM}$ and $V_{Hd}$ are both unitary
matrices containing three real angles and six complex phases.
Varying independently the phases of ordinary up- and down-quark states
allows us to rotate five phases away from $V_\text{CKM}$ (an over-all phase
change of all the quark states leaves $V_\text{CKM}$ invariant).
In rotating phases away from $V_{Hd}$,  one can still act on only three
mirror states, thus obtaining for $V_{Hd}$ a parameterization in terms of
three mixing angles and three phases.

The six parameters of $V_{Hd}$ have to be determined in flavour violating
processes. In \cite{BBPTUW} we have outlined briefly this
determination in the context of particle-antiparticle
mixing. Including rare $K$, $B_d$ and $B_s$ decays will further help
to determine these parameters.

\newsection{Rare $\bm{K}$ and $\bm{B}$ Decays beyond MFV}
\label{sec:XYZ}

\subsection{Preliminaries}
\label{subsec:3.1}

Before presenting in Sections~\ref{sec:Teven} and~\ref{sec:Todd} the
details of the calculations of rare $K$ and $B$ decays in the LHT
model in question, it will be useful to have a general look at
rare decays within models with new flavour and CP-violating
interactions but with the same local operators of the SM or
more generally of constrained MFV (CMFV) models, as defined in
\cite{mfv1,BBGT}. While the presentation given below is tailored to the
subsequent sections, it can easily be  adapted to any model of this
type.

It should be emphasized that while the formulae given below
bear many similarities to the ones given in~\cite{BFRS}, they
differ from the latter ones in the following important
manner. In~\cite{BFRS} a simple beyond-MFV scenario of new
physics has been considered in which new physics affected
only the $Z^0$-penguin function $C$ that became a complex
quantity, but remained universal for $K$, $B_d$ and $B_s$ decays. In
this manner several CMFV relations involving only CP-conserving quantities remained valid and the main new
effects were seen in CP-violating quantities like $Br(K_L \to \pi^0
\nu \bar \nu)$ and the CP-asymmetries in $B \to X_s \ell^+ \ell^-$.
In particular, the full system of rare $K$,
$B_d$ and $B_s$ decays considered in this section could be
described by three complex functions
\be
X=|X|\, e^{i\,\theta_X}\,,\qquad Y=|Y|\, e^{i\,\theta_Y}\,,\qquad Z=|Z|\,
e^{i\,\theta_Z}\,,
\label{eq:XYZ}
\ee
with correlations between these functions resulting from the
universality of the $Z^0$-penguin function $C=|C| \exp{(i \theta_C)}$.
As a result the
CMFV correlations between observables in $K$, $B_d$ and $B_s$ were
only affected in the cases in which $\theta_i$ played a role. In the LHT model
the structure of new flavour violating interactions is much richer. 
Let us spell it out in explicit terms.

\subsection{\boldmath $X_i$, $Y_i$, $Z_i$ functions \unboldmath}
\label{susec:3.2}

In the CMFV models the new physics contributions enter for
all practical purposes only through the functions $X$, $Y$ and $Z$
that multiply the CKM factors $\lambda_t^{(i)}$
\be
\lambda_t^{(K)} = V_{ts}^*\,V_{td}\,,\qquad
\lambda_t^{(d)} = V_{tb}^*\,V_{td}\,,\qquad
\lambda_t^{(s)} = V_{tb}^*\,V_{ts}\,,
\label{eq:lambdas}
\ee
for $K$, $B_d$ and $B_s$ systems respectively.

It will be useful to keep this structure in the LHT model and
absorb all new physics contributions in the functions $X_i$,
$Y_i$, $Z_i$ with $i=K,d,s$ defined as follows: \bea \label{eq:Xi}
X_i &=& X_\text{SM} + \bar X_\text{even} + \frac{1}{\lambda_t^{(i)}}
\bar
X_i^\text{odd} \equiv |X_i|\, e^{i\, \theta_X^i}\,,\\
Y_i &=& Y_\text{SM} + \bar Y_\text{even} + \frac{1}{\lambda_t^{(i)}}
\bar Y_i^\text{odd} \equiv |Y_i|\, e^{i\, \theta_Y^i}\,, \label{eq:Yi}\\
Z_i &=& Z_\text{SM} + \bar Z_\text{even} + \frac{1}{\lambda_t^{(i)}}
\bar Z_i^\text{odd} \equiv |Z_i|\, e^{i\, \theta_Z^i}\,.
\label{eq:Z_i} \eea 
Here $X_\text{SM}$, $Y_\text{SM}$ and $Z_\text{SM}$ are the SM contributions for which explicit
expressions can be found in Appendix \ref{sec:appC}.  $\bar
X_\text{even}$, $\bar Y_\text{even}$ and $\bar Z_\text{even}$ are the contributions from the T-even sector, that
is the contributions of $T_+$ and of $t$ at order $v^2/f^2$
necessary to make the  GIM mechanism \cite{GIM} work. The latter contributions, similarly to
$X_\text{SM}$, $Y_\text{SM}$ and $Z_\text{SM}$, are real and independent of $i=K,
d, s$. They can be extracted from~\cite{BPUB} and will be given in
Section~\ref{sec:Teven}. The main result of the present paper is the
calculation of the functions $\bar X_i^\text{odd}$,  $\bar
Y_i^\text{odd}$ and $\bar Z_i^\text{odd}$, that represent the T-odd
sector of the LHT model and are obtained from penguin and box
diagrams with internal mirror fermions. The details of this
calculation can be found in Section~\ref{sec:Todd}. In what follows
we will present the most interesting branching ratios in terms of
$X_i$ and $Y_i$. The CKM elements that we will use are those
determined from tree level decays.

\subsection{\boldmath${K\to\pi\nu\bar\nu}$\unboldmath}

Generalizing the formulae in \cite{BFRS} we have
\bea\label{eq:BrK+}
Br(K^+\to\pi^+\nu\bar\nu) &=& \kappa_+\left[ \tilde r^2 A^4 R_t^2 |X_K|^2 +
2\tilde r \bar P_c(x) A^2 R_t|X_K| \cos\beta_X^K +\bar
P_c(x)^2\right],\\
\label{eq:BrKL} Br(K_{L}\to\pi^0\nu\bar\nu) &=&
\kappa_L\tilde r^2 A^4 R_t^2 |X_K|^2 \sin^2{\beta_X^K}\,, \eea
where \cite{BGHN06}
\begin{gather}
\tilde r = \left|\frac{V_{ts}}{V_{cb}}\right|\simeq 0.98 \,,\quad \kappa_+ =
  (5.08\pm0.17)\cdot 10^{-11}\,,\quad \kappa_L=(2.22\pm0.07)\cdot 10^{-10}\,,\\
\bar P_c(x)=\left(1-\frac{\lambda^2}{2}\right)P_c(x)\,,\quad
P_c(x)=0.42\pm0.05\,,
\end{gather}
with $P_c(x)$ including both the NNLO corrections \cite{BGHN06} and long distance
contributions \cite{Isidori05}. Finally
\be
\beta_X^K=\beta-\beta_s-\theta_X^K\,.
\ee
The values of $A$, $R_b$, $\beta$ and $\beta_s$ are collected in Table
\ref{tab:input} of Section~\ref{sec:numerics}.

Of particular interest is the relation
\be\label{eq:betarel}
\sin{2(\beta+\varphi_{B_d})}=\sin{2\beta_X^K}\,, 
\ee 
that for $\varphi_{B_d}=0,\,\theta^K_X=0$ reduces to the MFV relation
of  \cite{BB94,BF01}. A violation of this relation would
signal the presence of new complex phases and generally non-MFV
interactions. In this context the ratio 
\be
\frac{Br(K_{L}\to\pi^0\nu\bar\nu)}{Br(K_{L}\to\pi^0\nu\bar\nu)_\text{SM}}
=
\left|\frac{X_K}{X_\text{SM}}\right|^2\left[\frac{\sin\beta_X^K}{\sin{(\beta-\beta_s)}}\right]^2
\ee 
is very useful, as it is very sensitive to $\theta_X^K$ and is
theoretically very clean.

The most recent SM predictions for the branching ratios read
  \cite{BGHN06} 
\be
Br(K^+\to\pi^+\nu\bar\nu)=(8.0 \pm 1.1) \cdot 10^{-11}\,,\qquad
Br(K_{L}\to\pi^0\nu\bar\nu)=(2.9 \pm 0.4) \cdot 10^{-11}\,,
\label{eq:SMKpinunu}
\ee
to be compared with the present experimental measurements \cite{expK+,Ahn:2006uf}
\be
Br(K^+\to\pi^+\nu\bar\nu)=(1.47^{+1.30}_{-0.89})\cdot 10^{-10} \,,\quad
Br(K_{L}\to\pi^0\nu\bar\nu)<2.1\cdot 10^{-7}\, (90\% \text{C.L.}) \,.
\ee
Recent reviews of the $K\to\pi\nu\bar\nu$ decays can be found in \cite{BSU,Isidori}.

\subsection{\boldmath${B_{s,d}\to\mu^+\mu^-}$\unboldmath}

Here, we will mainly be interested in the following ratios
\bea
\frac{Br(B_s\to\mu^+\mu^-)}{Br(B_s\to\mu^+\mu^-)_\text{SM}}&=&
\left|\frac{Y_s}{Y_\text{SM}}\right|^2,\\
\frac{Br(B_d\to\mu^+\mu^-)}{Br(B_d\to\mu^+\mu^-)_\text{SM}}&=&
\left|\frac{Y_d}{Y_\text{SM}}\right|^2,\\\label{eq:dostomumu}
\frac{Br(B_d\to\mu^+\mu^-)}{Br(B_s\to\mu^+\mu^-)}&=&
\frac{\tau(B_d)}{\tau(B_s)}\frac{m_{B_d}}{m_{B_s}}
\frac{F_{B_d}^2}{F_{B_s}^2}\left|\frac{V_{td}}{V_{ts}}\right|^2
\left|\frac{Y_d}{Y_s}\right|^2,
\eea
where the departure of the
last factor from unity signals non-MFV interactions. In obtaining
these formulae we assume that the CKM parameters have been determined
in tree level decays independently of new physics so that they cancel
in the ratios in question.

In the LHT model \cite{BBPTUW}
\be\label{eq:DMdos}
\frac{\Delta M_d}{\Delta
M_s} = \frac{m_{B_d}}{m_{B_s}} \frac{\hat  B_{B_d} F_{B_d}^2}{\hat
B_{B_s} F_{B_s}^2}
\left|\frac{V_{td}}{V_{ts}}\right|^2\frac{C_{B_d}}{C_{B_s}}\,, \ee
where
\be
C_{B_q}=\frac{\Delta M_q}{(\Delta M_q)_\text{SM}}\qquad (q=d,s)\,.
\ee
Consequently, using \eqref{eq:dostomumu} and \eqref{eq:DMdos},
the golden relation between $Br(B_{d,s}\to\mu^+\mu^-)$ and $\Delta
M_d/\Delta M_s$ valid in CMFV models \cite{MFVrel} gets modified as
follows:
 \be\label{eq:r}
\frac{Br(B_s\to\mu^+\mu^-)}{Br(B_d\to\mu^+\mu^-)}= \frac{\hat
B_{B_d}}{\hat B_{B_s}} \frac{\tau(B_s)}{\tau(B_d)} \frac{\Delta
M_s}{\Delta M_d}\,r\,,\quad r= \left|\frac{Y_s}{Y_d}\right|^2
\frac{C_{B_d}}{C_{B_s}}\,,
\ee
with $r$ being generally different from unity.

The most recent SM predictions read \cite{BBGT}
\be
Br(B_s\to\mu^+\mu^-)=(3.35 \pm 0.32) \cdot 10^{-9}\,,\qquad
Br(B_d\to\mu^+\mu^-)=(1.03 \pm 0.09) \cdot 10^{-10}\,,
\ee
to be compared with the experimental upper bounds from CDF \cite{CDF}
\be
Br(B_s\to\mu^+\mu^-)< 1 \cdot 10^{-7}\,,\qquad
Br(B_d\to\mu^+\mu^-)< 3 \cdot 10^{-8}\,.
\ee

\subsection{\boldmath${B\to X_{s,d}\nu\bar\nu}$\unboldmath}

We will also study the theoretical clean decay ${B\to
  X_{s,d}\nu\bar\nu}$ and look at the ratios
\bea
\frac{Br(B\to X_s\nu\bar\nu)}{Br(B\to X_s\nu\bar\nu)_\text{SM}}&=&
\left|\frac{X_s}{X_\text{SM}}\right|^2,\\
\frac{Br(B\to X_d\nu\bar\nu)}{Br(B\to X_d\nu\bar\nu)_\text{SM}}&=&
\left|\frac{X_d}{X_\text{SM}}\right|^2,\\
\frac{Br(B\to X_d\nu\bar\nu)}{Br(B\to X_s\nu\bar\nu)}&=&
\left|\frac{X_d}{X_s}\right|^2 \left|\frac{V_{td}}{V_{ts}}\right|^2,
\label{eq:BrXdXs}
\eea
with $\vtd$ and $\vts$ obtained from tree level decays. Note that for
$X_d\ne X_s$ the relation of the last ratio to $|V_{td}/V_{ts}|$ is
modified with respect to MFV models.

\newsection{Results for the T-even Sector}\label{sec:Teven}

The contribution from the T-even sector, denoted by
$\bar X_\text{even}$ and $\bar Y_\text{even}$ in \eqref{eq:Xi} and
\eqref{eq:Yi}, respectively, can be extracted from \cite{BPUB},
where the functions $X$ and $Y$ have been calculated in the LH
model without T-parity. Imposing
  T-parity implies
\be s=c=s'=c'=\frac{1}{\sqrt{2}} \ee and vanishing of the diagrams
in the classes $1$, $2$, $4$ and $6$ in \cite{BPUB}. Moreover there are no
corrections from the breakdown of custodial symmetry, and the
left-over divergence, discussed in detail in \cite{BPUB}, is also
absent.

We find then
\bea\label{eq:Xeven}
\bar X_\text{even} &=& x_L^2 \frac{v^2}{f^2}\left[
  U_3(x_t,x_T)+\frac{x_L}{1-x_L} \frac{x_t}{8} \right],\\
\label{eq:Yeven}
\bar Y_\text{even} &=& x_L^2 \frac{v^2}{f^2}\left[
  V_3(x_t,x_T)+\frac{x_L}{1-x_L} \frac{x_t}{8} \right]\,,
\eea with the two terms on the r.h.s.~coming from class $3$ and $5$ in
\cite{BPUB}, respectively. The functions $U_3(x_t,x_T)$ and $V_3(x_t,x_T)$
are given in Appendix \ref{sec:appC}.

\newsection{Results for the T-odd Sector}\label{sec:Todd}

\subsection{Preliminaries}\label{sec:5.1}

In Figs.~\ref{fig:ping} and \ref{fig:box} we show the diagrams contributing to
$\bar X_i^\text{odd}$ in \eqref{eq:Xi}. Similar diagrams but with external
charged leptons contribute to $\bar Y_i^\text{odd}$ in \eqref{eq:Yi}. 
They can be divided into two classes. The first class involves in the unitary gauge $W_H^\pm$ and
the mirror fermions $u^i_H$ exchanges, while the second class involves 
$Z_H,\;A_H$ and $d^i_H$ exchanges. In the renormalizable $R_\xi$ gauges also
diagrams with Goldstone bosons have to be included. 
We will first discuss the calculation in the unitary gauge.
Subsequently we will turn to the calculation in the 't
Hooft-Feynman gauge, verifying our result in the unitary gauge.

\subsection{$\bm{Z_L}$-Penguin Diagrams in the Unitary Gauge}
\label{sec:5.2}

Only $Z_L$-penguin diagrams contribute to the decays considered because the couplings of
$Z_H$ and $A_H$ to $\nu\bar\nu$ and $\mu^+\mu^-$ vanish due to
T-parity. We note that the diagrams with internal $W_H^\pm$ are
fully analogous to the corresponding SM diagrams with internal
$W_L^\pm$. Moreover the diagrams with triple gauge boson
vertices vanish in the case of internal $A_H$ and $Z_H$ contributions.

\begin{figure}
\center{\epsfig{file=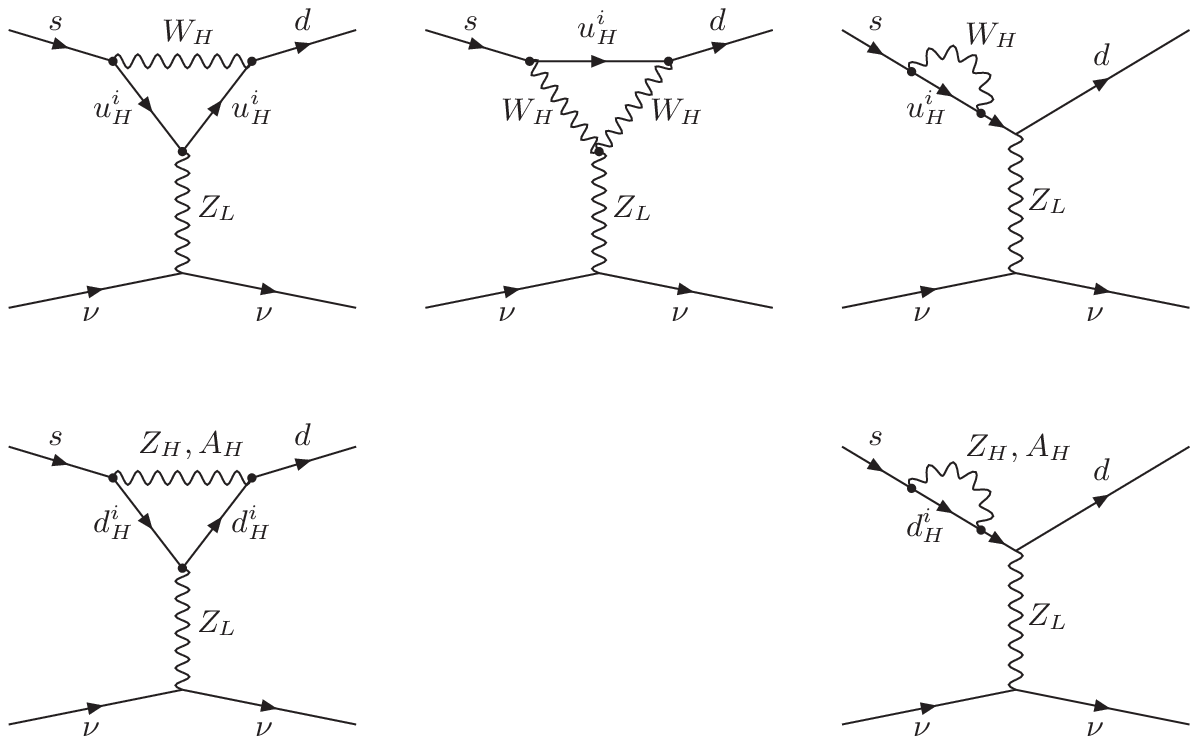}}
\caption{\it $Z_L$-penguin diagrams contributing in the T-odd sector.}
\label{fig:ping}
\end{figure}

There are two additional features with respect to the SM
calculation and the box diagram calculation presented in
\cite{BBPTUW} and below:
\begin{itemize}
\item The diagrams in Fig.~\ref{fig:ping}
  appear first to be $\ord{(1)}$, that is they are not suppressed by
  $v^2/f^2$.
\item The couplings of mirror fermions to $Z_L$ are vectorial
  ($\gamma^\mu$) in contrast to the SM couplings that have both
  $\gamma^\mu$ and $\gamma^\mu\gamma_5$ components.
\end{itemize}
Clearly the $\ord(1)$ contributions have to vanish as otherwise it
would not be possible to decouple the mirror fermions in the limit
$f\to\infty$. This is assured by the vectorial coupling of $Z_L$
to the mirror fermions. The missing of diagrams with triple gauge
boson vertices in the neutral gauge boson case is compensated by
the difference between $\bar d^i_H Z_L^\mu d^i_H$ and $\bar u^i_H
Z_L^\mu u^i_H$ couplings, so that the charged ($W_H^\pm$) and
neutral ($Z_H,\,A_H$) gauge boson contributions of $\ord(1)$ to the
$Z_L$-penguin vanish independently of each other in the unitary
gauge.

As the inclusion of $v^2/f^2$
  corrections to the neutral gauge boson interactions leads only to an
  overall factor multiplying the $Z_H$ and $A_H$ contributions, which
  vanish independently of each other,  we find that there is no contribution
from mirror fermions to $Z_L$-penguin diagrams in the unitary
gauge. The inclusion of $v^2/f^2$ corrections to the relations between
the masses of $u^i_H$ and $d^i_H$ and to the gauge boson masses does
not change this result. 

\subsection{Box Diagrams in the Unitary Gauge}

\begin{figure}
\begin{center} \epsfig{file=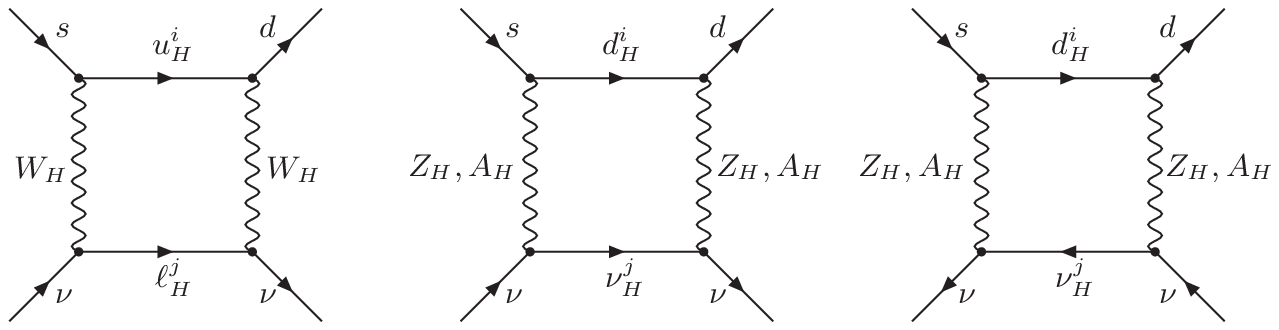}\end{center}
\caption{\it Box diagrams  in the unitary gauge.}\label{fig:box}
\end{figure}

In order to simplify the formulae, we will present the results for the
box
diagrams shown in Fig.~\ref{fig:box} in the limit of degenerate mirror leptons. As the box
contributions vanish in the limit of degenerate mirror quarks, the
inclusion of mass splittings in the lepton spectrum is a higher order
 effect. We have numerically verified that it can be neglected for the range
 of mirror fermion masses considered in the analysis.

Let us begin with the neutral gauge boson contributions. Similarly to
$\Delta F=2$ transitions considered in \cite{Hubisz,BBPTUW}, only the $g^{\mu\nu}$ part of the gauge
boson propagators is relevant. The contributions involving $k^\mu
k^\nu/M_{W_H}^2$ cancel each other between the two last sets of box diagrams
in Fig.~\ref{fig:box}. Consequently the neutral gauge
boson box contributions to $\bar X_i^\text{odd}$ and  $\bar
Y_i^\text{odd}$ are gauge independent. This means that the neutral
gauge boson contributions to $Z_L$-penguins must vanish in an
arbitrary gauge which is confirmed through an explicit calculation, as
discussed below.

The result for the box contributions involving $W_H^\pm$ turns out
to be divergent. As box contributions in a renormalizable gauge,
like the 't Hooft-Feynman gauge, are finite by power counting, the
box diagram contributions involving $W_H^\pm$ must then be gauge
dependent. Before giving the result for the unitary gauge
calculation, including also neutral gauge boson contributions, let
us repeat the calculation in the 't Hooft-Feynman gauge.

\subsection{Calculation in the 't Hooft-Feynman Gauge}

In the 't Hooft-Feynman gauge also the diagrams with Goldstone bosons have
to be included. Let us first compute the $Z_L$-penguin diagrams. The $\mathcal{O}\left(1\right)$ contributions
vanish as expected and we have to consider
$\mathcal{O}\left(v^{2}/f^{2}\right)$ corrections. Here, as in the
unitary gauge, there are no contributions from diagrams involving
only gauge bosons. On the other hand diagrams with Goldstone
bosons contribute at $\mathcal{O}\left(v^{2}/f^{2}\right)$. To
this end we had to generalize the Feynman rules of \cite{Hubisz}
to include $\mathcal{O}\left(v^{2}/f^{2}\right)$ corrections to
vertices involving Goldstone bosons. It turns out that
$\mathcal{O}\left(v^{2}/f^{2}\right)$ corrections to 
quark--mirror quark--Goldstone boson vertices cancel in the
calculation, which implies that the neutral gauge boson
contributions to the $Z_{L}$-penguin, not having triple gauge
boson vertices and corresponding vertices with Goldstone bosons,
vanish also in the 't Hooft-Feynman gauge. This was to be expected
as the box contributions from neutral gauge bosons are gauge
independent.

\begin{figure}
\begin{center} \epsfig{file=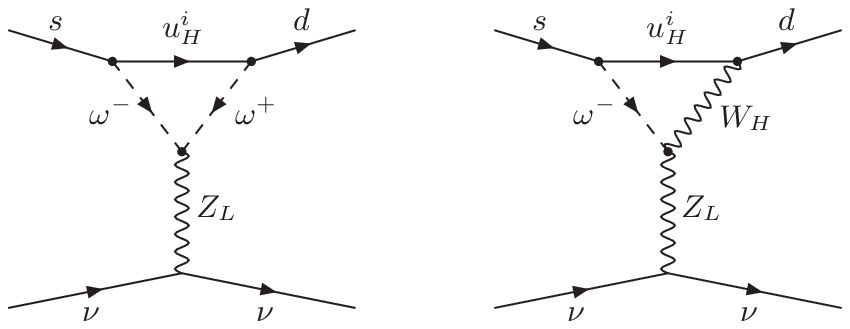}\end{center}
\caption{\it $\mathcal{O}\left(v^{2}/f^{2}\right)$ contributions to
  $Z_L$-penguin in the 't Hooft-Feynman gauge.}\label{fig:feyn}
\end{figure}

Thus in the 't Hooft-Feynman gauge only two diagrams at
$\mathcal{O}\left(v^{2}/f^{2}\right)$, shown in Fig.~\ref{fig:feyn},
contribute to the $Z_{L}$-penguin vertex. Using the Feynman rules
of Appendix \ref{sec:appB} we find that the first diagram in
Fig.~\ref{fig:feyn} is divergent with the divergence precisely equal to the
one found in box diagrams with $W_{H}^{\pm}$ exchanges calculated
in the unitary gauge. 

Including next the finite contributions from
the diagrams in Fig.~\ref{fig:feyn} and the finite contributions from box
diagrams with $W_{H}^{\pm}$ and Goldstone boson exchanges in the
't~Hooft-Feynman gauge, we confirm the final results for
$\bar{X}_{i}^\text{odd}$ and $\bar{Y}_{i}^\text{odd}$ obtained in
the unitary gauge.

\subsection{Final Results for the T-odd sector}

As described above, we have performed the calculation of the functions $\bar{X}_{i}^\text{odd}$ and $\bar{Y}_{i}^\text{odd}$ in the unitary
gauge and in the 't Hooft-Feynman gauge obtaining the same results. In
 particular, we found that the left-over divergence obtained in the
unitary gauge was not an artifact of a non-renormalizable gauge
but a physical gauge independent result. A similar divergence has
been found in the $Z_{L}$-penguin calculation in the LH model
without T-parity \cite{BPUB}. We will recall the interpretation of
this divergent contribution given in \cite{BPUB} in the next section.

The final results for $\bar{X}_{i}^\text{odd}$ and $\bar{Y}_{i}^\text{odd}$ in the LHT model are then given as follows:
\begin{eqnarray}
\bar{X}_{i}^\text{odd}&=& \left[
\xi_2^{(i)}\big(J^{\nu\bar\nu}(z_2,y)-J^{\nu\bar\nu}(z_1,y)\big)
+\xi_3^{(i)}\big(J^{\nu\bar\nu}(z_3,y)-J^{\nu\bar\nu}(z_1,y)\big)
\right],
\label{Xodd}\\
\bar{Y}_{i}^\text{odd}&=& \left[
\xi_2^{(i)}\big(J^{\mu\bar\mu}(z_2,y)-J^{\mu\bar\mu}(z_1,y)\big)
+\xi_3^{(i)}\big(J^{\mu\bar\mu}(z_3,y)-J^{\mu\bar\mu}(z_1,y)\big)
\right] ,\label{Yodd}
\end{eqnarray}
where
\begin{eqnarray}
J^{\nu \bar{\nu}}\left(z_{i}, y\right)&=&
\frac{1}{64}\frac{v^2}{f^2}\bigg[z_i S_\text{odd} +F^{\nu
    \bar{\nu}}(z_i,y;W_H)\nn\\
&& \qquad +4
\Big(G(z_i,y;Z_H)+G_1(z'_i,y';A_H)+G_2(z_i,y;\eta)\Big)\bigg],
\label{Znunu}\\
J^{\mu\bar\mu}\left(z_{i}, y\right)&=&
\frac{1}{64}\frac{v^2}{f^2}\bigg[{z_i} S_\text{odd} +F^{\mu\bar\mu}(z_i,y;W_H)\nn\\
&& \qquad -4
\Big(G(z_i,y;Z_H)+G_1(z'_i,y';A_H)-G_2(z_i,y;\eta)\Big)\bigg]
,\label{Zmumu}\\
S_\text{odd}&=&\frac{1}{\varepsilon}+\log\frac{\mu^2}{M_{W_H}^2}\,,\label{eq:Sodd}
\end{eqnarray}
with the functions $F^{\nu \bar{\nu}}$, $F^{\mu \bar{\mu}}$, $G$, $G_{1}$ and $G_{2}$ given in Appendix
\ref{sec:appC} and the various variables defined as follows
\begin{gather}
z_{i} = \frac{m_{Hi}^{2}}{M_{W_{H}}^{2}} = \frac{m_{Hi}^{2}}{M_{Z_{H}}^{2}}\,,
\qquad z_{i}^{\prime}= a z_{i} \quad\text{with }\; a = \frac{5}{\tan^{2} \theta_{W}}\,,\\
y = \frac{m_{H \ell}^{2}}{M_{W_{H}}^{2}} =
\frac{m_{H \ell}^{2}}{M_{Z_{H}}^{2}}\,, \qquad y^{\prime}= y a\,,
\quad \eta = \frac{1}{a}\,.
\end{gather}

In the unitary gauge the results in (\ref{Xodd})-(\ref{Zmumu})
follow from box diagrams only, since the $Z_L$-penguin diagrams do not
contribute in this gauge, as discussed in Section \ref{sec:5.2}. The notation in (\ref{Znunu}) and
(\ref{Zmumu}) indicates which diagrams contribute to a given
function, with $G_2$ resulting from diagrams with both $Z_H$ and $A_H$
exchanges. In the 't Hooft-Feynman gauge the contribution of the
$Z_{L}$-penguin diagram is found to be
\begin{gather}
\Delta J^{\nu \bar{\nu}} = \Delta J^{\mu \bar{\mu}} \equiv \overline{\Delta J} \frac{1}{64}
\frac{v^{2}}{f^{2}}\,,\label{Feyngauge}\\
\overline{\Delta J} = {z_i}{S_\text{odd}} - 8 z_i R_2(z_i) +\frac{3}{2}
z_i + 2z_iF_2(z_i)\,,\label{DZbar}
\end{gather}
where the functions $R_{2}$ and $F_{2}$ are given in Appendix \ref{sec:appC}.

The box diagram contribution involving $W_{H}^{\pm}$ in the 't Hooft-Feynman
gauge can simply be obtained from (\ref{Xodd})-(\ref{Zmumu}) and
(\ref{Feyngauge}) using the gauge independence of $\bar
X_i^\text{odd}$ and $\bar
Y_i^\text{odd}$.

The formulae (\ref{Xodd})-(\ref{Zmumu}) are the main results of our paper.

\newsection{The Issue of left-over Singularities}\label{sec:sing}

It may seem surprising that FCNC amplitudes considered
in the previous section contain residual ultraviolet
divergences reflected by the non-cancellation of the $1/ \varepsilon$
poles at $\mathcal{O}\left(v^{2}/f^{2}\right)$ in our unitary gauge
calculation.
Indeed due to the GIM mechanism the FCNC processes considered here vanish at 
tree level both in the SM and in the LHT model in question. 
Therefore within the particle content of the low energy representation of 
the LHT model there seems to be no freedom to cancel the left-over divergences 
as the necessary tree level counter terms are absent.

At first sight one could worry that the remaining divergence is 
an artifact of the unitary gauge calculation. However, an additional
calculation in the 't
Hooft-Feynman gauge convinced us that the found
divergence is gauge independent. A similar result has been found
in the context of the LH model without T-parity in \cite{BPUB} and
understood as the sensitivity of the decay
amplitudes to the UV completion of the LH model.\footnote{A similar
   singularity has been found independently in \cite{tparity}, in the context
   of the study of electroweak precision constraints.} The same
interpretation can be made here. After all, the LHT model is a
non-linear sigma model, which is a non-renormalizable description of the low 
energy behavior of a symmetric
theory below the scale where the symmetry is dynamically broken.

We have found explicitly that in the 't Hooft-Feynman gauge 
the singularity followed entirely from the interactions of
the Goldstone bosons of the dynamically broken global symmetry
with the fermions.
As the Goldstone bosons in question are the only reminiscences of
the spontaneous symmetry breakdown present in the low energy
theory, the estimate of the size of the divergences through their
interactions with fermions in Fig.~\ref{fig:feyn} should in
principle be adequate. However, as emphasized in \cite{BPUB}, the
light fermions may have a more complex relation to the fundamental
fermions of the ultraviolet completion of the theory. We refer the
reader to \cite{BPUB}, where a discussion of this issue and a
comparison with QCD can be found. 

In what follows we will as in
\cite{BPUB} remove $1/\varepsilon$ terms from (\ref{eq:Sodd}) and set
$\mu=\Lambda$ to obtain
\begin{equation}
J^{\nu\bar\nu}_\text{div}=J^{\mu\bar\mu}_\text{div}=
z_i\frac{1}{64}\frac{v^2}{f^2}\log\frac{\Lambda^2}{M_{W_H}^2}\,,
\end{equation}
as a minimal estimate of the UV sensitivity of the model.
Setting
\begin{equation}
\Lambda=4\pi f\,,\qquad \qquad v=246\gev\,,
\end{equation}
we find that for $f=1000\gev$, implying $M_{W_H}=652\gev$,
\begin{equation}
J^{\nu\bar\nu}_\text{div}=J^{\mu\bar\mu}_\text{div}= z_i\cdot
0.006\,.
\end{equation}

\begin{figure}
\center{{\epsfig{file=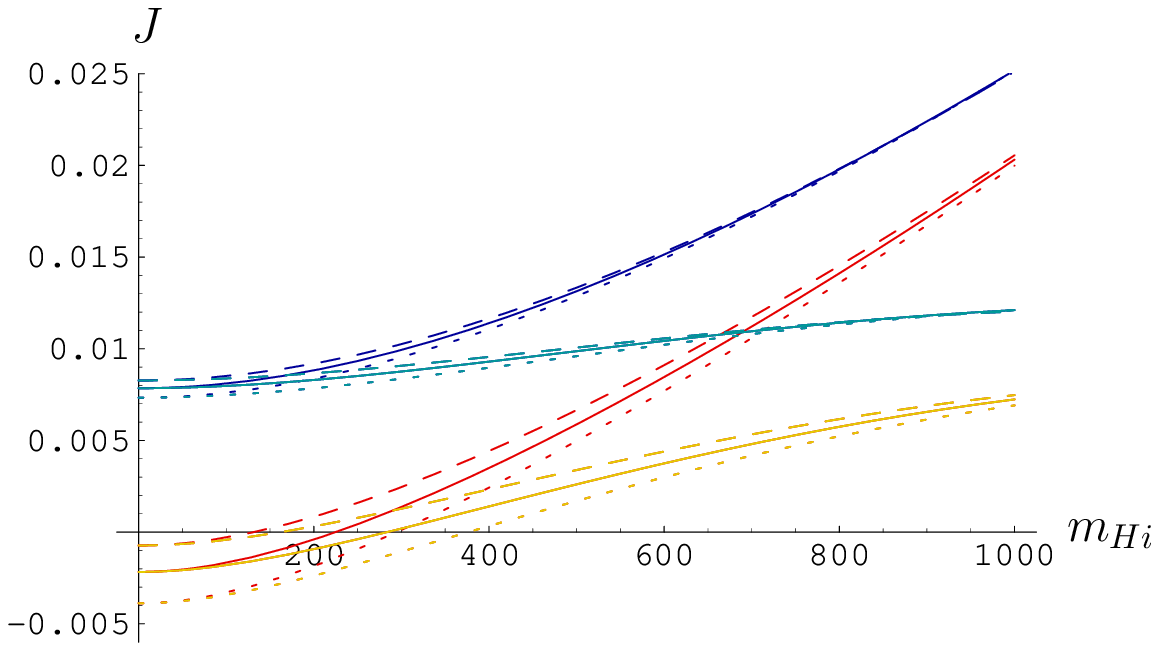,scale=0.8}}}
\caption{\textit{$J^{\nu\bar\nu}$ (lower)  and
$J^{\mu\bar\mu}$ (upper) as functions of $m_{Hi}$ for values of $m_{H \ell}=$400\,GeV(dotted), 500\,GeV (solid) and 600\,GeV (dashed) with (dark) and without (light) $J^{\nu\bar\nu}_\text{div}=J^{\mu\bar\mu}_\text{div}$ included.}\label{fig:Jdiv}}
\end{figure}

In Fig.~\ref{fig:Jdiv} we plot $J^{\nu\bar\nu}$ and
$J^{\mu\bar\mu}$ as functions of $m_{Hi}$ for three values of $m_{H \ell}$
with and without $J^{\nu\bar\nu}_\text{div}$ and
$J^{\mu\bar\mu}_\text{div}$ included. We observe that the
divergences constitute a sizable fraction of the total result. 
The coefficient of $z_i$ in the divergent terms $J^{\nu\bar\nu}_\text{div}$ and
$J^{\mu\bar\mu}_\text{div}$ is of the same order
of magnitude of the analogous linear coefficient in the convergent contributions, but
roughly four times larger.
Moreover, the linear contribution in the range of mirror fermion masses
considered is the dominant one, thus explaining the important impact of the divergences. 
At first sight this could imply the loss of the predictive power of
the theory as our estimate of the divergent contribution is
clearly an approximation. On the other hand the divergence found
has a \emph{universal} character and we can simply write
\begin{equation}
J^{\nu\bar\nu}_\text{div}=J^{\mu\bar\mu}_\text{div}= \delta_\text{div}\, z_i
\end{equation}
and treat $\delta_\text{div}$ as a free parameter. Assuming that
$\delta_\text{div}$ encloses all effects coming from the UV completion, which
is true if light fermions do not have a more complex relation to the 
fundamental fermions
of the UV completion that could spoil its flavour independence,
one can in principle fit $\delta_\text{div}$ to the data and trade it for one
observable.
At present this is not feasible, but could become realistic when more
data for FCNC processes will be available.

 On the other hand, implementing T-parity removes all divergences from the
  T-even sector. This is easy to understand. The only new T-even
  particle is $T_+$ which can be thought of as an arbitrary singlet
  field mixing with the SM top quark, independently of the non-linear
  sigma model. Of the ``pion'' matrix $\Pi$, only the SM Higgs doublet
is present in the T-even sector, and all modifications in its
couplings appear due to the mixing of $T_+$ with $t$. Thus the T-even
sector of the LHT model is effectively decoupled from the breaking
$SU(5)\to SO(5)$ of the non-linear sigma model, which has been the basic reason
for the appearance of the singularity described above and in \cite{BPUB}.

\newsection{\boldmath $B\to X_s\ell^+\ell^-$\unboldmath}
\label{sec:BSLL}

\subsection{Preliminaries}
 The branching ratio for the rare decay $B\to X_s \ell^+\ell^-$ depends in the SM 
 on the  functions $Y_{\rm SM}$, $Z_{\rm SM}$ and $D^\prime_{\rm SM}$ 
with the latter 
 relevant also for the $B\to X_s\gamma$ decay. The formulae for the branching 
 ratio are
very complicated and will not be presented here. They can be found in
\cite{BXsl+l-}, where also the formulae for the forward-backward
asymmetries are given.
 In the LHT model the function
$Y_{\rm SM}$ is generalized to $Y_s$ calculated in the previous section, 
whereas $D^\prime_{\rm LHT}$ has been calculated in \cite{BBPTUW} 
and is given for completeness in Appendix \ref{sec:appC}. 

What remains to be calculated is the function $Z_{s}$ that we defined
in \eqref{eq:Z_i}. The SM contribution can be written as
\be
Z_{\rm SM}=C_{0}+\frac{1}{4} D_{0}
\ee
with $C_{0}$ and $D_{0}$  given in the 't Hooft-Feynman gauge in Appendix \ref{sec:appC}.
$Z_{\rm SM}$ is gauge independent. We will also need the QCD-penguin function 
$E_0$ that can be found in Appendix \ref{sec:appC} as well.

\subsection{T-even Sector}
In the case of the T-even sector it is useful to work in the unitary gauge. 
The function $C_{\rm even}$ can be extracted from \cite{BPUB} by imposing 
 T-parity with the result 
\bea
C^{\rm
  even}_\text{unitary}&=&\frac{x_L^2}{8}\frac{v^2}{f^2}S_\text{even}\left(\frac{x_t-x_T}{2} - d_2 x_T \frac{v^2}{f^2}\right)\nn\\
&&-\frac{x_L^2}{16}\frac{v^2}{f^2}\bigg(\frac{-6-5x_t+5x_t^2-3x_T+3x_tx_T}{2(x_t-1)}\nn\\
&&\qquad +\frac{8x_t-10x_t^2+5x_t^3}{(x_t-1)^2}\log
x_t  -(4x_t+x_T)\log x_T\bigg)\nn\\
&&+\frac{x_L^2}{8}\frac{v^4}{f^4}x_T
\left( -\frac{3}{2}d_2+x_L^2+d_2 \log x_T\right)\,,
\eea
where
\be
S_\text{even}=\frac{1}{\varepsilon}+\log\frac{\mu^2}{M_{W_L}^2}\,,
\ee
and $d_2$ has been defined in \eqref{eq:xLd2}.

The function $D_{\rm even}$ in the LHT model can be obtained with the help
of $D_{\rm SM}$ in the unitary gauge. To our knowledge the latter function has 
never been given in this gauge in the literature. It can be found with the 
help of \cite{BPUB} as follows.

From the gauge independence of $Z_{\rm SM}$ we know that
\be
C_\text{SM}=C_0+\frac{1}{2}\bar\varrho_\text{SM}\,, \qquad D_\text{SM}=D_0-2\bar\varrho_\text{SM}\,,
\ee
where $\bar\varrho_\text{SM}$ is gauge dependent and vanishes in the 't Hooft-Feynman gauge.
It has been calculated in the unitary gauge in \cite{BPUB} with the result 
\be
\bar\varrho_\text{SM}=-\frac{1}{8}{x_t}{S_\text{even}}
-\frac{-3x_t^2+17x_t}{16(1-x_t)}-\frac{8x_t^2-x_t^3}{8(x_t-1)^2}\log x_t\,.
\ee
Consequently, using $D_0$ in Appendix \ref{sec:appC}, we find
\bea
D^\text{SM}_{\rm unitary}(x_t)&=& \frac{x_t}{4}S_\text{even}+\frac{-153
  x_t+383x_t^2-245x_t^3+27x_t^4}{72
  (x_t-1)^3}\nn\\
&&-\frac{16-64x_t+36x_t^2+93x_t^3-84x_t^4+9x_t^5}{36(x_t-1)^4} \log x_t\,.
\eea

Proceeding then as in the case of $B\to X_s\gamma$ in \cite{BPUB} 
but including also $v^4/f^4$ corrections to the diagrams with internal
$T_+$ exchanges we find
\be
D^{\rm even}_\text{unitary}=\frac{v^2}{f^2} x_L^2
\left[\left(1+2d_2\frac{v^2}{f^2}\right)D^\text{SM}_{\rm unitary}(x_T)-D^\text{SM}_{\rm unitary}(x_t)\right]\,,
\ee
and, dropping $\mathcal{O}(v^4/f^4)$ terms,
\bea
D^{\rm even}_\text{unitary}&=&\frac{v^2}{f^2} x_L^2
\left[ \frac{x_T}{4} \left(1+2d_2\frac{v^2}{f^2}\right)S_{\rm even}-
  D^\text{SM}_{\rm unitary}(x_t)\right]\nn\\
&&+\frac{v^2}{f^2} x_L^2
\left[-\frac{41-24\log x_T}{18}+\frac{x_T}{8}\left(1+2d_2\frac{v^2}{f^2}\right)(3-2\log x_T) \right]\,,
\eea
and subsequently
\be\label{7.9}
\bar Z_{\rm even}=C^{\rm even}_{\rm unitary}+\frac{1}{4} D^{\rm even}_{\rm unitary}
\ee
which is gauge independent.
The divergence $S_\text{even}$ cancels in \eqref{7.9} so that
$\bar Z_\text{even}$ is finite, in agreement with our statement in
Section \ref{sec:sing}.

\subsection{T-odd Sector}

In the T-odd sector it is useful to work in the 't Hooft-Feynman gauge.
Let us denote
\be
Z_{\rm odd}(z_i)=C_{\rm odd}(z_i)+\frac{1}{4} D_{\rm odd}(z_i)\,,
\ee
then, from (\ref{Feyngauge}) and (\ref{DZbar}), we find in the 't Hooft-Feynman gauge
\be
C_{\rm odd}(z_i)=\Delta J^{\mu\bar\mu}= \frac{1}{64}\frac{v^2}{f^2}
\left[ {z_i}{S_\text{odd}} - 8 z_i R_2(z_i) +\frac{3}{2}
z_i + 2z_iF_2(z_i)\right]  
\ee
with $S_\text{odd}$ defined in \eqref{eq:Sodd} and the functions $R_2$
and $F_2$ given in Appendix \ref{sec:appC}.

$D_{\rm odd}(z_i)$ is then found in analogy to our calculation 
of the $B\to X_s\gamma$ decay in \cite{BBPTUW}. We obtain
\be
D_{\rm odd}(z_i)=\frac{1}{4}\frac{v^2}{f^2}\left[ D_0(z_i)-\frac{1}{6}E_0(z_i)-\frac{1}{30}E_0(z'_i) \right]
\ee
with $D_0$ and $E_0$ given in Appendix \ref{sec:appC}. Finally we have
\be
\bar Z_s^{\rm odd}=\bigg[ \xi^{(s)}_2 \big(Z_{\rm odd}(z_2)-Z_{\rm
    odd}(z_1)\big)+ \xi^{(s)}_3 \big(Z_{\rm odd}(z_3)-Z_{\rm
    odd}(z_1)\big) \bigg] \,.
\ee
As $D_{\rm odd}(z_i)$ is finite, the divergence in $C_{\rm odd}(z_i)$
remains in $Z_\text{odd}$. Its estimate can be done as outlined in
Section \ref{sec:sing}.

\newsection{\boldmath $K_L\to\pi^0\ell^+\ell^-$\unboldmath}
\label{sec:KpiLL}

The rare decays $K_L\to\pi^0e^+e^-$ and $K_L\to\pi^0\mu^+\mu^-$ are
dominated by CP-violating contributions. In the SM the main
contribution comes from the indirect (mixing-induced) CP violation and
its interference with the direct CP-violating contribution
\cite{D'Ambrosio:1998yj,Buchalla:2003sj,Isidori:2004rb,Friot:2004yr}. 
The direct
CP-violating  contribution to the branching ratio is in the ballpark of 
$4\cdot 10^{-12}$, while the CP conserving contribution is at
most $3\cdot 10^{-12}$. Among the rare $K$ meson decays, the decays in
question belong to the theoretically cleanest, but certainly cannot
compete with the $K\to\pi\nu\bar\nu$ decays. Moreover,
the dominant indirect CP-violating contributions are practically
determined by the measured decays $K_S\to\pi^0\ell^+\ell^-$ and the
parameter $\eps_K$. Consequently they are not as sensitive as the 
$K_L\to\pi^0\nu\bar\nu$ decay to new
physics contributions, present only in the subleading direct
CP violation. However, as pointed out 
in \cite{BFRS}, in
the presence of large new CP-violating phases the direct CP-violating
contribution can become the dominant contribution and the branching
ratios for $K_L\to\pi^0\ell^+\ell^-$ can be enhanced by a factor of
2--3, with a stronger effect in the case of
$K_L\to\pi^0\mu^+\mu^-$ \cite{Isidori:2004rb,Friot:2004yr}.

Adapting the formulae in \cite{Buchalla:2003sj,Isidori:2004rb,Friot:2004yr,MST}
with the help of \cite{BFRS} to the LHT model we find
\be\label{eq:BrKpiLL}
Br(K_L\to\pi^0\ell^+\ell^-)=\left(C_\text{dir}^\ell\pm
  C_\text{int}^\ell\left|a_s\right| +
  C_\text{mix}^\ell\left|a_s\right|^2+C_\text{CPC}^\ell\right)\cdot
10^{-12}\,,
\ee
where
\begin{align}
&C_\text{dir}^e = (4.62\pm0.24)(\omega_{7V}^2+\omega_{7A}^2)\,,&\qquad&
C_\text{dir}^\mu =(1.09\pm0.05)(\omega_{7V}^2+2.32\omega_{7A}^2)\,,\\
&C_\text{int}^e = (11.3\pm0.3)\omega_{7V}\,,&\qquad&
C_\text{int}^\mu = (2.63\pm0.06)\omega_{7V}\,,\\
&C_\text{mix}^e = 14.5\pm0.05\,,&\qquad&
C_\text{mix}^\mu = 3.36\pm0.20\,,\\
&C_\text{CPC}^e \simeq 0\,,&\qquad&
C_\text{CPC}^\mu = 5.2\pm1.6\,,\\
&&&\hspace{-1.5cm}\left|a_s\right|=1.2\pm0.2
\end{align}
with
\bea
\omega_{7V} &=& \frac{1}{2\pi}\left[P_0+\frac{|Y_K|}{\sin^2\theta_W}
  \frac{\sin\beta^K_Y}{\sin(\beta-\beta_s)}-4|Z_K|
  \frac{\sin\beta^K_Z}{\sin(\beta-\beta_s)}\right]\left[\frac{\IM
  \,\lambda_t}{1.4\cdot10^{-4}}\right]\,,\\
\omega_{7A} &=& -\frac{1}{2\pi}\frac{|Y_K|}{\sin^2\theta_W}
  \frac{\sin\beta^K_Y}{\sin(\beta-\beta_s)}\left[\frac{\IM
  \,\lambda_t}{1.4\cdot10^{-4}}\right]\,,
\eea
where $P_0=2.88\pm 0.06$ \cite{BLMM} includes NLO QCD corrections and
\be
\beta^K_Y=\beta-\beta_s-\theta^K_Y\,,\qquad 
\beta^K_Z=\beta-\beta_s-\theta^K_Z
\ee
with $Z_K$ defined in \eqref{eq:Z_i} and obtained from $Z_s$ by
changing $\xi^{(s)}_i$ to $\xi^{(K)}_i$.

The effect of the new physics contributions is mainly felt in
$\omega_{7A}$, as the corresponding contributions in $\omega_{7V}$
cancel each other to a large extent.

The present experimental bounds
\be
Br(K_L\to\pi^0e^+e^-)<28\cdot10^{-11}\quad\text{\cite{Alavi-Harati:2003}}\,,\qquad
Br(K_L\to\pi^0\mu^+\mu^-)<38\cdot10^{-11}\quad\text{\cite{Alavi-Harati:2000}}
\ee
are still by one order of magnitude larger than the SM predictions
\cite{MST}
\begin{gather}
Br(K_L\to\pi^0e^+e^-)_\text{SM}=
3.54^{+0.98}_{-0.85}\left(1.56^{+0.62}_{-0.49}\right)\cdot 10^{-11}\,,\label{eq:KLpee}\\
Br(K_L\to\pi^0\mu^+\mu^-)_\text{SM}= 1.41^{+0.28}_{-0.26}\left(0.95^{+0.22}_{-0.21}\right)\cdot 10^{-11}\label{eq:KLpmm}
\end{gather}
with the values in parentheses corresponding to the ``$-$'' sign in
\eqref{eq:BrKpiLL}.

\newsection{Benchmark Scenarios for New Parameters}
\label{sec:benchmark}
\subsection{Preliminaries}
\label{subsec:6.1}

In what follows, we will consider as in \cite{BBPTUW} several scenarios for the
structure of the $V_{Hd}$ matrix and the mass spectrum of mirror
fermions with the hope to gain a global view about the
possible signatures of mirror fermions in the processes
considered and of $T_+$ present in the T-even contributions.

In the most interesting scenario considered in \cite{BBPTUW}
(Scenario $4$ below), the mixing matrix $V_{Hd}$ differed significantly 
from $V_\text{CKM}$.
It could have a large non-vanishing complex phase 
$\delta_{13}^d$, while the phases $\delta_{12}^d$ and $\delta_{23}^d$,
  with smaller phenomenological impact, were set to zero. In this scenario
large CP-violating effects in $B_s$
decays have been found. In particular, the CP asymmetries
$S_{\psi\phi}$ and $A^s_\text{SL}$ could be enhanced by an order of
magnitude with respect to the SM expectations.

In the next section we will be  primarily interested in calculating
the observables considered in the previous sections in the scenarios
defined in \cite{BBPTUW}. In particular, it will be
interesting to see how the CMFV correlations between $K^0$,
$B^0_d$ and $B_s^0$ systems \cite{BBGT,MFVlectures} are modified when new sources of
flavour and CP violation are present. The parameterization of various
decays in terms of the functions $X_i$, $Y_i$ and $Z_i$ that we
defined in Section \ref{sec:XYZ} is very useful for such tests.

The main purpose of our numerical analysis is to have a closer look 
at six scenarios, with the first five considered already in our
previous study of particle-antiparticle mixing and $B\to
X_s\gamma$. We will recall these five scenarios and
introduce a sixth one which has particularly interesting effects in
$K\to\pi\nu\bar\nu$ and $K_L\to\pi^0\ell^+\ell^-$ decays.
In all these six scenarios we set the phases $\delta_{12}^d$ and
$\delta_{23}^d$ to zero.
The values of the observables that can only be
produced by allowing these phases to vary freely, are covered
by our general scan anyway. This simplification therefore
does not restrict the generality of the analysis. 
We will see that Scenarios $4$ and $6$ turn out to be the most appealing, since
they respectively provide large enhancements in the $B_s$ and $K$ systems.
Spectacular effects in both $B$ and $K$ systems, however, are not
simultaneously allowed in a single scenario.
Therefore, we complete the numerical analysis with a general scan over mirror
fermion masses and $V_{Hd}$ parameters, with also 
$\delta_{12}^d$ and $\delta_{23}^d$ free to differ from zero, finding some 
interesting points where
significant enhancements in both $B$ and $K$ systems occur.

\subsection{Different Scenarios}
\label{Scenarios}
Here we just list the scenarios in question:

\vspace{0.2cm}
{\bf{Scenario 1:}}

\noindent
In this scenario the mirror fermions will be degenerate in mass 
\be
m_{H1}=m_{H2}=m_{H3}
\ee
and only the T-even sector will contribute. This is the MFV limit of the LHT
model.

\vspace{0.2cm}
{\bf Scenario 2:} 

\noindent
In this scenario the mirror
fermions are not degenerate in mass and
\be
V_{Hd}=V_\text{CKM}\,.
\label{eq:VHdVCKM}
\ee
In this case there are no contributions of mirror fermions
to $D^0 - \bar D^0$ mixing and flavour violating $D$ meson decays, and
\be
\xi_2^{(q)}=\lambda_c^{(q)}\,, \qquad \xi_3^{(q)}=\lambda_t^{(q)}\,,
\label{eq:xi2xi3}
\ee
with $q=d,s$ and no index $q$ in the $K$ system. 
As discussed in
\cite{BBPTUW} and below this scenario differs from the MFV case.

\vspace{0.2cm}
{\bf Scenario 3:}

\noindent
 In this scenario we will choose a linear spectrum for mirror quarks
\be\label{S3}
m_{H1}=400\gev, \qquad  m_{H2}=500\gev, \qquad m_{H3}=600\gev\,,
\ee
 set $m_{H \ell}=500\gev$ and take an arbitrary matrix $V_{Hd}$ but with the
 phases $\delta_{12}^d$ and $\delta_{23}^d$ set to zero. 
We stress that similar
results are obtained by changing the values above by $\pm 30\gev$,
with similar comments applying to \eqref{S4} below.
In the remaining scenarios we will modify the mirror quark spectrum but
  keep  $m_{H \ell}=500\gev$.

\vspace{0.2cm}
{\bf Scenario 4:}

\noindent
This was our favorite scenario in which large departures from the 
SM and MFV in $B_s$ decays could be obtained and some problems
addressed in \cite{BBPTUW} could
be solved, with small effects in the experimentally well measured
quantities $\Delta M_K$ and $\eps_K$. In this scenario
\be\label{S4}
m_{H1}\approx m_{H2}= 500\gev\,,  \qquad     m_{H3}= 1000 \gev\,,
\ee
\be\label{S4a}
  \frac{1}{\sqrt{2}} \le s_{12}^d \le 0.99\,, \qquad 
             5\cdot 10^{-5}\le   s_{23}^d \le  2 \cdot 10^{-4}\,,
\qquad       4\cdot 10^{-2}\le  s_{13}^d \le 0.6\,,
\ee
$\delta_{12}^d$ and $\delta_{23}^d$ are set to zero, while the phase $\delta^d_{13}$ is arbitrary. 
The hierarchical structure of the CKM matrix
\be\label{CKMH}
s_{13}\ll s_{23}\ll s_{12}\,, \qquad (\text{CKM})
\ee
is changed in this scenario to
\be
s^d_{23}\ll s^d_{13}\le s^d_{12}\,, \qquad (V_{Hd})
\ee
so that $V_{Hd}$ looks as follows: 
\be\addtolength{\arraycolsep}{3pt}
V_{Hd} =  \begin{pmatrix}
 c^d_{12} & s^d_{12} & s^d_{13}e^{-i\delta^d_{13}}\\
-  s^d_{12} &  c^d_{12} &  s^d_{23}\\
- c^d_{12}  s^d_{13}e^{i\delta^d_{13}} & -  s^d_{12}  
s^d_{13}e^{i\delta^d_{13}} &1 \end{pmatrix}.
\label{eq:VHd4}
\ee
We would like to stress that with the degeneracy $m_{H1}\approx
m_{H2}$ the T-odd contributions in $\eps_K$ proportional to
$\IM(\xi_2)$ and $\RE(\xi_2)$ vanish, and only the T-odd term
proportional to $\IM(\xi_3)\RE(\xi_3)$ contributes. Being
$\IM(\xi_3)=s^d_{13}c^d_{23}s^d_{23}\sin\delta^d_{13}$, the hierarchy
chosen in this scenario for $V_{Hd}$, with $s^d_{23}\ll 1$, has the
advantage of suppressing mirror fermion effects in $\eps_K$, allowing
at the same time large CP-violating effects in the $B^0_s-\bar
B^0_s$ system \cite{BBPTUW}. Furthermore $\Delta M_s$ can be smaller than its SM value in this scenario,
and interesting effects in the $B^0_d-\bar B^0_d$ system are also
found.

It will be interesting to see whether in this scenario large
departures from the SM expectations for rare decays can be obtained.

\vspace{0.2cm}
{\bf Scenario 5:}

\noindent
In all the previous scenarios we will choose the first solution for
 the angle $\gamma$ from tree level decays as given in
 (\ref{eq:gamma}) below so 
 that only small departures from the SM in the $B^0_d-\bar B^0_d$ system 
 will be consistent with the data. In the present scenario one assumes
 the second solution for $\gamma$ in (\ref{eq:gamma}) in contradiction 
 with the SM and MFV. We have shown in \cite{BBPTUW} that the presence of 
 new flavour violating interactions could still bring the theory to
 roughly agree with the available data, in particular with the asymmetry $
S_{\psi K_S}$. 
In spite of that, the combined measurements on $A^d_\text{SL}$ and $\cos(2
 \beta + 2 \varphi_{B_d})$ and the indirect experimental estimate of 
 $A^s_\text{SL}$ make this scenario very
unlikely \cite{BBPTUW,UTFIT}, such that we will not consider it any further.

\vspace{0.2cm}
{\bf Scenario 6:}

\noindent
In studying this scenario we aim to enhance mirror fermion contributions to
 rare $K$ decays, keeping negligible effects in the experimentally well
measured quantities $\Delta M_K$ and $\eps_K$.
To this purpose we choose the mirror fermion masses as in Scenario 4
(see \eqref{S4})
since the near degeneracy between $m_{H1}$ and $m_{H2}$ helps to suppress
mirror fermion effects in $\Delta M_K$.

Concerning $\eps_K$, we recall that with the degeneracy $m_{H1}\approx
m_{H2}$ the T-odd contributions proportional to $\IM(\xi_2)$ and  $\RE(\xi_2)$ vanish, and only
the T-odd term proportional to $\IM(\xi_3) \RE(\xi_3)$ contributes.
In Scenario $4$ the hierarchical structure of $V_{Hd}$ is chosen as to satisfy
$\IM(\xi_3) \simeq 0$.
Here in Scenario $6$, instead, we suppress mirror fermion effects in $\eps_K$ 
due to the second and third generations, by requiring $\RE(\xi_3)=0$.
Setting also in this scenario the phases $\delta_{12}^d$ and $\delta_{23}^d$ to zero, the explicit expression of the real part reads
\be\label{rexi3}
\RE(\xi_3)=-c_{12}^d s_{12}^d \left({s_{23}^d}^2 - {c_{23}^d}^2 {s_{13}^d}^2\right) +
(\cos \delta^d_{13}) \, c_{23}^d s_{23}^d s_{13}^d \left({c_{12}^d}^2 - {s_{12}^d}^2\right)\,,
\ee
which vanishes for $\theta_{12}^d$, $\theta_{23}^d$ and $\theta_{13}^d$
(chosen in the first quadrant) satisfying
\begin{gather}
c_{12}^d=s_{12}^d=\frac{1}{\sqrt{2}}\,,\label{s12d}\\
s_{23}^d=\frac{s_{13}^d}{\sqrt{1+{s_{13}^d}^2}}\label{s13ds23d}\,.
\end{gather}

We note that while the value of $\theta_{12}^d$ is fixed to $45^\circ$ by
(\ref{s12d}), $\theta_{23}^d$ and $\theta_{13}^d$ have no specified value nor
order of magnitude, but (\ref{s13ds23d}) implies that only one of them is a
free parameter.
The matrix $V_{Hd}$ can then be expressed in terms of the two free parameters
$\theta_{13}^d$ and $\delta^d_{13}$  as
\be\addtolength{\arraycolsep}{3pt}\renewcommand{\arraystretch}{1.5}
V_{Hd} =  \left(\begin{array}{ccc}
 \frac{c_{13}^d}{\sqrt{2}} &
 \frac{c_{13}^d}{\sqrt{2}} & s_{13}^d e^{-i\delta^d_{13}}\\
- \frac{1}{\sqrt{2}\sqrt{1+{s_{13}^d}^2}}(1+{s_{13}^d}^2e^{i\delta^d_{13}}) &
\frac{1}{\sqrt{2}\sqrt{1+{s_{13}^d}^2}}(1-{s_{13}^d}^2e^{i\delta^d_{13}}) &
\frac{s_{13}^d c_{13}^d}{\sqrt{1+{s_{13}^d}^2}}\\
 \frac{s_{13}^d}{\sqrt{2}\sqrt{1+{s_{13}^d}^2}}(1-e^{i\delta^d_{13}}) &
-\frac{s_{13}^d}{\sqrt{2}\sqrt{1+{s_{13}^d}^2}}(1+e^{i\delta^d_{13}}) &
\frac{c_{13}^d}{\sqrt{1+{s_{13}^d}^2}}
\end{array}\right).\renewcommand{\arraystretch}{1.0}
\ee
Its structure becomes much simpler if the angle $\theta_{13}^d$ is
sufficiently small,
i.\,e., $s_{13}^d \le 0.1$, and reads
\be\addtolength{\arraycolsep}{3pt}
V_{Hd} \approx  \begin{pmatrix}
 \frac{1}{\sqrt{2}} &
 \frac{1}{\sqrt{2}} & s_{13}^d e^{-i\delta^d_{13}}\\
- \frac{1}{\sqrt{2}} &
\frac{1}{\sqrt{2}} &
s_{13}^d\\
 \frac{s_{13}^d}{\sqrt{2}}(1-e^{i\delta^d_{13}}) &
-\frac{s_{13}^d}{\sqrt{2}}(1+e^{i\delta^d_{13}}) & 1
\label{eq:VHd6}
\end{pmatrix}.
\ee
As we will see in Section \ref{sec:numerics} the very different structure of $V_{Hd}$
when compared with $V_{\rm CKM}$ implies enhancements in 
rare $K$ decays, without introducing problematic effects in 
$\Delta M_K$ and $\varepsilon_K$.
Moreover, as $V_{Hd}$ in (\ref{eq:VHd6}) has a different structure also
  from the (\ref{eq:VHd4}) one of Scenario 4, the new physics effects in the
$B^0_d-\bar B^0_d$ and mainly in the $B^0_s-\bar B^0_s$ system, turn out to be 
 small although visible.

\vspace{0.2cm}
{\bf General Scan:}

\noindent
As shown in Section \ref{sec:numerics}, Scenarios $4$ and $6$ turn out to be the
most interesting ones with, respectively, large new physics effects in the
$B_s$ and $K$ systems.
Such visible enhancements follow from the structure of $V_{Hd}$, primarily
required to satisfy the $\varepsilon_K$ and $\Delta M_K$ constraints, through
$\IM(\xi_3)\approx 0$ in Scenario $4$ and through $\RE(\xi_3)=0$ in Scenario $6$.
A further consequence of the $V_{Hd}$ structure is that in Scenario $4$
spectacular effects can be obtained in the $B_s$ system but not in the $K$
system and vice versa in Scenario $6$.
An even more interesting picture would be the simultaneous manifestation of
large enhancements in both $B$ and $K$ observables.
In order not to miss such a possibility, in addition to the scenarios
described above, we have performed a general scan over mirror fermion masses
and $V_{Hd}$ parameters.
To have a global view of the most general LHT effects, we have allowed
  here the phases $\delta_{12}^d$ and $\delta_{23}^d$ to differ from zero.
Qualitatively their effect is not significant, although they can help in
achieving very large effects in certain observables. 
We find that there exist some sets of masses and $V_{Hd}$
parameters where the new physics effects turn out
to be spectacular in both $B$ and $K$ systems.
We note that they do not really constitute a scenario, they rather appear in 
the plots shown in the next section as isolated (blue) points.
In contrast to previous scenarios, in fact, the blue
points corresponding to large new physics effects are quite sensitive to the
particular configuration of mirror fermion masses and $V_{Hd}$ parameters.

\newsection{Numerical Analysis }
\label{sec:numerics}
\subsection{Preliminaries}
\label{subsec:5.1}

In our numerical analysis we will set $|V_{us}|$, $|V_{cb}|$ and
$|V_{ub}|$ to their central values measured in tree level
decays~\cite{BBpage,CKM2005} and collected in
Table~\ref{tab:input}.

\begin{table}[ht]
\renewcommand{\arraystretch}{1}\setlength{\arraycolsep}{1pt}
\center{\begin{tabular}{|l||l|}
\hline
{\small$|V_{ub}|=0.00429(29)$} &{\small $G_F=1.16637(1)\cdot 10^{-5} \gev^{-2}$}\\
{\small $\vcb = 0.0416(9)$\hfill\cite{BBpage}} &{\small$\mw= 80.425(38)\gev$}\\\cline{1-1}
{\small$\lambda=|V_{us}|=0.225(1)$ \hfill\cite{CKM2005}} &{\small$\alpha=1/127.9$}\\\cline{1-1}
 {\small$|V_{ts}|=0.0409(9)$ \hfill\cite{UTFIT}} &{\small$\sin^2 \theta_W=0.23120(15)$}\\\cline{1-1}
{\small$A=0.822(16)$}&{\small$m_{K^0}= 497.65(2)\mev$} \\
{\small$R_b=0.447(31)$}&{\small$m_{B_d}= 5.2794(5)\gev$} \\
{\small$\beta=26.3(21)^\circ$}&{\small$m_{B_s}= 5.370(2)\gev$} \\
{\small$\beta_s=-1.28(7)^\circ$}&{\small
  $F_K=160(1)\mev$\hfill\cite{PDG}}\\\hline
{\small$\mcb= 1.30(5)\gev$}&{\small$F_{B_d}=189(27)\mev$ } \\
{\small$\mtb= 163.8(32)\gev$}&{\small$F_{B_s}=230(30)\mev$\qquad\hfill\cite{Hashimoto}} \\\hline
\end{tabular}  }
\caption {\textit{Values of the experimental and theoretical
    quantities used as input parameters.} }
\label{tab:input}
\renewcommand{\arraystretch}{1.0}
\end{table}

As the fourth parameter we will choose the angle $\gamma$ of the
standard unitarity triangle that to an excellent approximation equals the
phase $\delta_\text{CKM}$ in the CKM matrix. The angle $\gamma$ has been 
extracted
from $B \to D^{(*)} K$ decays without the influence of 
new physics with the
result~\cite{UTFIT}
\be
\gamma=(71 \pm 16)^\circ\,, \qquad \gamma=-(109 \pm 16)^\circ\,.
\label{eq:gamma}
\ee
Only the first solution agrees with the SM analysis of the unitarity triangle,
while the consistency of the second solution with data has
been investigated within Scenario 5 in our previous LHT analysis \cite{BBPTUW}.
It turns out that the combined measurements on $A^d_\text{SL}$
and $\cos(2 \beta + 2 \varphi_{B_d})$ and the indirect experimental estimate
of  $A^s_\text{SL}$ make this scenario very unlikely \cite{BBPTUW,UTFIT}.

We will consider here only the first
solution, whose uncertainty is sufficiently large to allow for significant
contributions from new physics.
The value of $\beta$ in Table \ref{tab:input} is obtained from $R_b$
and the first solution of $\gamma$, i.e. from tree level decays only and is
not affected by an eventual new physics phase.
Its difference from the value of $\beta$ obtained from the $S_{\psi K_S}$
asymmetry, $\beta(\psi K_S)=21.2 \pm 1.0$,  constitutes the ``$\sin 2 \beta$
problem'' which can be solved only in Scenarios $3-6$ \cite{BBPTUW}.

For the non-perturbative parameters entering the analysis of
particle-antiparticle mixing we choose and collect in Table~\ref{tab:input}
their lattice averages given in~\cite{Hashimoto}, which combine unquenched
results obtained with different lattice actions.  Other parameters
relevant for particle-antiparticle mixing can be found in \cite{BBPTUW}.

In order to simplify our numerical analysis we will, as in \cite{BBPTUW}, set all non-perturbative 
parameters to their central values and instead we will allow $\Delta M_K$, 
$\varepsilon_K$, $\Delta M_d$, $\Delta M_s$ and $S_{\psi K_S}$ to differ from 
their experimental values by $\pm 50\%$, $\pm 40\%$, $\pm 40\%$, $\pm 40\%$ 
and $\pm 8\%$, respectively. In the case of $\Delta M_s/\Delta M_d$ we
will choose $\pm 20\%$ as the error on the relevant parameter, $\xi$, is
smaller than in the case of $\Delta M_d$ and $\Delta M_s$ separately. The relevant expressions are given in
\cite{BBPTUW}. These uncertainties could appear rather conservative, but we 
do not want to miss any interesting effect by choosing too optimistic 
non-perturbative uncertainties.
The constraints from $B\to X_s\gamma$ and $B\to
X_s\ell^+\ell^-$ are also taken into account.
They turn out to be easily satisfied, within the present uncertainties, and
therefore to have only a minor impact.

In Scenarios $3-6$, the parameters $f$ and $x_L$ will be fixed to
$f=1000\gev$ and $x_L=0.5$ in accordance with electroweak precision
tests~\cite{mH}. 
Varying the breaking scale $f$ would obviously modify our results. The 
effect, however, turns out to be much smaller
   than one would naively expect from the $v^2/f^2-$``scaling". 
In other words, lower values of $f$ do not allow arbitrarily large NP contributions, since the
constraints imposed from the available data become more stringent in this case.

\subsection{The MFV Scenario 1}
\label{subsec:5.2}
Let us consider first the case of totally degenerate mirror 
fermions. In this case the odd contributions vanish due to the GIM 
mechanism~\cite{GIM}, the only new particle contributing is $T_+$ 
and the LHT model in this limit belongs to the class of MFV 
models. As only the T-even sector contributes, the new 
contributions to all FCNC processes are entirely dependent on only 
two parameters 
\be
x_L\,,\qquad f\,. \ee
Moreover, all the dependence on new physics contributions is
encoded in the functions
\be
X_{\rm even}=X_{\rm SM}+\bar X_{\rm even}\,, \qquad
Y_{\rm even}=Y_{\rm SM}+\bar Y_{\rm even}\,, \qquad
Z_{\rm even}=Z_{\rm SM}+\bar Z_{\rm even}\,.
\ee

There exist strong correlations between various processes
that are characteristic for models with MFV.

It should be emphasized that in this scenario  the ``$\sin 2\beta$ problem''
cannot be solved as it is a MFV scenario and that
$\Delta M_s\ge (\Delta M_s)_{\rm SM}$, which is not
favored by the CDF measurement \cite{CDFDMs}, as well as 
$\Delta M_d\ge (\Delta M_d)_{\rm SM}$.
In \cite{BB} the relations $\Delta M_{s,d}\ge (\Delta M_{s,d})_{\rm SM}$ have been proven to be valid
in constrained MFV, where flavour violation is governed entirely by the Yukawa
interactions and there are no new operators beyond the SM ones, and,
therefore, have been expected for this scenario.
We specify that in the numerical analysis of this scenario the $S_{\psi K_S}$
constraint is left out while the $\Delta M_{d,s}$ ones are taken into account.

\begin{figure}
\begin{minipage}{7.5cm}
\center{\epsfig{file=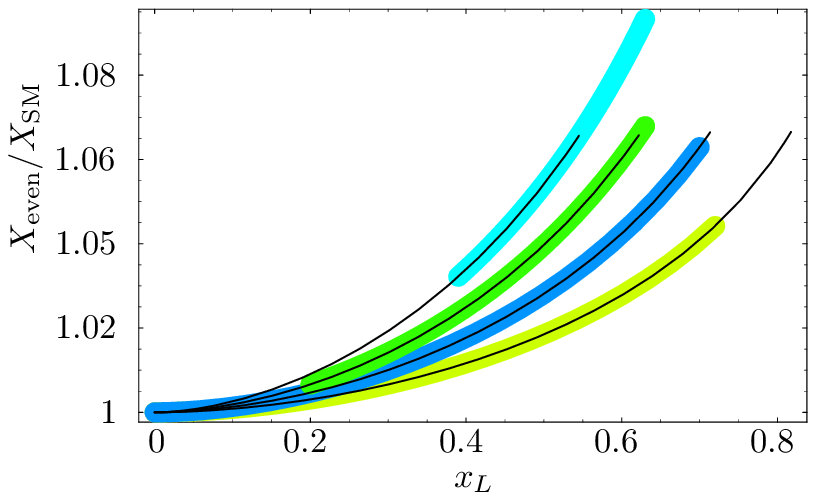,scale=0.8}}
\end{minipage}
\begin{minipage}{7.5cm}
\center{\epsfig{file=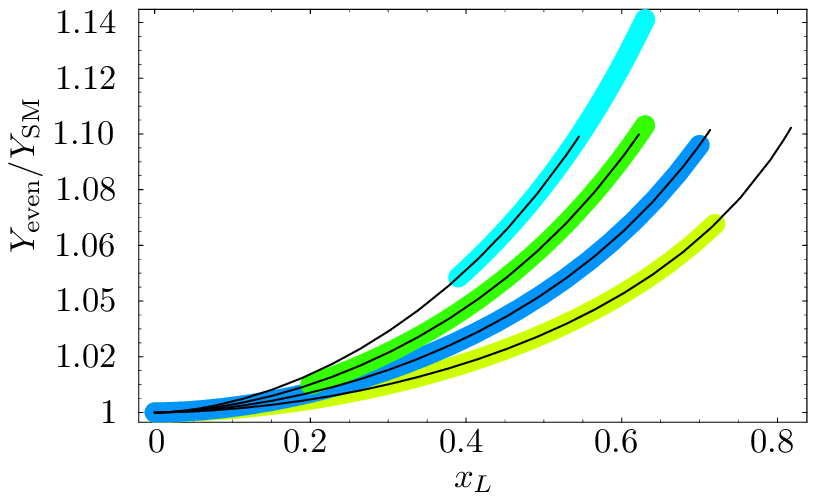,scale=0.8}}
\end{minipage}
\caption{\it $X_{\rm even}/X_\text{SM}$ (left) and $Y_{\rm even}/Y_\text{SM}$ (right) as functions of
  $x_L$,  for various values of $f=1,1.2,1.5$ and $2 \tev$ from top to bottom.
The bands underlying the curves show the allowed ranges after applying
electroweak precision constraints \cite{mH}.}
\label{fig:XYfirst}
\end{figure}

\begin{figure}
\begin{minipage}{7.5cm}
\center{\epsfig{file=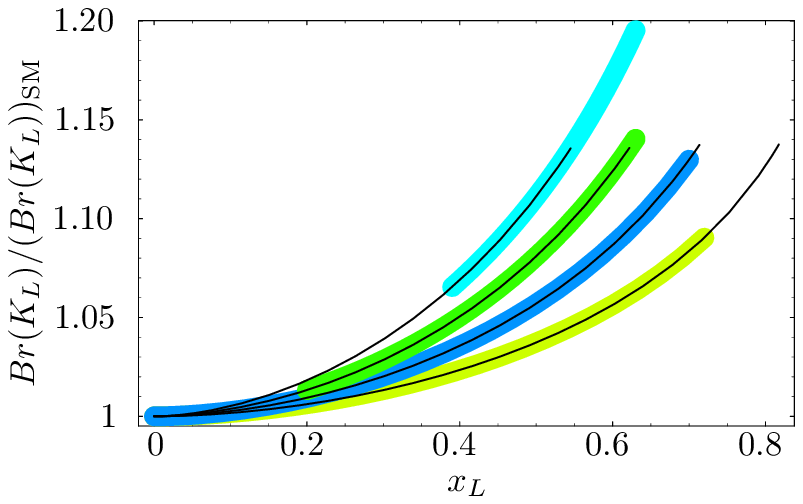,scale=0.8}}
\end{minipage}
\begin{minipage}{7.5cm}
\center{\epsfig{file=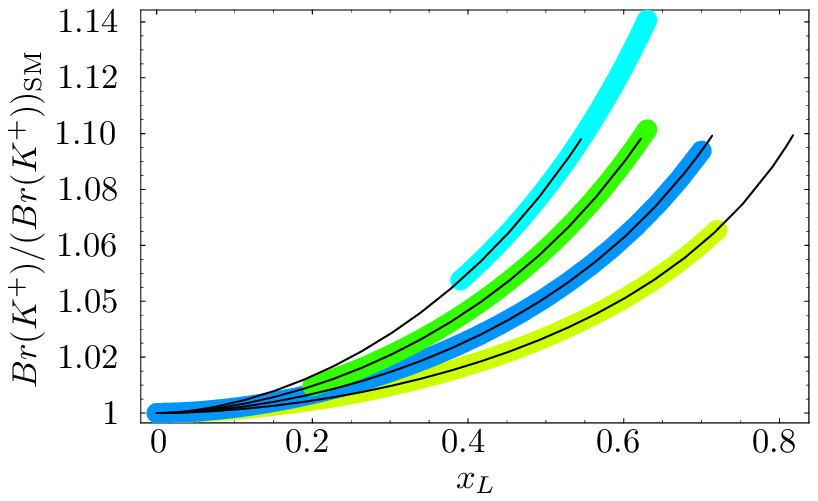,scale=0.8}}
\end{minipage}
\caption{\it $Br(\klpn)/Br(\klpn)_\text{SM}$ (left) and $Br(\kpn)/Br(\kpn)_\text{SM}$ (right) in Scenario $1$, as
functions of $x_L$ for different values of $f=1,1.2,1.5$ and $2 \tev$ from top to bottom.
The bands underlying the curves show the allowed ranges after applying
electroweak precision constraints \cite{mH}.}
\label{fig:KLKpfirst}
\end{figure}

In Fig.~\ref{fig:XYfirst} we plot $X_{\rm even}/X_\text{SM}$ and $Y_{\rm even}/Y_\text{SM}$ as
  functions of $x_L$ for various values of $f$.
We observe that the new physics contributions amount to modifications 
of the SM functions by at most $9\%$ and $14\%$, respectively, and of the
  corresponding rare decay branching ratios by at most $18\%$ and $28\%$.
As an example we show in Fig.~\ref{fig:KLKpfirst} $Br(\klpn)/Br(\klpn)_\text{SM}$ and $Br(\kpn)/Br(\kpn)_\text{SM}$ as
functions of $x_L$ for different values of $f$. 
We observe that slightly larger effects can occur in $Br(\klpn)$ relative to 
$Br(\kpn)$.

\subsection{Scenario 2}

\label{subsec:5.3}

The new contributions to FCNC processes are in this scenario entirely 
dependent on only six parameters
\be
x_L\,,\qquad f\,,\qquad 
m_{H1}\,,\qquad m_{H2}\,, \qquad m_{H3}\,, \qquad m_{H \ell}\,,
\ee
in addition to $m_t$ and the CKM parameters that we set to the central 
values obtained from tree level decays. 

In spite of the fact that in this scenario $V_{Hd}=V_{\rm CKM}$, 
it does not belong to the class of MFV models.
The point is that breaking the
    degeneracy of mirror fermion masses introduces a new source of 
 flavour violation that has nothing to do with the top Yukawa couplings. 
Only if accidentally the contributions proportional to $\xi^{(q)}_3=\lambda_t^{(q)}$  dominate the new physics contributions, one would again end up with a 
scenario that effectively looks like MFV.
However, as the mirror spectrum can
be 
generally different from the quark spectrum and not as hierarchical as the 
latter one, the terms involving $\xi^{(q)}_2=\lambda_c^{(q)}$ in the formulae
of the previous sections  cannot be neglected, while 
this can be done in the T-even contributions. 
Moreover, as  
$\lambda_c^{(q)}$ are different from $\lambda_t^{(q)}$, that dominate the 
SM contributions, even in this simple scenario the usual 
MFV relations between $K$, $B_d$ and $B_s$ systems will be violated.

Specifically, for $q=d,s$, $\lambda_c^{(q)}$ are of the same order of magnitude as
$\lambda_t^{(q)}$ and the MFV relations between $B_d$ and $B_s$ systems turn
out to be only weakly violated.
On the other hand, $\lambda_c^{(K)}/\lambda_t^{(K)}=\mathcal{O}(4 \cdot
10^{2})$, implying that not only the MFV relations between $K$ and $B_{d,s}$
systems are strongly violated, but also the rate
$Br(\kpn)$ can be
significantly enhanced in this scenario, relative to the SM and Scenario $1$.
In $Br(\kpn)$, in fact, the T-odd contribution proportional to 
$\RE(\lambda_c^{(K)})$ can have a significant effect, since 
$\RE(\lambda_c^{(K)})$ is much larger than 
$\IM(\lambda_t^{(K)})$ and $\RE(\lambda_t^{(K)})$.
In $Br(\klpn)$, instead, only the imaginary part of the CKM contributions
enters and, since $\IM(\lambda_c^{(K)})=- \IM(\lambda_t^{(K)})$, the T-odd
contribution can only yield a slight suppression.
\begin{figure}
\begin{minipage}{7.5cm}
\center{\epsfig{file=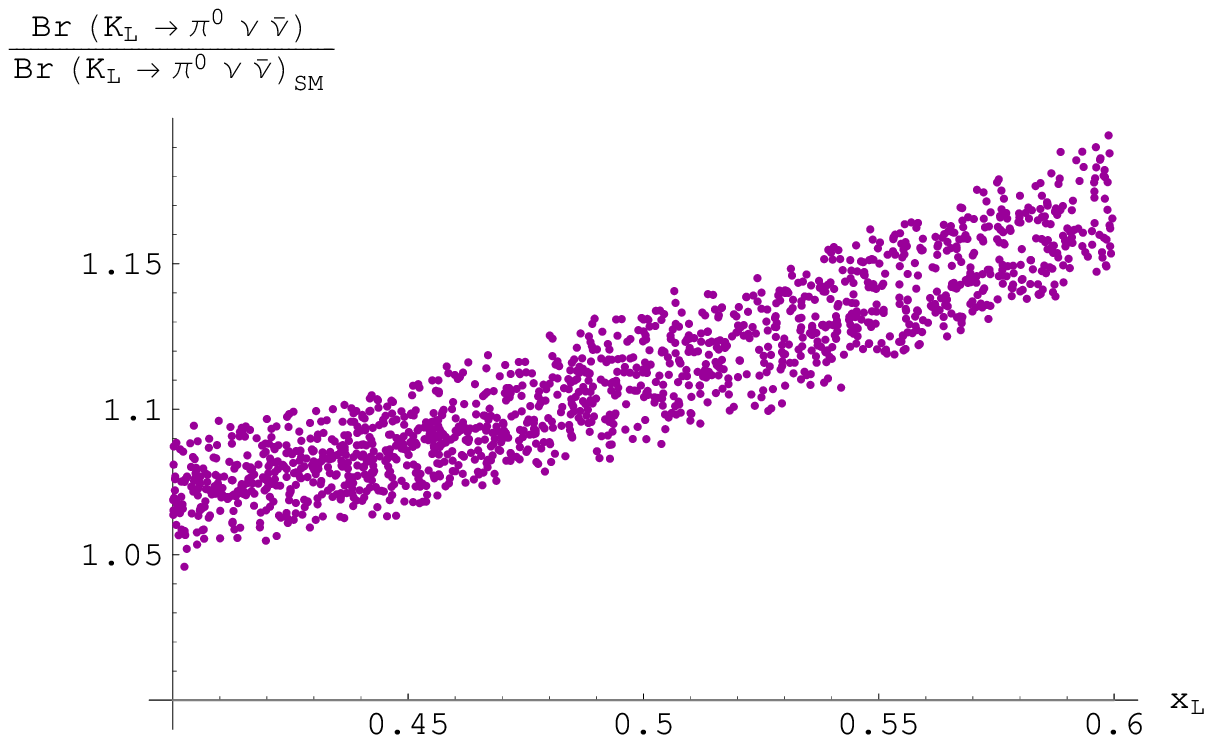,scale=0.6}}
\end{minipage}
\begin{minipage}{7.5cm}
\center{\epsfig{file=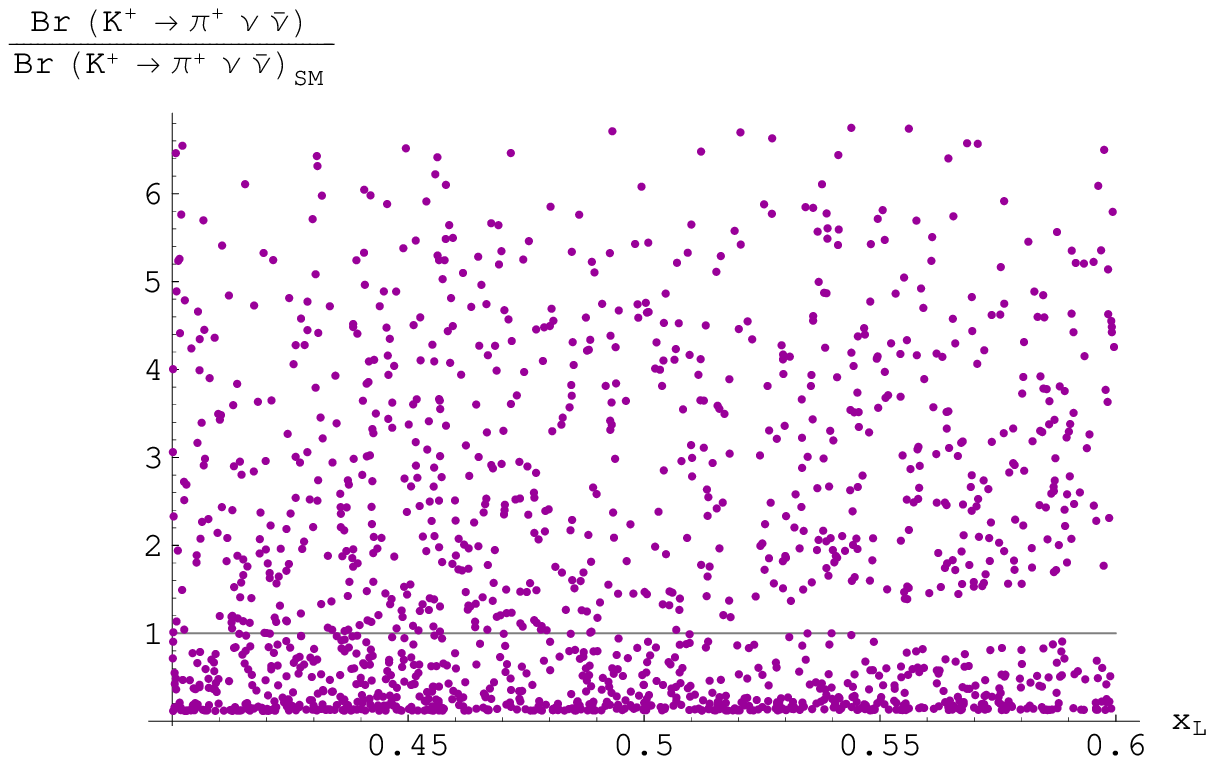,scale=0.6}}
\end{minipage}
\caption{\it $Br(\klpn)/Br(\klpn)_\text{SM}$ (left) and $Br(\kpn)/Br(\kpn)_\text{SM}$ (right) in Scenario $2$, as
functions of $x_L$, choosing $f=1 \tev$ and scanning over mirror fermion
masses. The solid line in the right plot represents the SM prediction and separates the regions where $Br(\kpn)$ is suppressed or enhanced relative to
the SM.}
\label{fig:KLKpsecond}
\end{figure}

As an example, we show in Fig.~\ref{fig:KLKpsecond} $Br(\klpn)/Br(\klpn)_\text{SM}$ and $Br(\kpn)/Br(\kpn)_\text{SM}$ as
functions of $x_L$ , choosing $f=1 \tev$ and scanning over mirror fermion
masses.
We point out that the central values of the SM predictions appearing in the 
ratios shown in Fig.~\ref{fig:KLKpsecond} differ from those quoted in 
(\ref{eq:SMKpinunu}) and read 
$Br(K^+\to\pi^+\nu\bar\nu)=8.7 \cdot 10^{-11}$, 
 $Br(K_{L}\to\pi^0\nu\bar\nu)=4.1 \cdot 10^{-11}$.
This difference comes from the CKM inputs that in the present analysis are
taken from tree-level decays only.
We observe that $Br(\klpn)$ can be enhanced at most by $17 \%$ relative to 
the SM prediction, with stronger new physics effects at higher values of the 
$x_L$  parameter.
In $Br(\kpn)$, instead, no clear dependence on the $x_L$ parameter can be
seen, while larger (of a factor $5$) enhancements as well as suppressions of
an order of magnitude can be obtained.

\subsection{Breakdown of the Universality}
\label{subsec:UB}
In MFV models the functions $X_i$, $Y_i$ and $Z_i$ are independent of the
index  $i=K,d,s$. Consequently, they are universal quantities implying strong 
correlations between observables in $K$, $B_d$ and $B_s$ systems. The 
presence of mirror fermions in the LHT model generally breaks this
universality, as we have already seen in Scenario $2$.
\begin{figure}
\center{\epsfig{file=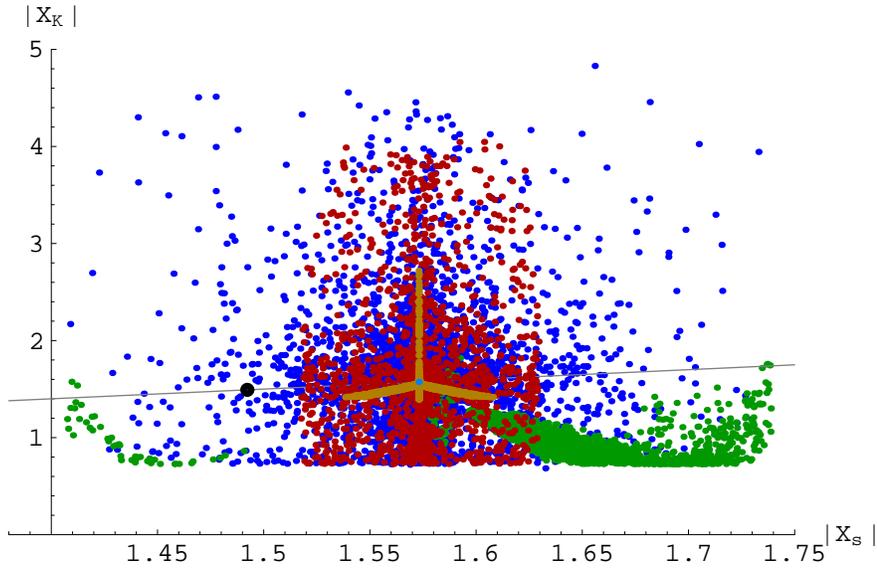}}
\caption{\it Breakdown of the universality between $|X_K|$ and $|X_s|$.}
\label{fig:XsXK}
\end{figure}

In Fig.~\ref{fig:XsXK} we show the ranges allowed in different
scenarios in the space $(|X_s|,|X_K|)$. 
Here and in all the following plots, Scenarios $3$, $4$ and $6$ are
represented by red, green and brown points, respectively, while blue points
stand for the general scan.
The solid line represents the MFV
relation $|X_s|=|X_K|$, with the black point corresponding to the SM prediction
$X_\text{SM}=1.49$ and the light blue point showing, for illustrative
purposes, the Scenario $1$ result.
The departure from the solid line gives the size of non-MFV 
contributions allowed in the various, differently coloured, scenarios.
We observe that roughly
\be
1.40 \le |X_s| \le 1.75\,, \qquad 0.7 \le |X_K| \le 4.7\,,
\ee
implying that the CP-conserving effects in the $K$ system can be 
much larger than in the 
$B_s$ system.
A very similar behavior is found for the 
functions $Y_i$ and $Z_i$ with larger effects in the $K$ system 
relative to the $B_{d,s}$ systems. For instance
\be
0.89\le |Y_s|\le 1.17\,, \qquad     0.41\le |Y_K|\le 3.9\,,
\end{equation}
to be compared with $|Y_s|=|Y_K|=0.95$ in the SM.

\begin{figure}
\center{\epsfig{file=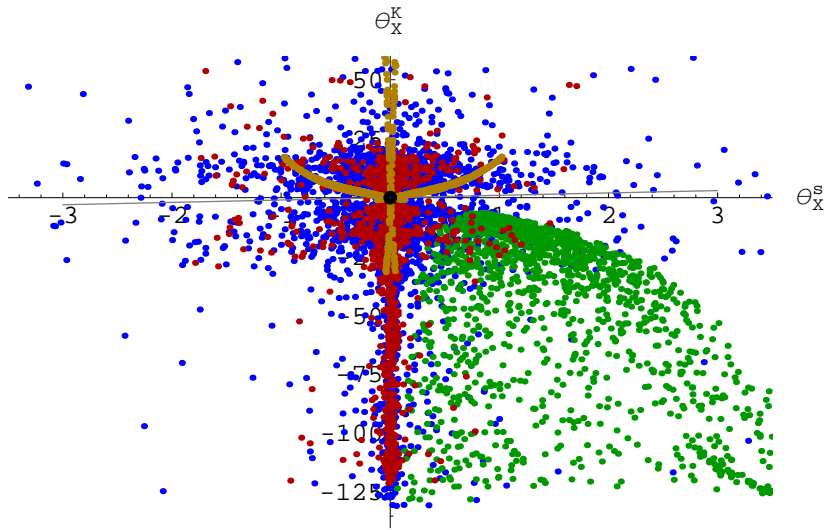}}
\caption{\it Breakdown of the universality between $\Theta_X^K$ and $\Theta_X^s$.}
\label{fig:TsTK}
\end{figure}
In Fig.~\ref{fig:TsTK}, then, we show the allowed ranges in the space
$(\Theta_X^s,\Theta_X^K)$, which roughly turn out to be
\begin{equation}
-3.5^\circ \le \Theta_X^s \le 3.5^\circ\,, \qquad 
-130^\circ \le \Theta_X^K\le 55^\circ\,,
\label{eq:ThetaKs}
\end{equation}
implying that the CP--violating effects in the $b\to s$ transitions
are tiny, while those in $K_L$ decays can be very large.
An analogous pattern is found for the $Y_{s,K}$ and $Z_{s,K}$ functions.
For the absolute values, the ranges in the $B_d$ and $B_s$ systems are
similar, while in the $B_d$ system larger values for the phase ($\pm
13^\circ$)  can be reached. 

From the discussion of the previous plots it is evident that mirror fermion contributions
break universality, and in a scenario-dependent way.
Furthermore, we would like to stress and explain the origin of larger effects
found in the $K$ system relative to the $B_{d,s}$ systems.
Looking at the $X_i$ expression in~(\ref{eq:Xi}) one sees that the mirror
fermion contribution is enhanced by a factor $1/\lambda_t^{(i)}$.
As  $\lambda_t^{(K)}\simeq 4\cdot 10^{-4}$, whereas
$\lambda_t^{(d)}\simeq 1\cdot 10^{-2}$ and $\lambda_t^{(s)}\simeq
4\cdot 10^{-2}$, we naively expect the deviation from $X_\text{SM}$ in the
$K$ system to be by more than an order of magnitude larger than in the
$B_d$ system, and even by two orders of magnitude larger than in the
$B_s$ system. Analogous statements are valid for the $Y_i$
and $Z_i$ functions.

In view of the smallness of the new physics contributions in $b\to s$ 
transitions it is easy to satisfy the constraints from $B\to
X_s\gamma$ and $B\to X_s \ell^+\ell^-$ that turn out to be close to the SM 
expectations. Therefore we will not further discuss them.

\boldmath
\subsection{The $K\to \pi\nu\bar\nu$ System}
\unboldmath

\begin{figure}
\center{\epsfig{file=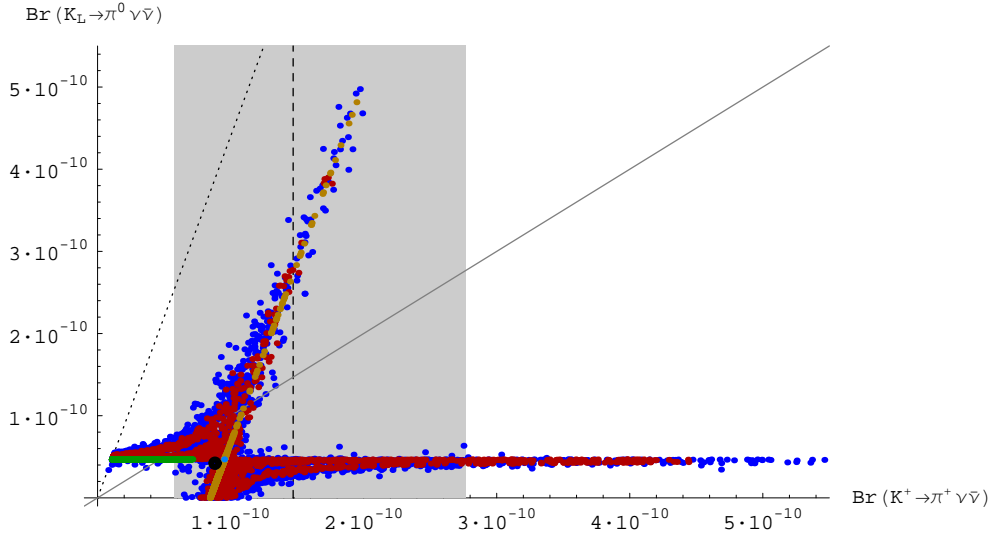,scale=0.8}}
\caption{\it $Br(\klpn)$ as a function of $Br(\kpn)$. The shaded
    area represents the experimental $1\sigma$-range for $Br(\kpn)$. The
    GN-bound is displayed by the dotted line, while the solid line
    separates the two areas where $Br(\klpn)$ is larger or smaller than
    $Br(\kpn)$.}
\label{fig:KLKp}
\end{figure}
In Fig.~\ref{fig:KLKp} we show the correlation between $Br(\kpn)$ and 
$Br(\klpn)$ for the scenarios in question. The experimental
$1\sigma$-range for $Br(\kpn)$ \cite{expK+} and the
model-independent Grossman-Nir (GN) bound \cite{GNbound} are also
shown. We observe that even for the most general case, there are two
branches of possible points. The first one is parallel to the GN-bound
and leads to possible huge enhancements in $Br(\klpn)$ so that values
as high as $5\cdot 10^{-10}$ are possible, being at the same time
consistent with the measured value for $Br(\kpn)$. Within Scenario $6$ (brown points),
this branch reduces to a straight line. The second branch corresponds
to values for $Br(\klpn)$ being rather close to its SM prediction,
while  $Br(\kpn)$ is allowed to vary in the range $[1\cdot
10^{-11},5\cdot 10^{-10}]$, however, values above $4\cdot 10^{-10}$
are experimentally not favored. We note that within Scenario $4$ (green points),
$Br(\klpn)$ is fixed to the T-even contribution and close to the SM
value, and $Br(\kpn)$ is always smaller than in the SM so that the
GN-bound can be reached.

\begin{figure}
\begin{minipage}{7.5cm}
\center{\epsfig{file=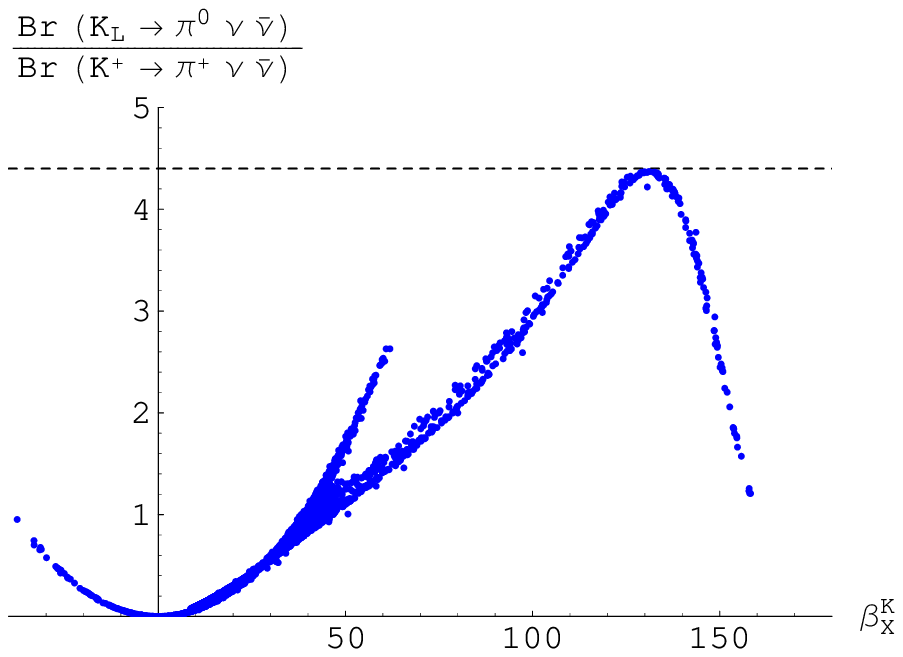,scale=0.7}}
\end{minipage}
\begin{minipage}{7.5cm}
\center{\epsfig{file=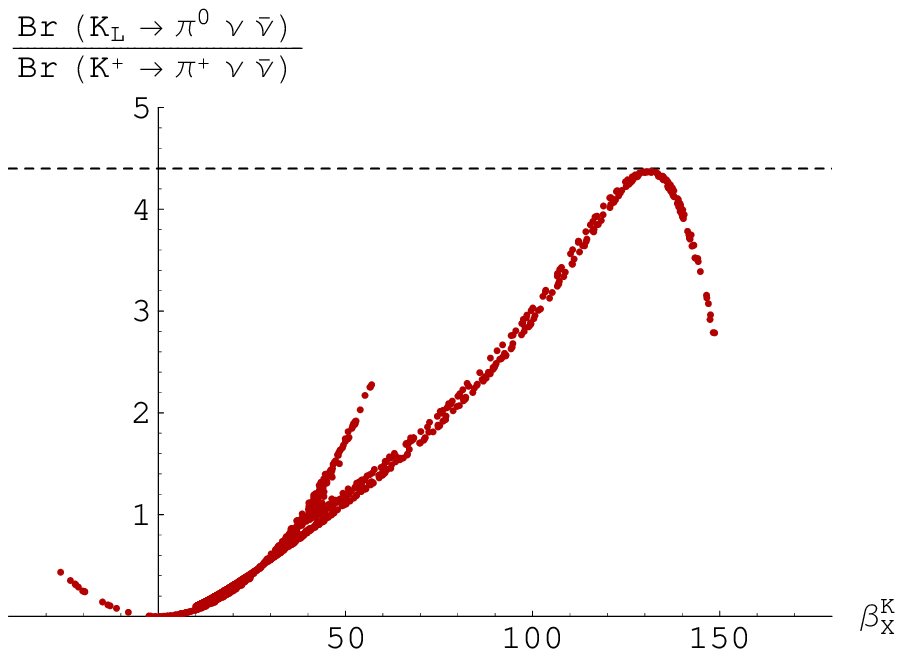,scale=0.7}}
\end{minipage}
\vspace{0.3cm}

\begin{minipage}{7.5cm}
\center{\epsfig{file=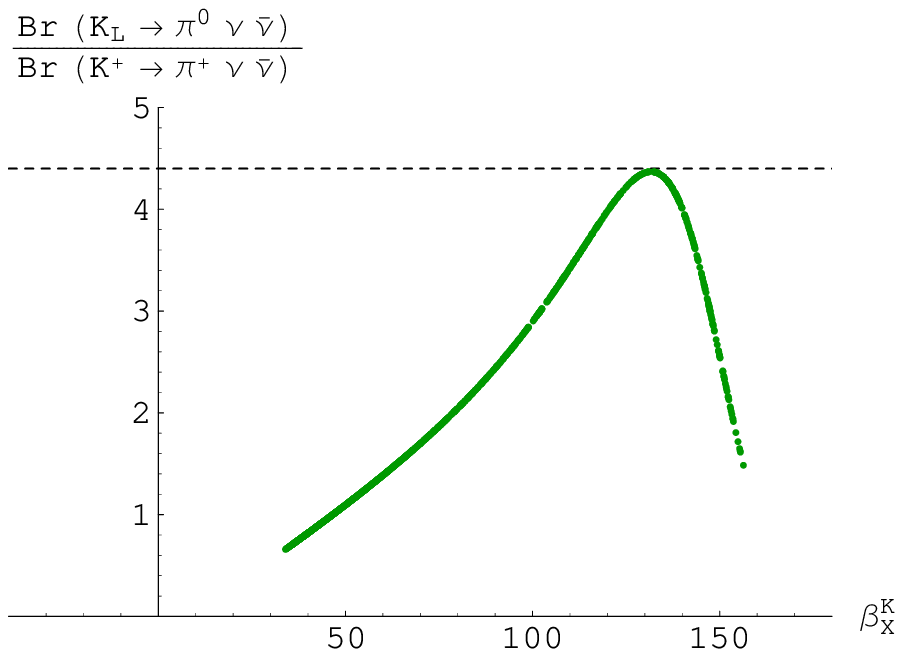,scale=0.7}}
\end{minipage}
\begin{minipage}{7.5cm}
\center{\epsfig{file=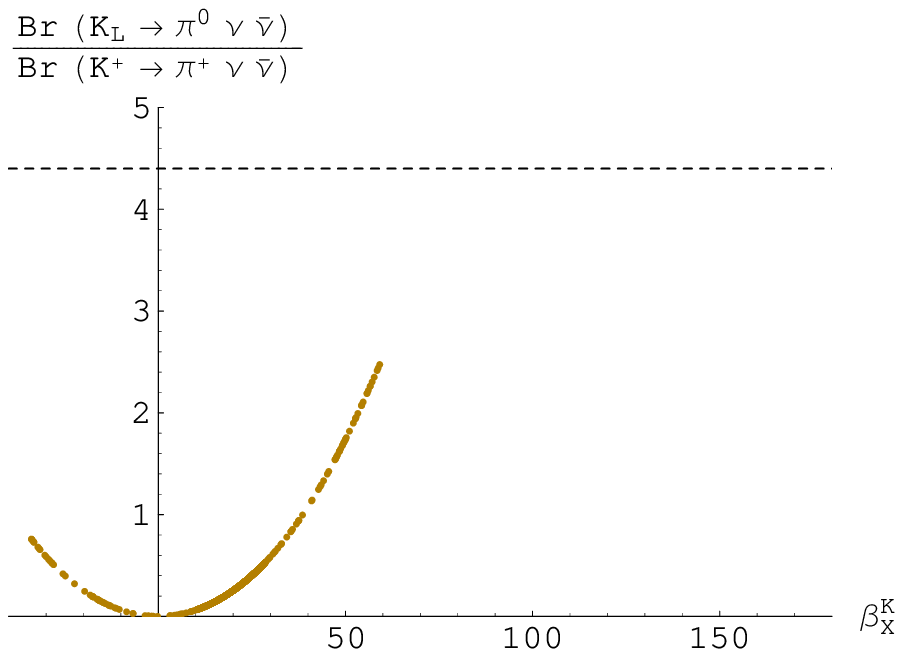,scale=0.7}}
\end{minipage}
\caption{\it $Br(\klpn)/Br(\kpn)$ as a function of $\beta_X^K$, in the
  general scan and Scenarios $3$, $4$ and $6$, respectively. 
  The dashed line represents the GN-bound.}
\label{fig:KrTK}
\end{figure}

In Fig.~\ref{fig:KrTK} we show the ratio $Br(\klpn)/Br(\kpn)$ as a function of
the phase $\beta_X^K$, displaying again the GN-bound.
We observe that the ratio can be significantly different from the SM
prediction, with a possible enhancement of an order of magnitude.
The two branches of Fig.~\ref{fig:KLKp} can be distinguished also here. In
particular, points generated in Scenario $6$ appear only in the left branch
and can not reach the GN-bound, while points belonging to Scenario $4$ populate
the right branch and approach this bound.

The most interesting implications of this analysis are:
\begin{itemize}
\item
If $Br(\kpn)$ is found sufficiently above the SM prediction but below 
$2.3\cdot 10^{-10}$, basically only two values for $Br(\klpn)$ are 
possible within the LHT model. One of these values is very close to
the SM value in (\ref{eq:SMKpinunu}) and the second much larger.
\item
If $Br(\kpn)$ is found above $2.3\cdot 10^{-10}$, then only 
$Br(\klpn)$ with a value close to the SM one in (\ref{eq:SMKpinunu})
is possible.
\item
If $Br(\kpn)$ is found above the SM value, Scenario $4$ will be
ruled out.
\end{itemize}
As Scenario $4$ was our favorite scenario in the analysis of $\Delta F=2$ 
processes in \cite{BBPTUW}, with spectacular new physics effects in
$S_{\psi\phi}$ and  $A^s_{\rm SL}$ asymmetries, let us next have a closer look
at the correlations between $S_{\psi\phi}$ and the $K\to\pi\nu\bar\nu$ decays.

\boldmath
\subsection{$S_{\psi\phi}$ and $K\to\pi\nu\bar\nu$}
\begin{figure}
\begin{minipage}{7.5cm}
\center{\epsfig{file=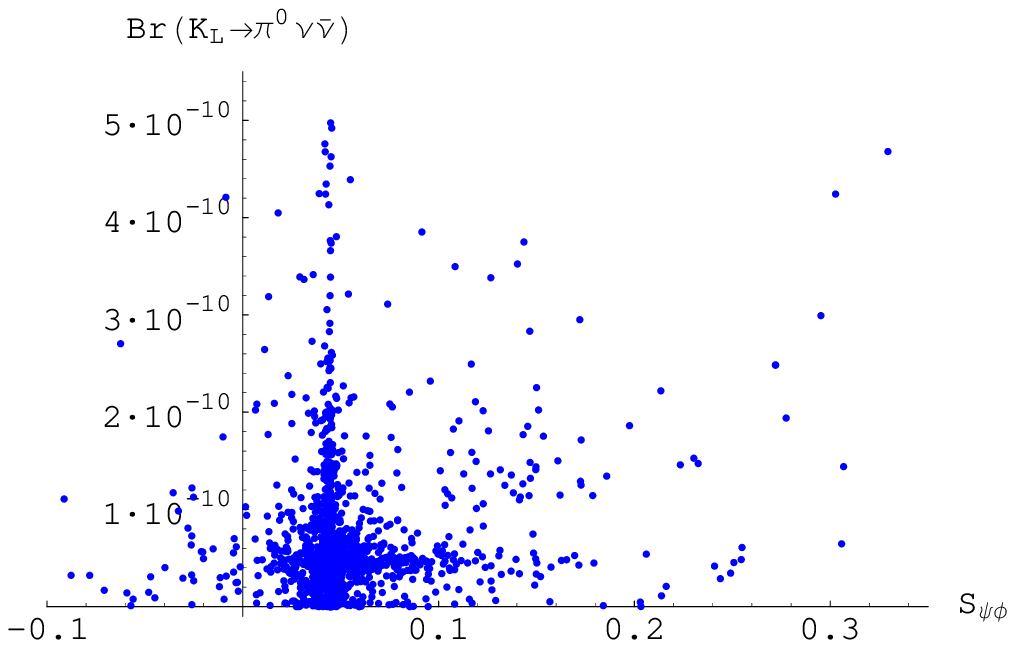,scale=0.7}}
\end{minipage}
\begin{minipage}{7.5cm}
\center{\epsfig{file=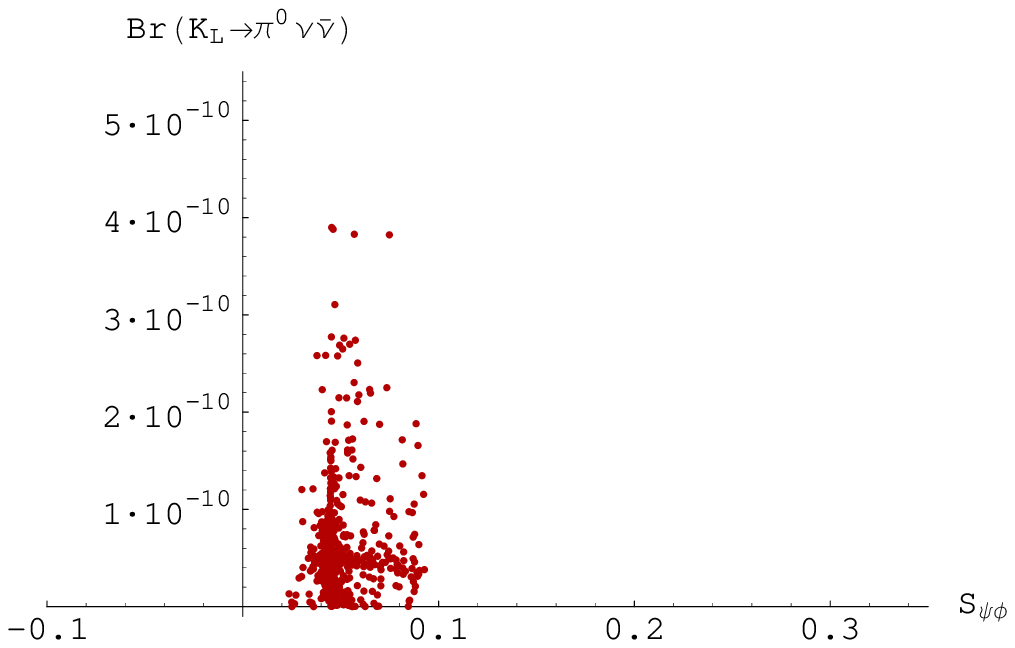,scale=0.7}}
\end{minipage}
\vspace{0.3cm}

\begin{minipage}{7.5cm}
\center{\epsfig{file=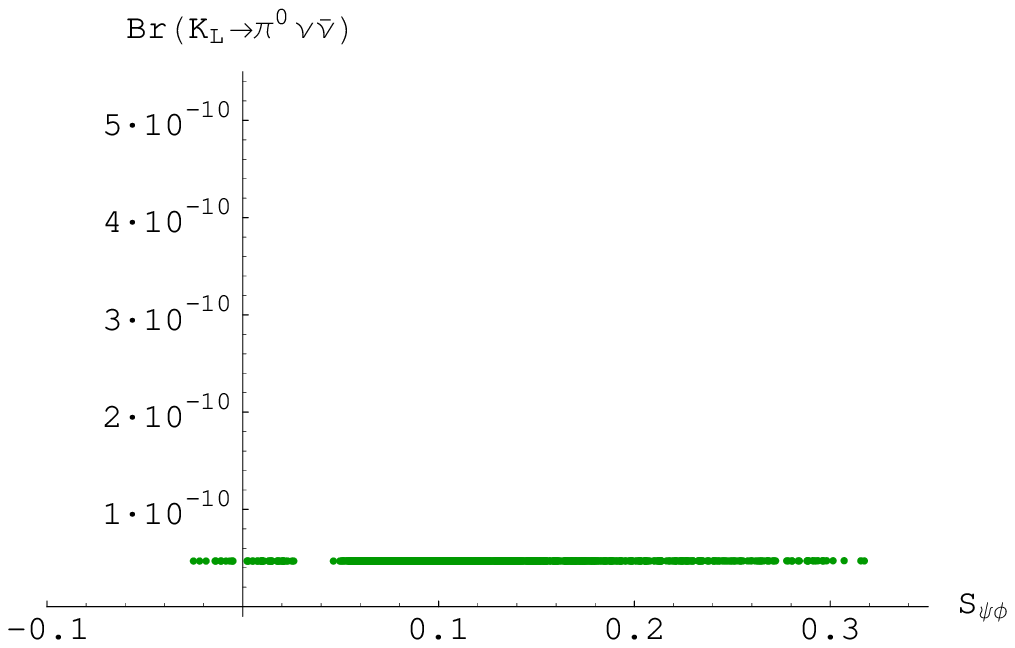,scale=0.7}}
\end{minipage}
\begin{minipage}{7.5cm}
\center{\epsfig{file=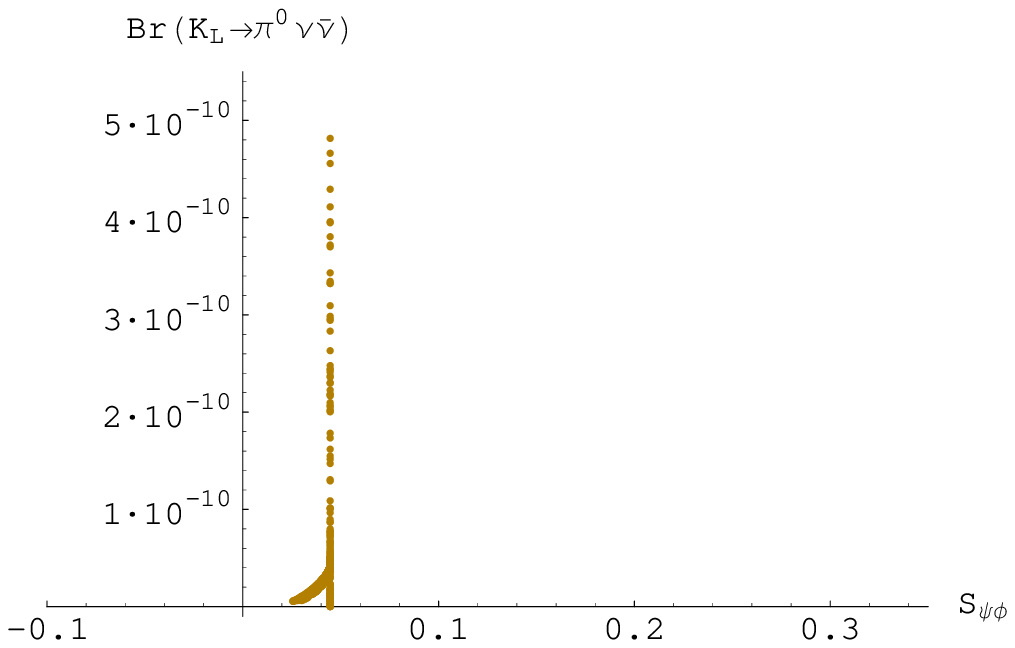,scale=0.7}}
\end{minipage}
\caption{\it $Br(K_L \to \pi^0
\nu \bar \nu)$ as a
  function of $S_{\psi \phi}$, in the general scan and Scenarios $3$, $4$
  and $6$, respectively.}
\label{fig:KLS}
\end{figure}

\begin{figure}
\center{\epsfig{file=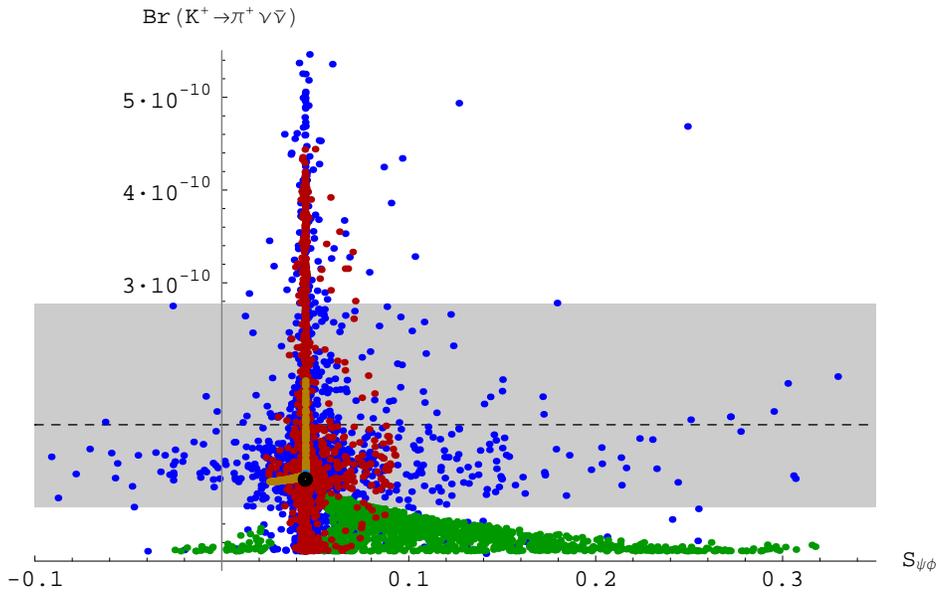,scale=0.9}}
\caption{\it $Br(K^+ \to \pi^+
\nu \bar \nu)$ as a
  function of $S_{\psi \phi}$. The shaded area represents the experimental 
             $1\sigma$-range for 
             $Br(\kpn)$.}
\label{fig:KpS}
\end{figure}

\begin{figure}
\center{\epsfig{file=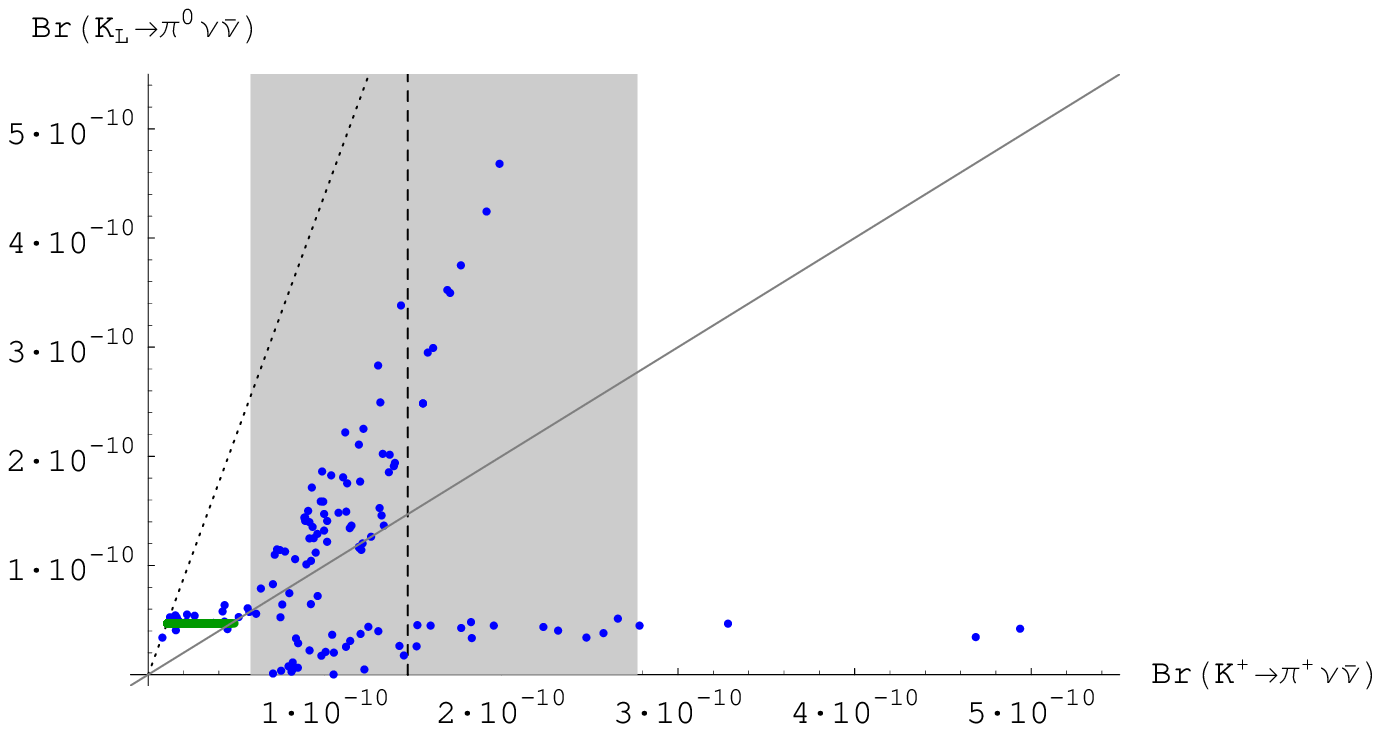,scale=0.9}}
\caption{\it $Br(\klpn)$ as a function of $Br(\kpn)$, showing only the points 
             that satisfy $S_{\psi \phi} \ge 0.1$. Like in
             Fig.~\ref{fig:KLKp}, the shaded area represents the experimental 
             $1\sigma$-range for 
             $Br(\kpn)$, the GN-bound is displayed by the dotted line, while
             the solid line separates the two areas where $Br(\klpn)$ is 
             larger or smaller than
    $Br(\kpn)$.}
\label{fig:KLKpS}
\end{figure}

\unboldmath
In Figs.~\ref{fig:KLS} and~\ref{fig:KpS} we show the correlation between $S_{\psi\phi}$ and 
$Br(\klpn)$ and $Br(\kpn)$, respectively. We observe that in Scenario $4$ 
(green points), in which $S_{\psi\phi}$ can be significantly enhanced,
$Br(\klpn)$ is very close to the SM value, while $Br(\kpn)$ is suppressed 
as already found previously. On the other hand in Scenario $6$ (brown points), both
$Br(\klpn)$ and $Br(\kpn)$ can be strongly enhanced, while $S_{\psi\phi}$ 
is very close to the SM value. 
Only the general scan (blue points) can yield a simultaneous enhancement of
these three observables.
In order to see the triple correlation 
in question even better for all the scenarios, we show in
Fig.~\ref{fig:KLKpS}  only those points of 
Fig.~\ref{fig:KLKp} for which $S_{\psi\phi}\ge 0.1$. It is evident that it is rather 
difficult to obtain simultaneously large values of the three observables 
in question. Still some sets of parameters belonging to the general scan exist 
for which this is possible.

\boldmath
\subsection{$B \to X_{s,d}\nu\bar\nu$}
\unboldmath
From the discussion of Section \ref{subsec:UB} we conclude that the branching
ratios for the $B \to X_{s,d}\nu\bar\nu$ decays can be enhanced by at most
$35\%$ over the SM predictions.
Moreover, we find that
\be
0.64 \le \left | \frac{X_d}{X_s} \right |^2 \le 1.56\,,
\ee
implying that the MFV relation between the ratio of the branching ratios in
question and the CKM parameters given in (\ref{eq:BrXdXs}) can be significantly violated. 

\boldmath
\subsection{$B_{d,s}\to \mu^+\mu^-$ versus $K\to\pi\nu\bar\nu$}
\unboldmath
\begin{figure}
\center{\epsfig{file=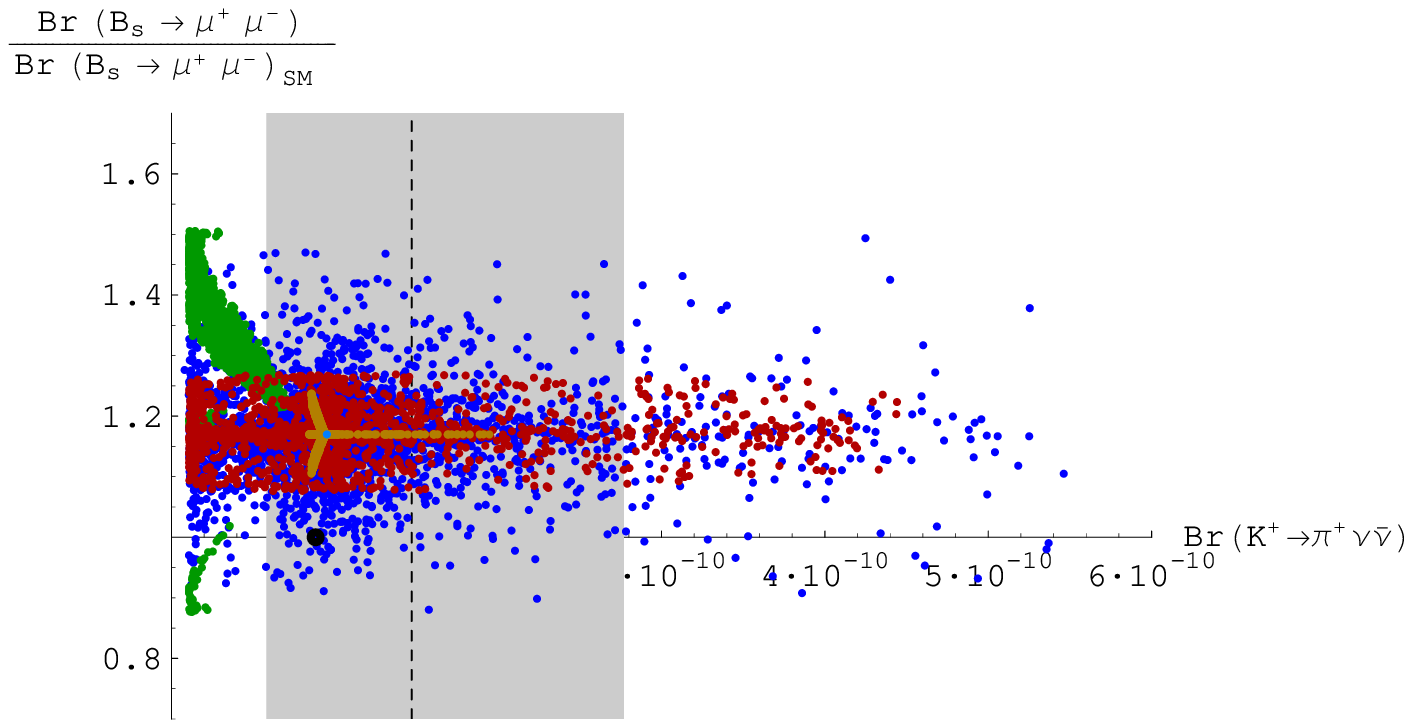,scale=0.90}}
\caption{\it $Br(B_s\to \mu^+\mu^-)/Br(B_s\to \mu^+\mu^-)_\text{SM}$ as a
  function of $Br(\kpn)$. The shaded area represents the experimental
  $1\sigma$-range for $Br(\kpn)$ and the dark point shows the SM prediction.}
\label{fig:BsKp}
\end{figure}
We next investigate possible correlations between
$B_{d,s}\to \mu^+\mu^-$ and $K\to\pi\nu\bar\nu$. In particular, we 
show in Fig.~\ref{fig:BsKp} the first correlation that will be accessible to
future experiments: $Br(B_s\to \mu^+\mu^-)/Br(B_s\to \mu^+\mu^-)_\text{SM}$ 
as a function of $Br(\kpn)$.
The experimental
$1\sigma$-range for $Br(\kpn)$ \cite{expK+} is represented by the shaded area
and the SM prediction by the dark point.
$Br(B_s\to \mu^+\mu^-)$ can be about $50\%$ larger than in the SM, while
more pronounced effects are possible in $Br(\kpn)$. 
Scenarios $4$ (green points) and $6$ (brown points) turn out to be again 
distinguishable through this
correlation. As expected from our previous discussion, 
Scenario $4$ allows larger effects in the
$B_s$ system, while Scenario $6$ is characterized by significant (of a factor
$5$) enhancements in the $K$ system.

\boldmath
\subsection{The $K_L\to \pi^0\ell^+\ell^-$  System}
\unboldmath
\begin{figure}
\center{\epsfig{file=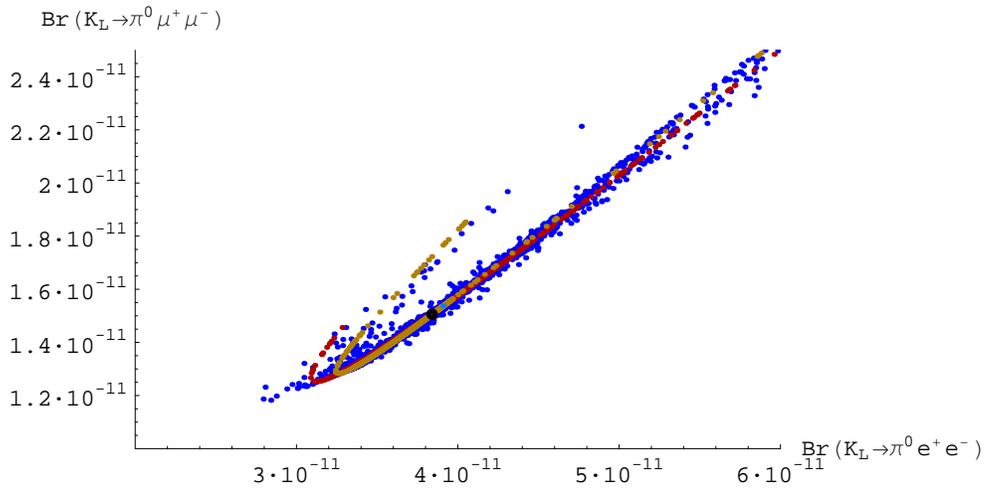,scale=0.90}}
\caption{\it $Br(K_L \to \pi^0 \mu^+\mu^-)$ as a
  function of $Br(K_L\to \pi^0 e^+e^-)$.}
\label{fig:KmuKe}
\end{figure}
In Fig.~\ref{fig:KmuKe} we show the correlation between 
$Br(K_L\to \pi^0 e^+e^-)$ and $Br(K_L \to \pi^0 \mu^+\mu^-)$
that has been first investigated in \cite{Isidori:2004rb,Friot:2004yr} in the framework of \cite{BFRS}.
This correlation is only moderately sensitive to the scenario considered.
We observe that both branching ratios can be enhanced up to a factor two, over
the SM values (black point) in~(\ref{eq:KLpee}) and~(\ref{eq:KLpmm}).

\boldmath
\subsection{$K_L\to \pi^0\ell^+\ell^-$  versus $\klpn$}
\unboldmath
\begin{figure}
\center{\epsfig{file=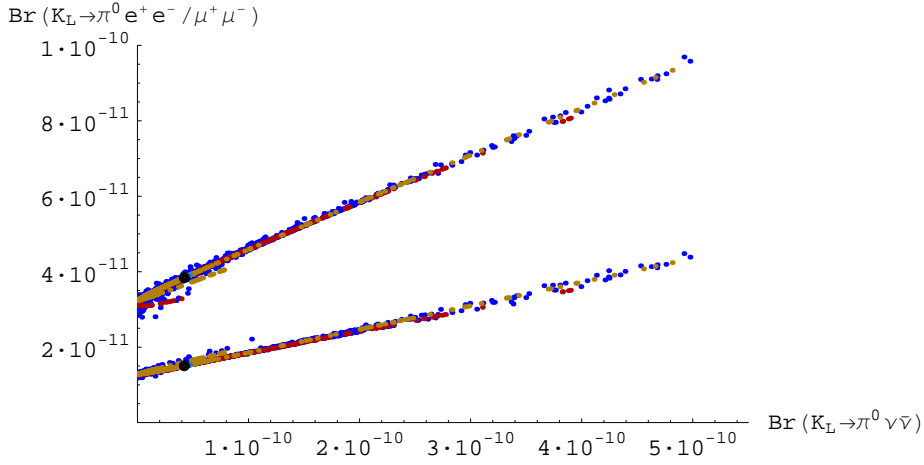,scale=0.90}}
\caption{\it $Br(K_L\to \pi^0 e^+e^-)$ (upper curve) and  $Br(K_L \to \pi^0
  \mu^+\mu^-)$ (lower curve) as functions of $Br(\klpn)$. The corresponding SM
  predictions are represented by dark points.}
\label{fig:KemuKp}
\end{figure}
In Fig.~\ref{fig:KemuKp}  we show
 $Br(K_L\to\pi^0 e^+e^-)$ and 
$Br(K_L\to\pi^0 \mu^+\mu^-)$ versus $Br(\klpn)$. 
We observe a strong correlation between $K_L\to \pi^0\ell^+\ell^-$ and 
$\klpn$ decays that we expect to be valid beyond the LHT model, at least
in models with the same operators present as in the SM. We note that
a large enhancement of $Br(\klpn)$ automatically implies significant 
enhancements of $Br(K_L\to \pi^0\ell^+\ell^-)$ and that
different models and their parameter sets can than be distinguished
by the position on the correlation curve.
Moreover, measuring $Br(K_L\to\pi^0\ell^+\ell^-)$ should 
allow a rather precise prediction of $Br(\klpn)$ at least in models
with the same operators as the SM.

\subsection{Violation of Golden MFV Relations}

\begin{figure}
\center{\epsfig{file=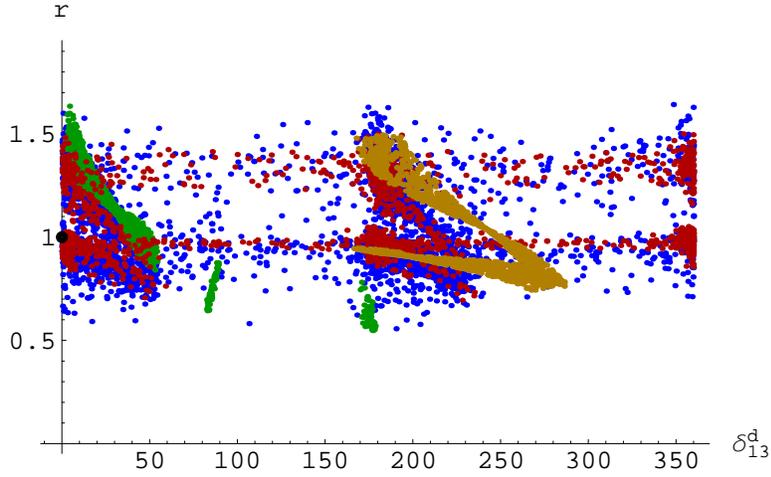,scale=1.0}}
\caption{\it The ratio $r$ as a function of $\delta_{13}^d$.}
\label{fig:rd13d}
\end{figure}
In Fig.~\ref{fig:rd13d} we show the ratio $r$ of (\ref{eq:r}) as a
function of $\delta_{13}^d$. Its 
departure from unity measures the violation of the golden MFV relation
between $B_{d,s}\to\mu^+\mu^-$ decays and $\Delta M_{d,s}$ in (\ref{eq:r}).
We observe that $r$ can vary in the range
\be
0.6\le r \le 1.7\,,
\ee
with, as expected, Scenario $4$ (green points) able to achieve these bounding 
values more easily than Scenarios $3$ (red points) and  $6$ (brown points).
Such departures from unity could easily be tested in view of a 
theoretically clean character of (\ref{eq:r}).

\begin{figure}
\center{\epsfig{file=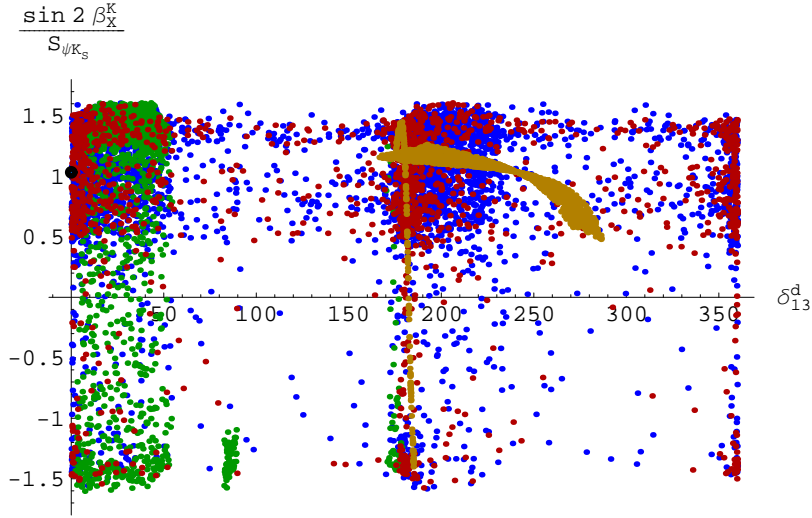,scale=0.9}}
\caption{\it $\sin2\beta_X^K/\sin(2\beta+2\varphi_{B_d})$ as a
  function of $\delta^d_{13}$.}
\label{fig:rbetad13d}
\end{figure}
Furthermore, in Fig.~\ref{fig:rbetad13d} we show the ratio of $\sin 2\beta_X^K$
over $\sin(2\beta+2\varphi_{B_d})$ as a function of $\delta_{13}^d$ 
for the scenarios considered. 
Similarly to $r$, the 
departure of this ratio from unity measures the violation of a 
golden MFV relation (\ref{eq:betarel}), this time between the CP-violating
phases in the $K\to\pi\nu\bar\nu$ system and in the 
$B_d^0-\bar B^0_d$ mixing. 
As $\varphi_{B_d}$ is constrained by the measured $S_{\psi K_S}$ asymmetry 
to be at most a few degrees \cite{BBGT,UTFIT}, large violations of the relation in question 
can only follow from the $K\to\pi\nu\bar\nu$ decays. As seen in Fig.~\ref{fig:rbetad13d}, they can be spectacular.

\boldmath
\subsection{The dependence on $\delta^d_{13}$}
\unboldmath
As seen in Figs.~\ref{fig:rd13d} and~\ref{fig:rbetad13d} there are two oases in the values of 
$\delta^d_{13}$
\be
-10^\circ \le \delta^d_{13}\le 50^\circ, \qquad
170^\circ \le \delta^d_{13}\le 250^\circ, 
\ee
with the desert between the two oases visibly but not densely populated. 
The origin of the oases is the constraint from $S_{\psi K_S}$. It is 
interesting to observe again a clear separation between Scenarios 4 and 
6. While in Scenario 4 (green points) the first oasis is dominantly populated, in 
Scenario 6 (brown points) the second oasis is occupied. Moreover in the latter case there
is an interesting excursion of points into the desert up to 
$\delta_{13}^d=290^\circ$. The desert, instead, in which large effects 
simultaneously in $K$ and $B$ systems can be found, is dominated 
by red and blue points corresponding to Scenario 3 and the general scan, 
respectively.

\subsection{Comparison of Various Scenarios}
The plots in Figs.~\ref{fig:KLKp}-\ref{fig:rbetad13d} are self-explanatory. Yet, we would like to make 
a few general observations:
\begin{itemize}
\item
There is a very clear distinction between scenarios $4$ and $6$. Scenario $4$ 
can be considered as ``$B_s$--scenario" as it gives 
interesting effects in the $B_s$ system.
\item
On the other hand Scenario $6$ can be considered as ``$K$--scenario"
as it admits spectacular effects in $K$ decays.
\item
Moreover, for certain sets of the LHT parameters obtained in the general scan, 
large departures from the SM predictions in $K$ and $B_{d,s}$ 
decays can be simultaneously found.
\end{itemize}

\boldmath
\subsection{$B \to \pi K$ Decays in the LHT Model}
\unboldmath
We have finally investigated the impact of new physics contributions on $B \to
\pi K$ decays that, for some time, signaled the presence of enhanced
electroweak penguin contributions with new large CP-violating phases.
In a simple new physics scenario in which the universality between 
$b \bar s$-penguins relevant for $B \to \pi K$ and $s \bar
d$-penguins relevant for $K \to \pi \nu \bar \nu$ has been assumed
\cite{BFRS}, the enhanced electroweak penguins required to fit the $B \to \pi K$ data implied
spectacular effects in the $K \to \pi \nu \bar \nu$ system, similar to those
shown in Figs.~\ref{fig:KLKp}-\ref{fig:KLS}.

Meanwhile, the data on $B \to \pi K$ decays significantly changed
\cite{babar,babar2,belle}
so that the SM predictions for the so-called $R_n$ and $R_c$ ratios
\cite{BFRS}  are nowadays within one standard deviations from the data.
Consequently, within the new physics scenario of \cite{BFRS} no large effects
in $K \to \pi \nu \bar \nu$ are expected.

The picture is quite different in the LHT model, where the universality
between $b \bar s$ and $s \bar d$ systems can be strongly broken.
Following the approach of \cite{BFRS} to determine the hadronic parameters of the
$B \to \pi K$ system from the $B \to \pi \pi$ data and calculating the
electroweak penguin contributions to $B \to \pi K$ decays in the LHT model, we
find that the new physics effects in these rare decays are smaller than the
theoretical uncertainties present in non-leptonic decays.
These small effects follow primarily from the smallness  of complex phases in
the $b \bar s$ penguins as given in (\ref{eq:ThetaKs}) to be compared with $\pm
90^\circ$ taken in past $B \to \pi K$ analyses.

In summary, in the LHT model the smallness of new physics in $B \to \pi K$
decays does not imply its smallness in $K \to \pi \nu \bar \nu$ decays as seen
in Figs \ref{fig:KLKp}-\ref{fig:KLKpS}.  

\boldmath
\subsection{What if the $\sin 2 \beta$ problem disappears?}
\unboldmath
Until now, our analysis was based on the tree level determination of the angle
$\beta$ that, due to the high value of $R_b$, is larger than the one measured
through the $S_{\psi K_S}$ asymmetry.
It is conceivable that future tree level determinations of $R_b$ will yield
lower values for $R_b$ and consequently for $\beta$, so that the $\sin 2
\beta$ problem will disappear.
We have repeated the whole analysis for such a scenario, finding that the
$\sin 2 \beta$ problem has a very minor impact on our analysis, in particular
in $B_s$ and $K$ decays.
For instance, large effects in $K \to \pi \nu \bar \nu$ decays and
simultaneously in $S_{\psi \phi}$ can still be found.

\boldmath
\subsection{What if the $B_d$ and $B_s$ Decays are SM-like?}
\unboldmath
Finally, we can consider a very pessimistic scenario where BaBar, Belle and 
LHC will confirm all the SM
expectations in $B_d$ and $B_s$ decays, finding in particular both the $S_{\psi
  K_S}$ and $S_{\psi \phi}$ asymmetries very close to the SM values.
This could  already happen at the end of this decade.
The question then arises whether in such a situation we could still expect
large departures from the SM values in rare $K$ decays, whose precise
measurements will be available only in the next decade.
It is evident from Figs.~\ref{fig:KLKp}-\ref{fig:KLKpS} that even in this, pessimistic
for $B$-physics, scenario, spectacular effects in $K \to \pi \nu \bar \nu$ and
also large effects in $K_L \to \pi^0 \ell^+ \ell^-$ will be possible in particular in Scenario 6.

\newsection{Summary and Outlook}\label{sec:summary}

In this paper we have analyzed for the first time rare $K$ and $B$ decays in
the Littlest Higgs model with T-parity \cite{oldLH,LHreview,l2h,tparity}.
Together with our previous work \cite{BBPTUW} on observables related to
particle-antiparticle mixing and $B \to X_{s,d} \gamma$, the results
of the present paper allow us to obtain a general description of FCNC
processes in this model.

On the technical side, we have presented a complete set of Feynman rules for
the LHT model including also vertices with Goldstone bosons.
The inclusion of $\mathcal{O}(v^2/f^2)$ corrections to some of the vertices
has been performed here for the first time.
These Feynman rules allow the calculations of $\mathcal{O}(v^2/f^2)$
contributions in arbitrary gauge and should turn out to be useful also 
for other observables.

Using these rules we have calculated in the LHT model the short distance
functions $X_i$, $Y_i$ and $Z_i$ ($i = K, d, s$).
In the LHT model these functions are complex quantities and carry the index
$i$ to signal the breakdown of the universality of FCNC processes, valid in
the  SM.
The new weak phases in $X_i$, $Y_i$ and $Z_i$, which are absent in the
SM and models with MFV,
imply potential new CP-violating effects beyond the SM ones.
We would like to emphasize that the new parameterization of rare
  decays in terms of non-universal functions $X_i$, $Y_i$ and $Z_i$
  can be applied to any model with new flavour and CP-violating 
  interactions but the same low energy operators.

With the functions $X_i$, $Y_i$ and $Z_i$ at hand, one can straightforwardly
calculate the branching ratios for a number of interesting rare decays.
In particular, we analyzed $K^+ \to \pi^+ \nu \bar \nu$, $K_L
\to \pi^0 \nu \bar \nu$, $B_{s,d} \to \mu^+ \mu^-$, $B
\to X_{s,d} \nu \bar \nu$, $B \to X_s \ell^+ \ell^-$,
$K_L\to\pi^0\ell^+\ell^-$ and $B \to \pi K$.
At all stages of our numerical analysis we took into account the existing
constraints from electroweak precision studies \cite{mH} and from particle-antiparticle
mixing and $B \to X_s \gamma$ studied by us in \cite{BBPTUW} and from $B \to
X_s \ell^+ \ell^-$ calculated here.

The main messages of our paper are as follows:
\begin{itemize}

\item {The most evident departures from the SM predictions are found for
    CP-violating observables that are strongly suppressed within this
    model. These are the branching ratio for $K_L \to \pi^0 \nu \bar \nu$ and
    the CP-asymmetry $S_{\psi \phi}$.}

\item {Large departures from SM expectations are also possible for $Br(K_L \to
  \pi^0 \ell^+ \ell^-)$ and $Br(K^+ \to \pi^+ \nu \bar \nu)$.}

\item{The branching ratios for $B_{s,d} \to \mu^+ \mu^-$ and $B \to X_{s,d}
    \nu \bar \nu$, instead,  are modified by at most $50\%$ and $35\%$,
    respectively, and the effects of new electroweak penguins in $B \to \pi K$
    are small, in agreement with the recent data.}

\item{Sizable departures from MFV relations between $\Delta M_{s,d}$ and
    $Br(B_{s,d} \to \mu^+ \mu^-)$ and between $S_{\psi K_S}$ and the $K \to
    \pi \nu \bar \nu$ decay rates are possible.}

\item{The universality of new physics effects, characteristic for MFV models,
    can be largely broken, in particular between $K$ and $B_{s,d}$ systems.}
\item
{The new physics effects in $B\to X_{s,d}\gamma$ and $B\to X_{s,d}\ell^+\ell^-$
turn out to be below $5\%$ and $15\%$, respectively so that agreement 
with the data can easily be obtained.}
\end{itemize}

One of the important findings of our paper is the presence of left-over
  singularities in the mirror sector that signals some sensitivity
  of the final results to the UV completion of the theory. 
This issue has been discussed in detail in the context of the LH model
  without T-parity in \cite{BPUB}, is known from the study of
  electroweak constraints \cite{tparity} and has been considered
  in Section \ref{sec:sing} of the present paper.
In estimating the contribution of these logarithmic
singularities,
 we have  assumed the UV completion of the theory not to have a complicate
  flavour pattern or at least that it has no impact below the cut-off.
Clearly, this
additional
 assumption lowers the predictive power of the theory. 
In spite of that, we believe
 that the general picture of FCNC processes presented here and in our
previous paper is only insignificantly shadowed by this general property of
 non-linear sigma models.

We conclude with probably one of the most important messages of this paper and of our previous analysis:
\begin{itemize}  
\item
In spite of an impressive agreement of the SM with the available data, 
large departures from the SM expectations in $B_s$ decays are still
possible. However, even if future Tevatron and LHC data
would not see any significant new physics effect in these decays, this 
will not imply necessarily that new physics is not visible in $\klpn$, $\kpn$ 
and $K_L\to\pi^0\ell^+\ell^-$. On the contrary, as seen in the case of Scenario $6$,
large departures in these three decays will still be possible. It  may 
then be that in the end, it will be $K$ physics and not $B$ physics 
that will offer the best information about the new phenomena at 
very short distance scales, in accordance with the arguments in \cite{Bryman:2005xp,Grinstein:2006cg}.
\end{itemize}

\subsection*{Acknowledgements}

We would like to thank Bj{\"o}rn Duling for checking the Feynman rules
given in Appendix \ref{sec:appB} and William A.~Bardeen, Ulrich Haisch and
Kazuhiro Tobe for useful discussions. 
This research was partially supported by the German `Bundesministerium f\"ur 
Bildung und Forschung' under contracts 05HT4WOA/3, 05HT6WOA.

\newpage
\begin{appendix}

\newsection{Non-leading contributions of $\bm{\Phi}$}\label{sec:appA}

In this Appendix we show that the scalar triplet $\Phi$ does not
contribute to the decays considered in the present paper at $\ord(v^2/f^2)$.

In principle, also the scalar triplet $\Phi$ could contribute to the
decays considered in the present analysis. The corresponding
diagrams can be obtained from the ones shown in
Figs.~\ref{fig:ping} and \ref{fig:box} by simply replacing
$W^\pm_H$ by $\phi^\pm$ and $Z_H,\,A_H$ by $\phi^0,\,\phi^P$.
Therefore we also derived the Feynman rules for the vertices
containing $\Phi$. We found them to have the following generic
structure:
\be
\bar q_H \Phi q \sim c
\frac{m^q_{H}}{M_\Phi}\frac{v^2}{f^2}P_L-c'\frac{m_{q}}{M_\Phi}P_R\,,\qquad
\Phi G_H G_L \sim \ord\left(\frac{v^2}{f}\right)\,,
\ee
where
$c,\,c'$ are $\ord(1)$ coefficients depending on which $q_H$, $q$ and
component of $\Phi$ are involved, and
$P_{R,L}=(1\pm\gamma_5)/2$ are the usual projectors. $q_H$ and $q$
denote mirror and SM fermions, respectively, and $G_H,\,G_L$ are
heavy and light gauge bosons.

As we have set the masses of the external quarks and leptons to
zero throughout our analysis, we obtain
\be\label{eq:FRphi}
\bar
q_H \Phi q \sim \ord\left(\frac{v^2}{f^2}\right)
\qquad\text{and}\qquad \Phi G_H G_L \sim
\ord\left(\frac{v^2}{f}\right)\,.
\ee

It can now be easily seen that each diagram contains at least two
such vertices, so that they are suppressed by $\ord(v^4/f^4)$.

Furthermore, in contrast to the LH model without T-parity, all
particles running in the loops have $\ord(f)$ masses, so that no
cancellation of $v/f$ factors due to large mass ratios, as
encountered in \cite{BPU2,BPUB} for diagrams with $T_+$ exchanges, can
appear.

Thus we find that $\Phi$ contributes to the decays in question --
and more in general to all decays with external SM fermions -- only
at $\ord(v^4/f^4)$. For this reason we do not give any Feynman
rule for the interactions of the scalar triplet $\Phi$.

\newsection{Relevant Feynman Rules}\label{sec:appB}

In this Appendix we list all Feynman rules relevant
for the analysis performed in the present paper, and describe
briefly how they have been derived.
We note that given the Feynman rule for a vertex, the rule for the conjugate
one can be obtained through the replacement (vertex)$\rightarrow -$(vertex)$^\dagger$.
There follow, in particular, the prescriptions $P_L \rightarrow - P_R$, $P_R \rightarrow -
P_L$ and $\gamma_\mu P_{L,R} \rightarrow \gamma_\mu P_{L,R}$.
 A similar, but more detailed
description for the LH model without T-parity can be found in
\cite{Logan} and in \cite{BPU,BPUB}, where some of the Feynman rules given
in \cite{Logan} have been corrected.

\subsection{Fermion--Gauge Boson Couplings}
\setlength{\arraycolsep}{2pt}

The fermion--gauge boson couplings can easily be obtained from
the fermion kinetic term \cite{LHTphen}
\be
\mathcal{L}_{\text{fermion}}=\bar\Psi_1\bar\sigma^\mu
D^1_\mu\Psi_1+\bar\Psi_2\bar\sigma^\mu D^2_\mu\Psi_2+\bar
t_1'\bar\sigma^\mu D'^{1}_\mu t_1'+\bar
t_2'\bar\sigma^\mu D'^{2}_\mu t_2'\,,
\ee
where
\bea
D^1_\mu &=& \partial_\mu
-\sqrt{2}igQ_1^aW_{1\mu}^a-\sqrt{2}ig'Y_1^{(\Psi_1)}B_{1\mu}-\sqrt{2}ig'Y_2^{(\Psi_1)}B_{2\mu}\,,\\
D^2_\mu &=& \partial_\mu
+\sqrt{2}igQ_2^{aT}W_{2\mu}^a-\sqrt{2}ig'Y_1^{(\Psi_2)}B_{1\mu}-\sqrt{2}ig'Y_2^{(\Psi_2)}B_{2\mu}\,,\\
D'^i_\mu &=& \partial_\mu-\sqrt{2}ig'Y_1^{(t'_i)}B_{1\mu}-\sqrt{2}ig'Y_2^{(t'_i)}B_{2\mu}\,,
\eea
 by inserting the mass
eigenstates of all particles involved.
The $U(1)$ charges can be obtained from gauge invariance of the Yukawa
couplings and T-parity \cite{LHTphen} and are given in Table \ref{tab:hypercharges}.

\begin{table}[ht]
\renewcommand{\arraystretch}{1}\setlength{\arraycolsep}{1pt}
\center{\begin{tabular}{|c|c||c|c|}
\hline
$q_1$ & $(1/30,2/15)$ & $q_2$ & $(2/15,1/30)$\\\hline
$t'_1$ & $(8/15,2/15)$ & $t'_2$ &  $(2/15,8/15)$\\\hline
$t'_{1R}$ & $(8/15,2/15)$ & $t'_{2R}$ &  $(2/15,8/15)$\\\hline
$u_R$ & $(1/3,1/3)$ & $d_R$ & $(-1/6,-1/6)$\\\hline
$\ell_1$ & $(-1/5,-3/10)$ & $\ell_2$ & $(-3/10,-1/5)$\\\hline
$e_R$ & $(-1/2,-1/2)$ & & \\
\hline
\end{tabular}  }
\caption {\textit{$U(1)_1\times U(1)_2$ quantum numbers of the fermion fields.}}
\label{tab:hypercharges}
\renewcommand{\arraystretch}{1.0}
\end{table}
Similar kinetic terms can be written down for the right-handed SM
fields and for $t'_{iR},\;i=1,2$.

The kinetic term for the right-handed mirror fermions can be written
down by using the CCWZ approach \cite{mirror,CCWZ} and is given in explicit terms
in \cite{LHTphen}.
 As the mirror fermions are
purely vector-like under $SU(2)_L  \times U(1)_Y$, their couplings to SM gauge bosons are
proportional to $\gamma^\mu$. 
Thus there is no need to consider the
CCWZ kinetic term any further, as the couplings of the right-handed mirror
fermions to SM gauge bosons have to be equal to the ones of the
left-handed mirror fermions.

\begin{center}
\begin{longtable}{|c|c|}  \hline
\multicolumn{2}{|c|}{\bf Fermion couplings to SM gauge bosons}\\\hline\hline
 $\bar f A_{L} f$ & $i
e Q_f \gamma^\mu$\\\hline
 $\bar u^i Z_{L} u^i$ &
$\frac{ig}{\cos{\theta_W}}\gamma^\mu\left[\left(\frac{1}{2}-\frac{2}{3}\sin^2{\theta_W}\right)P_L
  -\frac{2}{3}\sin^2{\theta_W} P_R\right]$\\\hline
$\bar d^i Z_L d^i$ & $\frac{ig}{\cos{\theta_W}}\gamma^\mu\left[\left(-\frac{1}{2}+\frac{1}{3}\sin^2{\theta_W}\right)P_L
  +\frac{1}{3}\sin^2{\theta_W} P_R\right]$ \\\hline
$\bar t Z_{L} t$ & $ \frac{ig}{\cos{\theta_W}}\gamma^\mu\left[\left(\frac{1}{2}\left(1-x_L^2\frac{v^2}{f^2}\right)-\frac{2}{3}\sin^2{\theta_W}\right)P_L
  -\frac{2}{3}\sin^2{\theta_W} P_R\right]$ \\\hline
$\bar T_+ Z_{L} T_+$ & $\frac{ig}{\cos{\theta_W}}\gamma^\mu\left[\left(\frac{1}{2}x_L^2\frac{v^2}{f^2}-\frac{2}{3}\sin^2{\theta_W}\right)P_L
  -\frac{2}{3}\sin^2{\theta_W} P_R\right]$ \\\hline
 $\bar T_+ Z_{L} t$ & $ \frac{ig}{\cos{\theta_W}}\frac{x_L}{2}\frac{v}{f} \left[1+\frac{v^2}{f^2}\left(d_2-\frac{x_L^2}{2}\right)\right]\gamma^\mu P_L$ \\\hline
$\bar u^i_H Z_{L} u^i_H$ &
$\frac{ig}{\cos{\theta_W}}\left[\left(\frac{1}{2}-\frac{2}{3}\sin^2{\theta_W}\right)\right]\gamma^\mu$
\\\hline
 $\bar d^i_H Z_{L} d^i_H$ & $\frac{ig}{\cos{\theta_W}}\left[\left(-\frac{1}{2}+\frac{1}{3}\sin^2{\theta_W}\right)\right]\gamma^\mu$ \\\hline
$\bar T_- Z_{L} T_-$ &
$-\frac{2ig}{3\cos{\theta_W}}\sin^2{\theta_W}\gamma^\mu$ \\\hline
$\bar u^i W^{+\mu}_{L} d^j$ & $\frac{ig}{\sqrt{2}}(V_\text{CKM})_{ij}\gamma^\mu P_L$\\\hline
$\bar t W^{+\mu}_{L} d^j$ & $\frac{ig}{\sqrt{2}}(V_\text{CKM})_{tj}\left(1-\frac{x_L^2}{2}\frac{v^2}{f^2}\right)\gamma^\mu P_L$\\\hline
$\bar T_+ W^{+\mu}_{L} d^j$ & $\frac{ig}{\sqrt{2}}(V_\text{CKM})_{tj}x_L\frac{v}{f}\left(1+\frac{v^2}{f^2}d_2\right)
\gamma^\mu P_L$\\\hline
$\bar u^i_{H} W^{+\mu}_{L} d^j_{H}$ & $\frac{ig}{\sqrt{2}}\delta_{ij} \gamma^\mu$\\\hline
$\bar \nu^i Z_{L} \nu^i$ &
$\frac{ig}{2\cos{\theta_W}}\gamma^\mu P_L$ \\\hline
$\bar \ell^i Z_{L} \ell^i$ & $\frac{ig}{\cos{\theta_W}}\gamma^\mu\left[\left(-\frac{1}{2}+\sin^2{\theta_W}\right)P_L
  +\sin^2{\theta_W} P_R\right]$\\\hline
$\bar \nu^i_H Z_{L} \nu^i_H$ &
$\frac{ig}{2\cos{\theta_W}}\gamma^\mu$\\\hline $\bar \ell^i_H
Z_{L} \ell^i_H$ &
$\frac{ig}{\cos{\theta_W}}\left[\left(-\frac{1}{2}+\sin^2{\theta_W}\right)\right]\gamma^\mu$\\\hline
$\bar \nu^i W^{+\mu}_{L} \ell^j$ &
$\frac{ig}{\sqrt{2}}(V_\text{PMNS})_{ij}\gamma^\mu P_L$\\\hline
$\bar \nu^i_{H} W^{+\mu}_{L} \ell^j_{H}$ &
$\frac{ig}{\sqrt{2}}\delta_{ij} \gamma^\mu$\\\hline
\end{longtable}
\end{center}

The couplings to the heavy gauge bosons have already been given in
\cite{Hubisz}. We confirm the findings of these authors, and
include also $v^2/f^2$ corrections to the couplings of the neutral
gauge bosons. These corrections turn out to be irrelevant in the
decays considered in the present paper, but could be of relevance
for other processes. 
While our paper was being completed, the rules for the quark sector appeared
also in \cite{0609179}, but without performing the $v/f -$~expansion and without taking into account the
flavour mixing matrices $V_{Hu}$ and $V_{Hd}$.

Note that the couplings to the heavy gauge bosons are purely
left-handed as there is no kinetic term including both heavy and
light right-handed fermions. The only exception are the couplings involving $T_-$ together with $T_+$ or $t$.

\begin{center}
\begin{longtable}{|c|c|} \hline
\multicolumn{2}{|c|}{\bf Fermion couplings to heavy gauge bosons}\\\hline\hline
 $\bar u^i_{H} A_{H} u^j$
& $\left(-\frac{ig'}{10}-\frac{ig}{2} x_H\frac{v^2}{f^2}
\right)(V_{Hu})_{ij}\gamma^\mu P_L$\\\hline $\bar u^i_{H} Z_{H}
u^j$ & $\left(\frac{ig}{2}-\frac{ig'}{10}x_H\frac{v^2}{f^2}
\right)(V_{Hu})_{ij}\gamma^\mu P_L$\\\hline $\bar d^i_{H} A_{H}
d^j$ & $\left(
  -\frac{ig'}{10}+\frac{ig}{2}x_H\frac{v^2}{f^2} \right)(V_{Hd})_{ij}\gamma^\mu P_L$\\\hline
$\bar d^i_{H} Z_{H} d^j$ & $\left(-\frac{ig}{2}-\frac{ig'}{10}x_H\frac{v^2}{f^2} \right)(V_{Hd})_{ij}\gamma^\mu
P_L$\\\hline
$\bar u^i_{H} A_{H} t$ & $\left[-\frac{ig'}{10}+\left(\frac{ig'}{20}{x_L^2}-\frac{ig}{2}x_H\right)\frac{v^2}{f^2} \right](V_{Hu})_{i3}  \gamma^\mu P_L$\\\hline
$\bar u^i_{H} Z_{H} t$ & $\left[\frac{ig}{2}-\left(\frac{ig}{4}x_L^2+\frac{ig'}{10}x_H\right)\frac{v^2}{f^2} \right](V_{Hu})_{i3}\gamma^\mu
P_L$\\\hline
$\bar u^i_{H} A_{H} T_+$ & $-\frac{ig'}{10}x_L\frac{v}{f}(V_{Hu})_{i3}
\gamma^\mu P_L$\\\hline
$\bar u^i_{H} Z_{H} T_+$ & $\frac{ig}{2}x_L\frac{v}{f}(V_{Hu})_{i3}
\gamma^\mu P_L$\\\hline
$\bar T_- A_H t$ & $ -\frac{2}{5}i
g'\gamma^\mu\left(x_L\frac{v}{f} P_L+\sqrt{x_L}\left(1-\frac{v^2}{f^2}(1-x_L)(\frac{1}{2}-x_L)\right)  P_R\right)$ \\\hline
$\bar T_- Z_H t$ & $ -\frac{2}{5}i
g'x_H\sqrt{x_L}\frac{v^2}{f^2}\gamma^\mu P_R$\\\hline
$\bar T_- A_H T_+$ & $ \frac{2}{5}i
g'\gamma^\mu\left[\left(1-\frac{x_L^2}{2}\frac{v^2}{f^2}\right) P_L+\sqrt{1-x_L}\left(1+\frac{v^2}{f^2}x_L(\frac{1}{2}-x_L)\right)P_R\right]$ \\\hline
$\bar T_- Z_H T_+$ & $ \frac{2}{5}i
g' x_H\frac{v^2}{f^2} \gamma^\mu (P_L+\sqrt{1-x_L}P_R)$ \\\hline
$\bar u^i_{H} W^{+\mu}_{H} d^j$ &  $\frac{ig}{\sqrt{2}}(V_{Hd})_{ij}\gamma^\mu P_L$\\\hline
$\bar d^i_{H} W^{-\mu}_{H} u^j$ &  $\frac{ig}{\sqrt{2}}(V_{Hu})_{ij}\gamma^\mu P_L$\\\hline
$\bar d^i_{H} W^{-\mu}_{H} t$ & $ \frac{ig}{\sqrt{2}}(V_{Hu})_{i3} \left(1-\frac{x_L^2}{2}\frac{v^2}{f^2}\right) \gamma^\mu P_L$\\\hline
$\bar d^i_{H} W^{-\mu}_{H} T_+$&
$\frac{ig}{\sqrt{2}}(V_{Hu})_{i3}x_L\frac{v}{f}\gamma^\mu P_L$\\\hline
$\bar \nu^i_{H} A_{H} \nu^j$ & $\left(\frac{ig'}{10}-\frac{ig}{2}x_H\frac{v^2}{f^2}
\right)(V_{H\nu})_{ij}\gamma^\mu P_L$\\\hline
$\bar \nu^i_{H} Z_{H} \nu^j$ & $\left(\frac{ig}{2}+\frac{ig'}{10}x_H\frac{v^2}{f^2}
\right)(V_{H\nu})_{ij}\gamma^\mu P_L$\\\hline
$\bar \ell^i_{H} A_{H} \ell^j$ & $\left(
  \frac{ig'}{10}+\frac{ig}{2}x_H\frac{v^2}{f^2} \right)(V_{H\ell})_{ij}\gamma^\mu P_L$\\\hline
$\bar \ell^i_{H} Z_{H} \ell^j$ & $\left(-\frac{ig}{2}+\frac{ig'}{10}x_H\frac{v^2}{f^2} \right)(V_{H\ell})_{ij}\gamma^\mu
P_L$\\\hline
$\bar \nu^i_{H} W^{+\mu}_{H} \ell^j$ & $\frac{ig}{\sqrt{2}}(V_{H\ell})_{ij}\gamma^\mu P_L$\\\hline
$\bar \ell^i_{H} W^{-\mu}_{H} \nu^j$ &  $\frac{ig}{\sqrt{2}}(V_{H\nu})_{ij}\gamma^\mu P_L$\\\hline
\end{longtable}
\end{center}

\subsection{Fermion--Goldstone Boson Couplings}

\label{sec:appB2}

The fermion couplings to T-even and T-odd Goldstone bosons can be
derived from the Yukawa interactions \eqref{eq:Ltop},
\eqref{eq:Lup} and \eqref{eq:Ldown} and from the mass term
\eqref{eq:Dirac}. Note that \eqref{eq:Ldown} with
$X=(\Sigma_{33})^{-1/4}$ corresponds to case A in \cite{Chen}. Using
instead $X=(\Sigma_{33}^\dagger)^{1/4}$ (case B in \cite{Chen}) does not modify the
fermion couplings to Goldstone bosons as required by gauge invariance.

As these terms have again to be expressed in mass eigenstates of
the particles involved, we have to take into account the
$\ord(v^2/f^2)$ mixing of the Goldstone bosons and scalars
\cite{mH}. Following the steps of \cite{mH}, we find the mass
eigenstates, denoted here with a subscript ``$m$'', to be
\bea
\pi^0_m &=& \pi^0\left( 1-\frac{v^2}{12f^2} \right)\,, \\
\pi^\pm_m &=&  \pi^\pm\left( 1-\frac{v^2}{12f^2} \right)\,,\\
h_m &=& h\,, \\
\phi^0_m &=& \phi^0\left( 1-\frac{v^2}{12f^2} \right) \,,\\
\phi^P_m &=& \phi^P -
\left(\sqrt{10}\eta-\sqrt{2}\omega^0+\phi^P\right)\frac{v^2}{12f^2}\,, \\
\phi^\pm_m &=& \phi^\pm\left(1-\frac{v^2}{24f^2}\right)\mp
i\omega^\pm\frac{v^2}{12f^2} \,,\\
\phi^{++}_m &=& \phi^{++} \,,\\
\eta_m &=& \eta - \frac{5 g' \eta -4
  \sqrt{5}[g'(\omega^0+\sqrt{2}\phi^P)-6g x_H \omega^0]}{24
  g'}\frac{v^2}{f^2}\,, \\
\omega^0_m &=& \omega^0
-\frac{5g(\omega^0+4\sqrt{2}\phi^P)-4\sqrt{5}\eta(5g+6g'x_H)}{120
g}\frac{v^2}{f^2}
\,,\\
\omega^\pm_m &=& \omega^\pm\left(1-\frac{v^2}{24f^2}\right)\mp
  \frac{i}{6}\phi^\pm \frac{v^2}{f^2}\,.
\eea
In what  follows, we will drop the subscript ``$m$'', as it
is clear that all Feynman rules are given in terms of mass
eigenstates.

Note that the mixing of Goldstone bosons and scalars in question affect the corresponding Feynman rules at
$\ord(v^2/f^2)$, thus without considering it, one would not be
able to get the right $\ord(v^2/f^2)$ corrections. Also the
$\ord(v^2/f^2)$ corrections to the particle masses have to be
taken into account.

We would like to caution the reader that this mixing has
not been taken into account in the derivation of the Feynman rules
for $\Phi$ given in \cite{LHTphen}.

\begin{center}
\begin{longtable}{|c|c|} \hline
\multicolumn{2}{|c|}{\bf Fermion couplings to SM Goldstone
bosons}\\\hline\hline $\bar u^i \pi^+ d^j$ &
$\frac{g}{\sqrt{2}M_{W_{L}}}\left(m^i_u P_L -m^j_d P_R\right)
(V_\text{CKM})_{ij}$\\\hline $\bar t \pi^+ d^j$ &
$\frac{g}{\sqrt{2}M_{W_{L}}}\left(1-\frac{x_L^2}{2}\frac{v^2}{f^2}\right)\left(m_t
P_L
  -m^j_d  P_R\right)(V_\text{CKM})_{tj}$\\\hline
$\bar T_+ \pi^+ d^j$ &
$\frac{gx_L}{\sqrt{2}M_{W_{L}}}\frac{v}{f}\left[m_{T_+}\left(1+\frac{v^2}{f^2}d_2\right) P_L
  -m^j_d  P_R\right] (V_\text{CKM})_{tj}$\\\hline
$\bar u^i \pi^0 u^j$&
$-\frac{gm^i_u}{{2}M_{Z_{L}}\cos\theta_W}\gamma_5\delta_{ij}$\\\hline
$\bar t \pi^0 t$ & $- \frac{g m_t
}{{2}M_{Z_{L}}\cos\theta_W}\left(1-x_L^2\frac{v^2}{f^2}\right)\gamma_5$\\\hline
$\bar T_+ \pi^0 T_+$ & $- \frac{g
m_{T_+}}{{2}M_{Z_{L}}\cos\theta_W}x_L^2\frac{v^2}{f^2}
\gamma_5$\\\hline $\bar T_+ \pi^0 t$ & $ \frac{g
x_L}{{2}M_{Z_{L}}{\cos\theta_W}}\frac{v}{f}
\left[m_{T_+}\left(1+\frac{v^2}{f^2}\left(d_2-\frac{x_L^2}{2}\right)\right)P_L-m_t
P_R \right]$\\\hline $\bar d^i \pi^0 d^j$ &
$\frac{gm^i_d}{{2}M_{Z_{L}}\cos\theta_W}\gamma_5\delta_{ij}$
\\\hline
$\bar u^i_{H} \pi^+ d^j_{H}$ & $
-\frac{g}{8\sqrt{2}M_{W_L}}\frac{v^2}{f^2}m^d_{Hi} \delta_{ij}$
\\\hline
$\bar \nu^i \pi^+ \ell^j$ &
$\frac{g}{\sqrt{2}M_{W_{L}}}\left(m^i_\nu P_L -m^j_\ell P_R\right)
(V_\text{PMNS})_{ij}$\\\hline $\bar \nu^i \pi^0 \nu^j$ & $
-\frac{g
m^i_\nu}{{2}M_{Z_{L}}\cos\theta_W}\gamma_5\delta_{ij}$\\\hline
$\bar \ell^i \pi^0 \ell^j$ & $\frac{g m^i_\ell
}{{2}M_{Z_{L}}\cos\theta_W}\gamma_5\delta_{ij}$
\\\hline $\bar \nu^i_{H} \pi^+ \ell^j_{H}$ & $
-\frac{g}{8\sqrt{2}M_{W_L}}\frac{v^2}{f^2}m^\ell_{Hi} \delta_{ij}$
\\\hline
\end{longtable}
\end{center}

The leading order contributions to the couplings of fermions to
the heavy Goldstone bosons have already been given in
\cite{Hubisz}. We included also $\ord(v^2/f^2)$ corrections,
necessary for the calculation of rare decays, and the
contributions proportional to the SM fermion masses. We note that
our Feynman rules for these vertices differ by sign from the ones
given in \cite{Hubisz}, as our sign conventions for the mass and
Yukawa terms \eqref{eq:Dirac}, \eqref{eq:Ltop}, \eqref{eq:Lup} and \eqref{eq:Ldown} are chosen such that we obtain
positive and real-valued masses, as also done in \cite{Chen}.

\begin{center}
\begin{longtable}{|c|c|}
\hline \multicolumn{2}{|c|}{\bf Fermion couplings to heavy
Goldstone bosons}\\\hline\hline $\bar u^i_{H} \omega^+ d^j$ &
$\frac{g}{\sqrt{2}M_{W_H}} \left(m^u_{Hi} P_L-m^j_d P_R \right)
(V_{Hd})_{ij}$\\\hline $\bar u^i_{H} \omega^0 u^j$ & $
\frac{g}{{2}M_{Z_H}} \left[m^u_{Hi}
\left(1+\frac{v^2}{f^2}\left(\frac{1}{8}-\frac{x_H}{\tan\theta_W}\right)\right)P_L
-m^j_u P_R \right] (V_{Hu})_{ij}$\\\hline $\bar u^i_{H} \omega^0
t$ & $ \frac{g}{{2}M_{Z_H}} \left[ m^u_{Hi}
\left(1+\frac{v^2}{f^2}\left(\frac{1}{8}-\frac{x_H}{\tan\theta_W}-\frac{x_L^2}{2}\right)\right)P_L
-  m_t P_R \right] (V_{Hu})_{i3}$\\\hline $\bar u^i_{H} \omega^0
T_+$ & $\frac{gx_L}{{2}M_{Z_H}} \frac{v}{f} \left( m^u_{Hi} P_L-
m_{T_+}  P_R \right) (V_{Hu})_{i3}$\\\hline $\bar T_- \omega^0 t$
& $\frac{g m_t v}{2 M_{Z_H} f}P_R$\\\hline $\bar T_- \omega^0 T_+$
& $\frac{g  m_{T_+} v^2}{2
 M_{Z_H} f^2}\left( 1-\frac{4x_H}{\tan\theta_W} \right)P_R$\\\hline
$\bar u^i_{H} \eta u^j$ & $-\frac{g'}{{10}M_{A_H}}
\left[m^u_{Hi}\left(1+\frac{v^2}{f^2}\left(\frac{5}{8}+x_H\tan\theta_W\right)\right)P_L
-m^j_uP_R\right](V_{Hu})_{ij}$
\\\hline
$\bar u^i_{H} \eta t$ & $-\frac{g'}{{10}M_{A_H}} \left[
  m^u_{Hi}  \left(1+\frac{v^2}{f^2}\left(\frac{5}{8}+x_H\tan\theta_W-\frac{x_L^2}{2}\right)\right) P_L
  -{m_t} P_R\right] (V_{Hu})_{i3}$ \\\hline
$\bar u^i_{H} \eta T_+$ & $-\frac{g'x_L}{{10}M_{A_H}}\frac{v}{f}
\left(  m^u_{Hi}P_L
  -m_{T_+} P_R\right)(V_{Hu})_{i3}$ \\\hline
$\bar T_- \eta t$ & $ -\frac{2 g'
m_t}{{5}M_{A_H}}\frac{f}{v}\left(1-\frac{v^2}{f^2}\left(\frac{x_L^2}{2}+\frac{1}{6}\right)\right)P_R$\\\hline
$\bar T_- \eta T_+$ & $ -\frac{2 g' x_L
m_{T_+}}{{5}M_{A_H}}\left(1-\frac{v^2}{f^2}\left(\frac{3}{2}x_L^2-2{x_L}+1\right)\right)P_R$\\\hline
$\bar d^i_{H} \omega^- u^j$ & $\frac{g}{\sqrt{2}M_{W_H}}
\left[m^d_{Hi}\left(1-\frac{v^2}{8f^2}\right)
  P_L-m^j_u P_R\right] (V_{Hu})_{ij}$\\\hline
$\bar d^i_{H} \omega^- t$ & $ \frac{g}{\sqrt{2}M_{W_H}} \left[
m^d_{Hi}\left(1-\frac{v^2}{f^2}\left(\frac{1}{8}+\frac{x_L^2}{2}\right)\right)
  P_L-{m_t} P_R\right] (V_{Hu})_{i3}$  \\\hline
$\bar d^i_{H} \omega^- T_+$ & $
\frac{g}{\sqrt{2}M_{W_H}}x_L\frac{v}{f} \left( m^d_{Hi} P_L-
m_{T_+} P_R\right)  (V_{Hu})_{i3}$  \\\hline $\bar d^i_{H}
\omega^0 d^j$ & $- \frac{g}{{2}M_{Z_H}}
\left[m^d_{Hi}\left(1+\frac{v^2}{f^2}\left(-\frac{1}{4}+\frac{x_H}{\tan\theta_W}\right)\right)
P_L-m^j_d P_R\right](V_{Hd})_{ij} $\\\hline $\bar d^i_{H} \eta
d^j$ & $ - \frac{g'}{{10}M_{A_H}} \left[m^d_{Hi}
\left(1-\frac{v^2}{f^2}\left(\frac{5}{4}+x_H\tan\theta_W\right)\right)
P_L -m^j_d P_R\right](V_{Hd})_{ij}$\\\hline $\bar \nu^i_{H}
\omega^+ \ell^j$ & $\frac{g}{\sqrt{2}M_{W_H}} \left(m^\nu_{Hi}
P_L-m^j_\ell P_R \right) (V_{H\ell})_{ij}$\\\hline $\bar \nu^i_{H}
\omega^0\nu^j$ &$ \frac{g}{{2}M_{Z_H}} \left[ m^\nu_{Hi}
\left(1+\frac{v^2}{f^2}\left(\frac{1}{8}-\frac{x_H}{\tan\theta_W}\right)\right)P_L
- m^j_\nu P_R \right] (V_{H\nu})_{ij}$\\\hline $\bar \nu^i_{H}
\eta \nu^j$ & $-\frac{g'}{{10}M_{A_H}}
\left[m^\nu_{Hi}\left(1+\frac{v^2}{f^2}\left(\frac{5}{8}+x_H\tan\theta_W\right)\right)P_L
-m^j_\nu P_R\right](V_{H\nu})_{ij}$ \\\hline $\bar \ell^i_{H}
\omega^- \nu^j$ & $\frac{g}{\sqrt{2}M_{W_H}}
\left[m^\ell_{Hi}\left(1-\frac{v^2}{8f^2}\right)
  P_L-m^j_\nu P_R\right] (V_{H\nu})_{ij}$\\\hline
$\bar \ell^i_{H} \omega^0 \ell^j$ & $- \frac{g}{{2}M_{Z_H}}
\left[m^\ell_{Hi}\left(1+\frac{v^2}{f^2}\left(-\frac{1}{4}+\frac{x_H}{\tan\theta_W}\right)\right)
P_L-m^j_\ell P_R\right](V_{H\ell})_{ij}$ \\\hline $\bar \ell^i_{H}
\eta \ell^j$ & $ - \frac{g'}{{10}M_{A_H}} \left[m^\ell_{Hi}
\left(1-\frac{v^2}{f^2}\left(\frac{5}{4}+x_H\tan\theta_W\right)\right)
P_L  -m^j_\ell P_R\right](V_{H\ell})_{ij}$\\\hline
\end{longtable}
\end{center}

\subsection{Triple Gauge Boson Couplings}

The self-interactions of the gauge bosons can be obtained, as usual,
from the gauge boson kinetic term
\be
\mathcal{L}_\text{gauge}=-\frac{1}{2}\text{Tr}(F^1_{\mu\nu}F^{1\mu\nu})-\frac{1}{2}\text{Tr}(F^2_{\mu\nu}F^{2\mu\nu})-\frac{1}{4}B^1_{\mu\nu}B^{1\mu\nu}-\frac{1}{4}B^2_{\mu\nu}B^{2\mu\nu}
\ee
by inserting the mass eigenstates of the gauge boson fields. Here,
$F^i_{\mu\nu}$ and $B^i_{\mu\nu}$ denote the field strength tensors of
the gauge groups $SU(2)_i$ and $U(1)_i$ $(i=1,2)$, respectively.

The momenta $k$, $p$ and $q$ are defined to be incoming.

\begin{center}
\begin{tabular}{|c|c|} \hline
\multicolumn{2}{|c|}{\bf Gauge boson self-interactions}\\\hline\hline
$W^{+\mu}_{L}(k)W^{-\nu}_{L}(p)Z^\rho_{L}(q)$ & $i g
\cos{\theta_W}\left[
  g^{\mu\nu}(k-p)^\rho + g^{\nu\rho}(p-q)^\mu + g^{\rho\mu}(q-k)^\nu
\right]$\\\hline
$W^{+\mu}_{L}(k)W^{-\nu}_{L}(p)A^\rho_{L}(q)$ & $i e\left[
  g^{\mu\nu}(k-p)^\rho + g^{\nu\rho}(p-q)^\mu + g^{\rho\mu}(q-k)^\nu
\right]$\\\hline
$W^{+\mu}_{H}(k)W^{-\nu}_{L}(p)Z^\rho_{H}(q)$ & $ i g \left[
  g^{\mu\nu}(k-p)^\rho + g^{\nu\rho}(p-q)^\mu + g^{\rho\mu}(q-k)^\nu
\right]$\\\hline
$W^{+\mu}_{H}(k)W^{-\nu}_{L}(p)A^\rho_{H}(q)$ & $i g x_H\frac{v^2}{f^2}\left[
  g^{\mu\nu}(k-p)^\rho + g^{\nu\rho}(p-q)^\mu + g^{\rho\mu}(q-k)^\nu
\right]$\\\hline
$W^{+\mu}_{H}(k)W^{-\nu}_{H}(p)Z^\rho_{L}(q)$ & $i g \cos{\theta_W}\left[
  g^{\mu\nu}(k-p)^\rho + g^{\nu\rho}(p-q)^\mu + g^{\rho\mu}(q-k)^\nu
\right]$\\\hline
$W^{+\mu}_{H}(k)W^{-\nu}_{H}(p)A^\rho_{L}(q)$  & $ i e\left[
  g^{\mu\nu}(k-p)^\rho + g^{\nu\rho}(p-q)^\mu + g^{\rho\mu}(q-k)^\nu
\right]$\\\hline
\end{tabular}
\end{center}

\subsection{Triple Gauge Boson--Goldstone Boson Couplings}

The kinetic term for the non-linear sigma model field $\Sigma$ is given by
\be\label{eq:Sigmakin}
\mathcal{L}=
\frac{f^2}{8}\text{Tr}\left[\left(D_\mu \Sigma\right)^\dagger
\left(D^\mu \Sigma\right)\right]\,,
\ee
where the covariant
derivative is defined through
\be
D_\mu \Sigma=\partial_\mu \Sigma
-\sqrt{2} i \sum_{j=1}^2\left[gW_{j \mu}^a\left(Q_j^a\Sigma+\Sigma
  Q_j^{aT}\right)+g'B_{j \mu}\left(Y_j\Sigma + \Sigma Y_j^T\right)\right].
\ee
From this term the interactions of the Goldstone boson fields with the
SM and heavy gauge bosons can be obtained. The mixing of the Goldstone
boson and scalar fields, as described in Appendix \ref{sec:appB2}, has
also to be taken into account.

All momenta are again defined to be incoming.

\begin{center}
\begin{tabular}{|c|c|} \hline
\multicolumn{2}{|c|}{\bf Gauge boson--Goldstone boson
interactions}\\\hline\hline $W_L^{+\mu} W_H^{-\nu} \eta$ & $
\frac{g}{4}M_{A_H}\left(\frac{5}{\tan\theta_W}+4
x_H\right)\frac{v^2}{f^2} g^{\mu\nu}$\\\hline $W_L^{+\mu}
W_H^{-\nu} \omega^0$ &$-g M_{Z_H}
\left(1-\frac{v^2}{4f^2}\right)g^{\mu\nu}$ \\\hline $W_H^{+\mu}
Z_L^{\nu} \omega^-$ & $- g M_{W_H} \cos{\theta_W} \left(1 -
\frac{v^{2}}{4 f^{2} \cos^{2}{\theta_W}}\right)
g^{\mu\nu}$\\\hline $W_H^{+\mu} A_L^{\nu} \omega^-$ & $- e M_{W_H}
g^{\mu\nu}$\\\hline $W_L^{+\mu} Z_H^{\nu} \omega^-$ & $g M_{W_H}
\left(1-\frac{v^2}{4f^2}\right)g^{\mu\nu}$ \\\hline $W_L^{+\mu}
A_H^{\nu} \omega^-$ & $\frac{ g  M_{W_H} v^{2}}{f^2
}\left(\frac{\tan{\theta_W}}{4}-x_H\right) g^{\mu\nu}$ \\\hline
$W_L^{+\mu} Z_L^{\nu} \pi^-$ & ${g
M_{W_{L}}}\frac{\sin^2{\theta_W}}{\cos\theta_W}
g^{\mu\nu}$\\\hline $W_L^{+\mu} A_L^{\nu} \pi^-$ & $-{e M_{W_{L}}}
g^{\mu\nu}$\\\hline $W_H^{+\mu} A_H^{\nu} \pi^-$ & $ g^{\prime}
M_{W_{L}} g^{\mu\nu}$\\\hline $\omega^+(p)\omega^-(q)Z_L$ & $ i g
\cos{\theta_W} \left(1 - \frac{v^{2}}{8 f^{2}
\cos^{2}{\theta_W}}\right)\left(p - q\right)^{\mu}$\\\hline
$\omega^+(p)\omega^-(q)A_L$ & $ i e\left(p -
q\right)^{\mu}$\\\hline $\pi^+(p)\pi^-(q)Z_L$ & $ \frac{i g
\left(1 - 2 \sin^2{\theta_W}\right)}{2 \cos{\theta_W}} \left(p -
q\right)^{\mu}$\\\hline $\pi^+(p)\pi^-(q)A_L$ & $ i e \left(p -
q\right)^{\mu}$\\\hline $\omega^+(p)\pi^-(q)A_H$ &
$-i\frac{g'v}{3f} \left(p - q\right)^{\mu} $ \\\hline
$\omega^+(p)\omega^0(q)W_L^{-\mu}$ & $-i
{g}\left(1-\frac{v^2}{8f^2}\right)(p-q)^\mu$ \\\hline
$\omega^+(p)\eta(q)W_L^{-\mu}$ & $i
\frac{25g+24g'x_H}{24\sqrt{5}}\frac{v^2}{f^2}(p-q)^\mu$\\\hline
$\pi^+(p)\omega^0(q)W_H^{-\mu}$ &$-i\frac{g}{2}\frac{v}{f}
(p-q)^\mu$ \\\hline $\pi^+(p)\eta(q)W_H^{-\mu}$
&$-i\frac{\sqrt{5}g}{6}\frac{v}{f} (p-q)^\mu$ \\\hline
$\pi^+(p)\pi^0(q)W_L^{-\mu}$ & $-i \frac{g}{2} (p-q)^\mu$ \\\hline
\end{tabular}
\end{center}

\newsection{Relevant Functions}\label{sec:appC}

In this Appendix we list the functions that entered the present
study of rare and CP-violating $K$ and $B$ decays. Both the SM
contributions and the new physics contributions coming from the
T-even and T-odd sectors are collected. The variables are defined
as follows: \begin{gather}
x_q=\dfrac{m_q^2}{M_{W_L}^2}\,,\qquad x_T=\dfrac{m_{T_+}^2}{M_{W_L}^2}\,\qquad(q=c,t)\,,\\
z_i=\dfrac{m_{Hi}^2}{M_{W_H}^2}\,,\quad
z'_i=\dfrac{m_{Hi}^2}{M_{A_H}^2}=z_i\,a \quad \text{with}\quad
a=\dfrac{5}{\tan^2 \theta_W}\,,\qquad(i=1,2,3)\,,\\
y = \frac{m_{H \ell}^{2}}{M_{W_{H}}^{2}} =
\frac{m_{H \ell}^{2}}{M_{Z_{H}}^{2}}\,, \qquad y^{\prime}= y a\,,
\qquad \eta = \frac{1}{a}\,. 
\end{gather}

\bea\label{X0}
X_\text{SM}(x_t)&=&\frac{x_t}{8}\;\left[\frac{x_t+2}{x_t-1}
+\frac{3 x_t-6}{(x_t -1)^2}\; \log x_t\right],\\
\label{Y0}
Y_\text{SM}(x_t)&=&\frac{x_t}{8}\; \left[\frac{x_t -4}{x_t-1}
+ \frac{3 x_t}{(x_t -1)^2} \log x_t\right],\\
Z_\text{SM}(x_t)&=&-\frac{1}{9}\log x_t+\frac{18 x_t^4-163x_t^3+259
  x_t^2-108x_t}{144(x_t-1)^3}\nn\\
&&+\frac{32x_t^4-38x_t^3-15x_t^2+18x_t}{72(x_t-1)^4}\log x_t\,.
\eea

\bea
C_0(y)&=&\frac{y}{8}\left[\frac{y-6}{y-1}+\frac{3y+2}{(y-1)^2}\log y\right],\\
D_0(y)&=&-\frac{4}{9}\log
y+\frac{-19y^3+25y^2}{36(y-1)^3}+\frac{y^2(5y^2-2y-6)}{18(y-1)^4}\log
y\,,\\
E_0(y)&=&-\frac{2}{3}\log y + \frac{y^2(15-16y+4y^2)}{6(y-1)^4}\log y
+ \frac{y(18-11y-y^2)}{12(1-y)^3}\,,\\
 D'_0(y)&=&-\dfrac{(3y^3-2y^2)}{2(y-1)^4}\log y +
\dfrac{(8y^3+5y^2-7y)}{12(y-1)^3}\,,\\
E'_0(y)&=&\dfrac{3y^2}{2(y-1)^4}\log y +
\dfrac{(y^3-5y^2-2y)}{4(y-1)^3}\,.
\eea

\bea
U_3(x_t,x_T)&=&\frac{-3+2x_t-2x_t^2}{8(-1+x_t)}-\frac{x_t\left(-4-x_t+2x_t^2\right)\log x_t}{8(-1+x_t)^2}+\frac{\left(3+2x_t\right)\log x_T}{8}\,,\\
V_3(x_t,x_T)&=&\frac{\left(3+2x_t-2x_t^2\right)}{8(-1+x_t)}-\frac{x_t\left(2-x_t+2x_t^2\right)\log x_t}{8(-1+x_t)^2}+\frac{\left(3+2x_t\right)\log x_T}{8}\,.
\eea

\bea
(D')_\text{SM}&=&D'_0(x_t)\,,\\
(D')_\text{LHT}&=&(D')_\text{even}+\frac{1}{\lambda_t^{(s)}}T_{D'}^\text{odd}\,,\\
(D')_\text{even}&=&D'_0(x_t)+\frac{v^2}{f^2}x_L^2\Big[D'_0(x_T)-D'_0(x_t)\Big]\,,\\
T_{D'}^\text{odd}&=&\dfrac{1}{4}\dfrac{v^2}{f^2}\left[\xi_2^{(s)}
\big(D'_\text{odd}(z_2)-D'_\text{odd}(z_1)\big)+\xi_3^{(s)}
\big(D'_\text{odd}(z_3)-D'_\text{odd}(z_1)\big)\right]\,,\label{eq:TDodd}\\
D'_\text{odd}(z_i)&=&D_0'(z_i)-\dfrac{1}{6}E_0'(z_i)-\dfrac{1}{30}E_0'(z_i')\,,\label{eq:Dodd}
\eea

\bea
R_2(z_i) &=& -\left[\frac{z_i \log z_i}{(1-z_i)^2}+\frac{1}{1-z_i}\right]\,,\\
F_2(z_i) &=& -\frac{1}{2}\left[\frac{z_i^2 \log
    z_i}{(1-z_i)^2}+\frac{1}{1-z_i}\right]\,.
\eea

\begin{eqnarray}
F^{\nu \bar{\nu}}\left(z_{i}, y; W_{H}\right) &=& \frac{3}{2} z_{i} - F_{5}\left(z_{i}, y\right) - 7 F_{6}\left(z_{i}, y\right) - 9 U\left(z_{i}, y\right)\,, \\
F^{\mu \bar\mu}\left(z_{i}, y; W_{H}\right) &=& \frac{3}{2} z_{i} - F_{5}\left(z_{i}, y\right)- 7 F_{6}\left(z_{i}, y\right) + 3 U\left(z_{i}, y\right)\,.
\end{eqnarray}

\begin{eqnarray}
F_{5}\left(z_{i}, y\right) &=& \frac{z_{i}^{3} \log z_{i}}{\left(1-z_{i}\right) \left(y-z_{i}\right)} + \frac{y^{3} \log y}{\left(1-y\right) \left(z_{i}-y\right)}\,, \\
F_{6}\left(z_{i}, y\right) &=& -\left[\frac{z_{i}^{2} \log z_{i}}{\left(1-z_{i}\right) \left(y-z_{i}\right)} + \frac{y^{2} \log y}{\left(1-y\right) \left(z_{i}-y\right)}\right]\,, \\
U\left(z_{i}, y\right) &=& \frac{z_{i}^{2} \log z_{i}}{\left(z_{i}-y\right) \left(1-z_{i}\right)^{2}} + \frac{y^{2} \log y}{\left(y-z_{i}\right) \left(1-y\right)^{2}} + \frac{1}{\left(1-z_{i}\right) \left(1-y\right)}\,.
\end{eqnarray}

\begin{eqnarray}
G\left(z_{i}, y; Z_{H}\right) &=&  -\frac{3}{4} U\left(z, y\right) \,,\\
G_{1}\left(z_{i}^{\prime},y_{i}^{\prime}; A_{H}\right) &=& \frac{1}{25 a} G\left(z_{i}^{\prime}, y_{i}^{\prime}; Z_{H}\right) \,,\\
G_{2}\left(z_{i}, y; \eta\right) &=& -\frac{3}{10 a} \left[\frac{z_{i}^{2} \log z_{i}}{\left(1-z_{i}\right) \left(\eta-z_{i}\right) \left(z_{i}-y\right)} \right.\nn \\
&& +  \left. \frac{y^{2} \log y}{\left(1-y\right) \left(\eta-y\right) \left(y-z_{i}\right)} + \frac{\eta^{2} \log \eta}{\left(1-\eta\right) \left(z_{i}-\eta\right) \left(\eta-y\right)}\right]\,.
\end{eqnarray}

\end{appendix}

\end{document}